\documentclass[preprint,12pt]{elsarticle}

\usepackage{amssymb}

\usepackage{siunitx}
\usepackage {./Figures/graphicx}
\usepackage{algorithm}
\usepackage{amsfonts}
\usepackage{amsmath}
\usepackage{wasysym}
\usepackage{algpseudocode}
\usepackage{ragged2e}
\usepackage{subfig}

\journal{(and accepted by) Chaos, Solitons and Fractals.}

\begin{document}

\begin{frontmatter}



\title{On Long-Term Species Coexistence in Five-Species Evolutionary Spatial Cyclic Games with Ablated and Non-Ablated Dominance Networks  }


\author{Dave Cliff} 

\affiliation{organization={School of Engineering Mathematics \& Technology},
            addressline={University of Bristol}, 
            postcode={BS8 1UB}, 
            country={U.K.}}

\begin{abstract}
I present a replication and, to some extent, a refutation of key results published by Zhong, Zhang, Li, Dai, \& Yang in their 2022 paper ``Species coexistence in spatial cyclic game of five species'' ({\em Chaos, Solitons and Fractals}, 156: 111806), where ecosystem species coexistence was explored via simulation studies of the evolutionary spatial cyclic game ({\sc Escg}) Rock-Paper-Scissors-Lizard-Spock ({\sc Rpsls}) with certain predator-prey relationships removed from the game's ``interaction structure'', i.e.\ with specific arcs ablated in the {\sc Escg}'s dominance network, and with the {\sc Escg} run for $10^5$ Monte Carlo Steps ({\sc mcs}) to identify its asymptotic behaviors. I replicate the results presented by Zhong et al.\ for interaction structures with one, two, three, and four arcs ablated from the dominance network. I then empirically demonstrate that the dynamics of the {\sc Rpsls} {\sc Escg} have sufficiently long time constants that the true asymptotic outcomes can often only be identified after running the ablated {\sc Escg} for $10^7${\sc mcs} or longer, and that the true long-term outcomes can be markedly less diverse than those reported by Zhong et al.\ as asymptotic.  Finally I demonstrate that, when run for sufficiently many {\sc mcs}, the original unablated {\sc Rpsls} system exhibits essentially the same asymptotic outcomes as the ablated {\sc Rpsls} systems, and in this sense the only causal effect of the ablations is to alter the time required for the system to converge to the long-term asymptotic states that the unablated system eventually settles to anyhow. 
\end{abstract}


\end{frontmatter}

\section{Introduction}
\label{sec:intro}

In 2022 a research paper by Zhong, Zhang, Li, Dai, \& Yang titled ``Species coexistence in spatial cyclic game of five species'' was published in this journal ({\em Chaos, Solitons and Fractals} 156: 111806) \cite{zhong_etal_2022_ablatedRPSLS}. In their paper, Zhong et al.\ explored the frequency distribution of outcomes from simulations studies of a minimal model of  ecosystem species coexistence, based on the evolutionary spatial cyclic game ({\sc Escg}) Rock-Paper-Scissors-Lizard-Spock ({\sc Rpsls}) with certain predator-prey relationships removed from the game's ``interaction structure'', i.e.\ with specific arcs ablated in the {\sc Escg}'s dominance network.\footnote{The terminology used in the literature varies: some authors such as \cite{laird_schamp_2015} refer to the dominance network as the {\em interaction graph}, others such as \cite{zhang_bearup_guo_zhang_liao_2022} refer to it instead as the {\em competitive matrix}.} Zhong et al.\ identified the asymptotic behaviors of the ablated  {\sc Escg} by running simulations for $10^5$ Monte Carlo Steps ({\sc mcs}: a major time-step in the simulation process, explained further in Section~\ref{sec:backgnd}, below). In this paper I replicate the results presented by Zhong et al.\ for interaction structures with one, two, three, and four arcs ablated from the dominance network. I then empirically demonstrate that the dynamics of the ablated {\sc Rpsls} {\sc Escg} have sufficiently long time constants that the true asymptotic outcomes can often only be identified after simulations running for $10^7${\sc mcs} or longer (i.e., $100\times$ the durations used by Zhong et al.), and that the true long-term outcomes can be markedly different from those reported by Zhong et al.\ as asymptotic.  Following this, I demonstrate that, when run for sufficiently many {\sc mcs}, the original {\em unablated} {\sc Rpsls} system exhibits almost exactly the same asymptotic outcomes as the ablated {\sc Rpsls} systems, and hence in this sense the only causal effect of the dominance-network arc  ablations is to reduce the time required for the system to converge on the long-term asymptotic states that the unablated system eventually settles to anyhow. 

Section~\ref{sec:backgnd} gives further details of the  background to this work, the text of that section being reproduced essentially verbatim from \cite{cliff_2024_noops}.
Section~\ref{sec:zhong_summary} then gives a detailed summary of the model and simulation methods as used by Zhong et al.\ in  \cite{zhong_etal_2022_ablatedRPSLS}.  After that, Section~\ref{sec:zhong_replication} shows visualization and analysis of results from my own simulation experiments which accurately replicate Zhong et al.'s results presented in \cite{zhong_etal_2022_ablatedRPSLS}. The next two sections then present a detailed critique of those results: Section~\ref{sec:howlong} argues that the observations made and conclusions drawn by Zhong et al.\ are based on data from experiments that had not been run for sufficiently many time-steps, and that some features in the graphs plotted by Zhong et al., features they highlighted as noteworthy, are in fact nothing more than non-asymptotic artefacts from decaying transients in the system dynamics which appear because the experiments have not been run for long enough; Section~\ref{sec:NA0} then goes on to present results from new simulation experiments, not reported by Zhong et al., where the original unablated {\sc Rpsls} dominance network is used, to give a proper baseline reference that the results from the ablated networks can be compared to: somewhat surprisingly, the original unablated {\sc Rpsls} system exhibits what are, qualitatively, essentially identical asymptotic behaviors to those of the ablated one-, two, and three-ablation {\sc Rpsls} systems reported by Zhong et al. Finally, in Section~\ref{sec:Z4}, results from the four-ablation system explored by Zhong et al.\ are discussed separately, because they seem not to show the same long transients as the one-, two-, and three-ablation cases, and in that sense the four-ablation results remain something of a mystery. All of these results are collectively discussed further in Section~\ref{sec:discussion}, and conclusions are drawn in Section~\ref{sec:conclusion}. 

For completeness, the appendices show numerous plots of data that more fully illustrate the points made in my critique but which did not fit naturally into the narrative of Sections~\ref{sec:howlong}  and~\ref{sec:NA0}. Python source-code developed for the simulation experiments reported here is being made freely available via the MIT Open-Source License, for download from Github.\footnote{See {\tt https://github.com/davecliff/ESCG\_Python}.} The data generated for this paper, in uncompressed {\sc csv}-format files, occupies approximately two terabytes of storage: a small amount of illustrative sample data is being made available on the GitHub source-code repository.

\section{Background}
\label{sec:backgnd}

There is a well-established body of peer-reviewed research literature which explores issues in ecosystems stability, biodiversity, and co-evolutionary dynamics via computationally intensive simulations of minimally simple models of multiple interacting biological species. Landmark papers in this field were published in 2007--09 by  Reichenbach, Mobilia, and Frey \cite{reichenbach_mobilia_frey_2007_nature,reichenbach_mobilia_frey_2007_physrevlet, reichenbach_mobilia_frey_2008_jtb} 
and by Laird \& Schamp \cite{laird_schamp_2006_RPSLS,laird_schamp_2008_RPSLS,laird_schamp_2009_coexistence_RPSLS}, who extended the previous non-spatial model of May and Leonard \cite{may_leonard_1975} by modeling each species as a time-varying number of discrete individuals, where at any one time each individual occupies a particular cell in a regular rectangular lattice or grid of cells, and can {\em move} from cell to cell over time --- that is, the individuals are {\em spatially located} and {\em mobile}. Individuals can also, under the right circumstances, {\em reproduce} (asexually, cloning a fresh individual of the same species into an adjacent empty cell on the lattice); and they can also {\em compete} with individuals in neighbouring cells. Different authors use different phrasings to explain the inter-species competition, but it is common to talk in terms of predator-prey dynamics: that is, each species is predator to (i.e., {\em dominates}) some specified set of other species, and is in turn also prey to (i.e., is {\em dominated} by) some set of other species. 

\begin{figure}
\begin{center}
\includegraphics[trim=0cm 0cm 0cm 0cm,clip=true,scale=0.35]{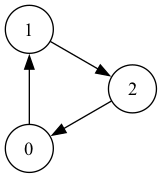}
\end{center}
\caption{Dominance network, a directed graph or {\em digraph}, for the three-species Rock-Paper-Scissors (RPS) game. Species $S_i$ are denoted by nodes numbered by index $i\in\{1, 2, 3\}$, with directed edges running from the dominator (``predator'') species to the dominated (``prey''') species. There are multiple {\em labelings} of this graph (e.g.: $(S_1$$=$$R, S_2$$=$$P, S_3$$=$$S); (S_1$$=$$P, S_2$$=$$S, S_3$$=$$R); \ldots$) but if one graph can be turned into another purely by rearranging the node labels then those two graphs are topologically equivalent or {\em isomorphic}. All possible labelings of the RPS digraph are isomorphic with each other, so there is only one isomorphically unique RPS digraph.}
\label{fig:RPSnet}
\end{figure}

The population dynamics are determined to a large extent by the model's {\em dominance network}, a directed graph (digraph) where each node in the network represents one of the species in the model, and a directed edge (i.e., an arc) from the node for species $S_i$ to the node for species $S_{j: j\neq i}$ denotes that $S_i$ dominates $S_j$. To exhibit interesting long-term dynamics, the dominance network must contain at least one cycle (i.e., a path from some species $S_i$ to some $S_j$ traced by traversing edges in the directions of the arrows, potentially passing through some number of intermediate species' nodes, where at $S_j$ there then exists an as-yet-untraversed edge back to $S_i$). Under the constraint that no species can be both predator and prey to another species at the same time, the smallest dominance network of interest is a minimal three-node cycle, which represents the intransitive dominance hierarchy of the simple hand-gesture game Rock-Paper-Scissors (RPS), as illustrated in Figure~\ref{fig:RPSnet}. In RPS-based models, the three species are $R$ (rock), $P$ (paper), and $S$ (scissors) and when two neighboring individuals compete the rules are as follows: if they are both the same species, the competition is a draw and nothing else happens; but otherwise $R$ kills $S$, $S$ kills $P$, and $P$ kills $R$, with the cell where the killed individual was located being set to empty, denoted by $\emptyset$. Because the individuals in these models are spatially located on a lattice, and because the inter-species competition is determined by having pairs of individuals play RPS-like games with cyclic dominance digraphs, this class of co-evolutionary population dynamics models is often referred to as  {\em evolutionary spatial cyclic games} ({\sc Escg}s). 

{\sc Escg}s are inherently stochastic and to generate rigorous results it is often necessary to simulate {\sc Escg} systems many times, aggregating results over many independent and identically distributed ({\sc iid}) repetitions of the system evolving over time.  At the core of these simulations is the {\em Elementary Step} (ES), in which one or two agents are chosen at random to either compete to the death, or to reproduce, or to move location. 
{\sc Escg} studies typically involve executing trillions of ESs and hence the computational efficiency of the core ES algorithm is a key concern: for further discussion of this, see \cite{cliff_2024_noops}.

Almost all simulations of co-evolutionary population dynamics via {\sc Escg}s are simple discrete-time systems that are technically unchallenging to write a program for, and are strongly reminiscent of -- but not identical to -- cellular automata (see e.g.\ \cite{wolfram_2002_book}). The lattice/grid needs to first be set up, i.e.\ its dimensions and initial conditions at the first time-step need to be specified. Each cell in the grid is either empty, or contains exactly one individual organism, and each individual is a member of exactly one of the model's set of species. If the number of species in the model is denoted by $N_s$, one common style of initialisation is to assign one individual to every cell in the grid, with that individual's species being an equiprobable choice from the set of available species (i.e., choose species $S_i$ with probability $1/N_s; \forall i$). The modeller also needs to specify the dimensionality of the lattice, and its {\em length} (i.e., number of cells) along each dimension. In almost all of the published work  in this field, the lattice is two-dimensional and square, so its extent is defined by a single system hyperparameter: the side-length (conventionally denoted by $L$). The total number of cells in the lattice (conventionally denoted by $N$) is hence $N=L^{2}$. Working with 2D lattices has the advantage that the global state of the system can be readily visualised as a snapshot at time $t$ as a color-coded or gray-shaded 2D image, with each species in the model assigned its own specific color or gray-scale value, and animations can easily be produced visualising the change in the system state over time. 

In the literature on {\sc Escg}s, authors often make the distinction between two scales of time-step in the simulation. At the very core of the simulation process is a loop that iterates over a number of {\em elementary steps} (ESs), the finest grain of time-step; and then some large number of consecutive ESs is counted as what is conventionally referred to as a {\em Monte Carlo Step} ({\sc MCS}).  

In a single ES, one individual cell (denoted $c_i$) is chosen at random, and then one of its immediately neighboring cells (denoted $c_n$) is also chosen at random: in almost all of the literature on 2D lattice model {\sc Escg}s, the set of neighbours is defined as the 4-connected von Neumann neighborhood rather than the 8-connected Moore neighborhood commonly used in cellular automata research (although e.g.\ \cite{laird_schamp_2006_RPSLS,laird_schamp_2008_RPSLS,laird_schamp_2009_coexistence_RPSLS} used the Moore neighborhood in their lattice models), and the work reported here uses von Nuemann. There seems to be no firm convention on whether to use periodic boundary conditions (also known as toroidal wrap-around) or ``walled garden'' no-flux boundary conditions (such that cells at the edges and corners of the lattice have a correspondingly reduced neighbour-count) -- some authors use periodic, others no-flux. The results presented in this paper come from simulations with no-flux boundary conditions. 

In each ES one of three possible actions occurs: {\em competition}, {\em reproduction}, or {\em movement}, and the probabilities of each of these three actions occurring per ES is set by system parameters $\mu, \sigma,$ and $\epsilon$, respectively (this is explained in more precise detail later, in Section~\ref{sec:zhong_summary}). {\em Competition} involves the individuals at $c_i$ and $c_n$ interacting according to the rules of the cyclic game, resulting in either a draw or one of the individuals losing, in which case it is deleted from its cell, replaced by  $\emptyset$; {\em reproduction} occurs when one of $c_i$ or $c_n$ holds $\emptyset$, the empty cell being filled by a new individual of the same species as the nonempty neighbor; and {\em movement} simply swaps the contents of $c_i$ and $c_n$.  

Because, in the original formulation, each ES involves only one of the three possible actions (competition, reproduction, or movement) occurring for a single cell, a Monte Carlo Step ({\sc mcs}) is conventionally defined as a sequence of $N$ consecutive ESs, the rationale being that, on the average, each cell in the lattice will be randomly chosen once per {\sc mcs}, and hence that, again on the average, every cell in the grid has the potential to change once between any two successive {\sc mcs}s. Most published research on this type of model uses {\sc mcs} as the unit of time when plotting time-series graphs illustrating the temporal evolution of the system, and I follow that convention here. Some authors (e.g.\ \cite{avelino_bazeia_losano_menezes_oliveira_2022}) don't refer to {\sc mcs} but instead talk of each sequence of $N$ consecutive ESs in their {\sc Escg} as one new {\em generation}.

In their seminal papers, Reichenbach et al.\ \cite{reichenbach_mobilia_frey_2007_nature,reichenbach_mobilia_frey_2007_physrevlet,reichenbach_mobilia_frey_2008_jtb} studied 2D lattice systems where interspecies competition was via $N_s$$=$$3$ RPS games, with $\mu$$=$$\sigma$$=$$1.0$, and where $L$ ranged from 100  to 500, and they showed and explained how the overall system dynamics result in emergence of one or more temporally and spatially coherent interlocked {\em spiral waves}.
The specific nature of the wave-patterning, i.e. the size and number of spiral waves seen in the system-snapshot images, depended on a {\em mobility} measure $M$$=$$\epsilon/2N$, which is proportional to the expected area of lattice explored by a single agent  per {\sc mcs}. 

In the years since publication of \cite{reichenbach_mobilia_frey_2007_nature,reichenbach_mobilia_frey_2007_physrevlet, reichenbach_mobilia_frey_2008_jtb}, many papers have been published that explore the dynamics of such co-evolutionary spatial RPS models. For examples of recent publications exploring a range of issues in the three-species RPS {\sc Escg}, see: \cite{
nagatani_ichinose_tainaka_2018_RPS,
kabir_tanimoto_2021_RPS,
mohd_park_2021_RPS,
bazeia_bongestab_deoliveira_2022_RPS,
park_2021_RPS,
menezes_batista_rangel_2022_RPS,
menezes_rangel_moura_2022_RPS,
zhang_bearup_guo_zhang_liao_2022_RPS,
menezes_barbalho_2023_RPS,park_jang_2023_RPS}; and \cite{kubyana_landi_hui_2024_RPS}.

\begin{figure}
\begin{center}
\includegraphics[trim=0cm 0cm 0cm 0cm, clip=true, scale=0.5]{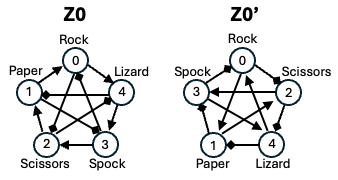}
\end{center}
\caption{
Network Z0 (left) is the {\sc Rpsls} dominance network as presented in Figure 1(a) of  Zhong et al.\ (2022): the outer pentagon subnetwork formed by triangle-headed arrows is referred to by Zhong et al.\ as the {\em spontaneous competition} dominance interactions while the inner pentagram subnetwork formed by diamond-headed arrows is referred to by Zhong et al.\ as the {\em alternative competition} dominance interactions. Network Z0' (right) is topologically equivalent to Z0, despite appearing superficially different. The rules of this game are: 
scissors cut paper;
paper covers rock; 
rock blunts scissors;
scissors decapitates lizard;
lizard eats paper;
paper disproves Spock;
Spock vaporizes rock;
rock crushes lizard;
lizard poisons Spock; and
Spock smashes scissors.
}
\label{fig:RPSLSDomNets}
\end{figure}

More recently, various authors have reported experiments with a closely related system where $N_s$=$5$: this game is known as Rock-Paper-Scissors-Lizard-Spock ({\sc Rpsls}), an extension of RPS introduced by \cite{kass_bryla_1998}  and subsequently featured in a 2012 episode of the popular US TV show {\em Big Bang Theory}. The dominance network for the {\sc Rpsls} game is illustrated in Figure~\ref{fig:RPSLSDomNets} and explained in the caption to that figure. This (and other five-species {\sc Escg}s) was first explored in the theoretical biology literature by \cite{laird_schamp_2006_RPSLS,laird_schamp_2008_RPSLS,laird_schamp_2009_coexistence_RPSLS}; and RPS-like {\sc Escg}s with $N_S$$\geq$$5$ were explored by \cite{avelino_deoliveria_trintin_2022_RPS_bigN}.

The rest of this paper focuses on one recently-published study of the evolutionary dynamics of the {\sc Rpsls} system when one or more of the arcs in the dominance network are ablated, reducing the number of inter-species interactions: this paper is Zhong et al.'s 2022 paper \cite{zhong_etal_2022_ablatedRPSLS}, which is explained in detail in the next section.

\newpage
\clearpage

\section{Summary of Zhong et al.\ (2022)}
\label{sec:zhong_summary}

\subsection{The Evolutionary Spatial Cyclic Game ({\sc Escg})}
\label{sec:escg_defn}

\begin{figure}[h]
\begin{center}
\includegraphics[trim=0cm 0cm 0cm 0cm, clip=true, scale=0.5]{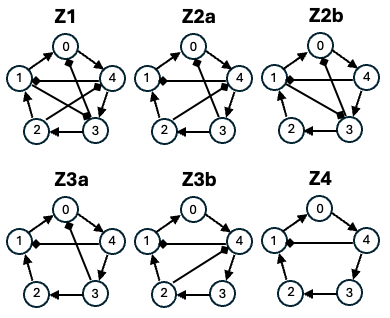}
\end{center}
\caption{
The ablated dominance networks explored in Zhong et al.\ (2022). 
Network Z1 is Zhong et al.'s Figure 1(b); 
Network Z2a is Zhong et al.'s Figure 1(c)-upper;
Network Z2b is Zhong et al.'s Figure 1(c)-lower;
Network Z3a is Zhong et al.'s Figure 1(d)-upper;
Network Z3b is Zhong et al.'s Figure 1(d)-lower;
and Network Z4 is Zhong et al.'s Figure 1(e).
}
\label{fig:ZhongDomNets}
\end{figure}

Figure~\ref{fig:ZhongDomNets} shows the set of ablated {\sc Rpsls} dominance networks explored by Zhong et al.\ in their paper: I have labelled each network with an identifier that includes the number of ablated arcs in that network; and for consistency I refer to the original unablated {\sc Rpsls} dominance network as `Z0'. 

The lattice is an $L$$\times$$L$ square Cartesian grid of cells. The contents of any cell in the lattice can be either the empty cell, denoted here by $\emptyset$, or a natural number $i$ representing that an individual agent of species-type $s_i \in \{0, 1, \dots, N_S-1\}$ occupies that cell.
Let ${\cal C}_{N_S}$ denote the set of all possible cell values in an {\sc Escg} with $N_S$ species: here, for the five-species {\sc Rpsls} game, 
${\cal C}_5=\{  0, 1, 2, 3, 4, \emptyset \}$.

In its simplest form, on each elementary step (ES), an individual cell $c_i$ is chosen at random and then one of its immediately neighboring cells $c_n$ is chosen, also at random: for instance, \cite[p.2]{park_jang_2023_RPS} state ``At each time step, we choose a pair of neighboring sites randomly''. Below, I give an initial definition of the {\sc Escg} on a 2D square lattice as  Algorithm~\ref{alg:game}, which uses this simple random selection of two neighboring cells; then later, in Section~\ref{sec:zhong_replication}, I'll discuss an improvement to this approach which is given in Algorithm~\ref{alg:nonemptycell}.

Let the contents of cell $c_i$ and the contents of cell $c_n$ be denoted by a pair of cell-content values, each in ${\cal C}_5$. The update rules for the {\sc Escg} using the {\sc Rpsls} dominance network $Z0$ (shown in Figure ~\ref{fig:RPSLSDomNets}) can then be expressed as follows: cell-to-cell {\em competition}, which occurs with frequency determined by the {\em competition rate} parameter $\sigma$, is defined by Equations~\ref{eq:RPSLS0} to~\ref{eq:RPSLS4}: 

\begin{align}
(0, s_i) \xrightarrow{\sigma} (0, \emptyset): s_i\in \{2, 4\} \label{eq:RPSLS0}\\
(1, s_i) \xrightarrow{\sigma} (1, \emptyset): s_i \in \{0, 3\} \\
(2, s_i) \xrightarrow{\sigma} (2, \emptyset): s_i \in \{1, 4\} \\
(3, s_i) \xrightarrow{\sigma} (3, \emptyset): s_i \in \{2, 0\} \\
(4, s_i) \xrightarrow{\sigma} (4, \emptyset): s_i \in \{1, 3\} \label{eq:RPSLS4}
\end{align}

{\em Reproduction}, where the individual of species $s_i$ in cell $c_i$ clones a copy of itself into an adjacent empty cell, an event which occurs with frequency determined by the {\em reproduction rate} parameter $\mu$, is as stated in Equation~\ref{eq:reproduction}: 

\begin{equation}
(s_i, \emptyset) \xrightarrow{\mu} (s_i, s_i): s_i \in \{0, \ldots, 4 \} 
\label{eq:reproduction} 
\end{equation}

And finally {\em movement}, where an individual of species $s_i$ in cell $c_i$ swaps location with whatever contents are in its neighbour $c_n$, 
an event which occurs with frequency determined by the {\em movement rate} parameter $\epsilon$, is expressed in Equation~\ref{eq:move}: 

\begin{equation}
(s_i, c_n) \xrightarrow{\epsilon} (c_n , s_i): s_i \in \{0, \ldots, 4\}; c_n \in {\cal C}_5 
\label{eq:move}
\end{equation}

Zhong et al.\ state that on each elementary step, only one of the three possible types of action occurs, and they also state that each of the three actions ``\ldots occur between the two selected nodes at the probabilities  $\sigma, \mu,$ and $\epsilon$, respectively'', but that is not correct: the three parameters $\sigma, \mu,$ and $\epsilon$ are relative rates, not probabilities. 
Strictly, they are each {\em hyperparameters} (i.e., exogenously imposed parameter-values that are constant during the course of an experiment): by convention, $\sigma$ and $\mu$ are constrained to take values in the range $[0.0,1.0]$ but $\epsilon$ is routinely set to values greater than one, and indeed both $\sigma$ and $\mu$ can also be set to values greater than one without loss of coherence. 
As is explained in more depth in \cite{cliff_2024_noops}, these three rate parameters are combined in a ``normalization'' process, not explicitly described in \cite{zhong_etal_2022_ablatedRPSLS} (but clearly explained by other authors e.g.\ \cite{park_jang_2019}), in which each of the three rates are divided by the sum of the three, giving {\em normalized probabilities} $P_\mu, P_\sigma,$ and $P_\epsilon$. That is: $P_\mu = \mu / (\sigma + \mu + \epsilon) \in [0,1], P_\sigma = \sigma / (\sigma + \mu + \epsilon) \in [0,1], P_\epsilon = \epsilon / (\sigma + \mu + \epsilon)\in [0,1]; P_\mu+P_\sigma+P_\epsilon=1.0$.  
Hence $\mu$ is {\em not} the probability of movement being the action chosen on any one ES, because the actual probability of choosing {\sc Move} as the next action is given by $P_\mu$ which depends  not only on $\mu$ but also on $\sigma$ and $\epsilon$.  As is common in the literature for this field of study, in all of the experiments reported by Zhong et al., and also in all of the experiments reported here in this paper, $\mu=\sigma=1.0$, which Zhong et al.\ state to be without loss of generality; and $\epsilon$ is a function of $M$, the {\sc Escg}'s {\em mobility} hyperparameter, which was introduced in Section~\ref{sec:backgnd}.

For the discussions that follow below, it is useful to introduce a few extra items of notation additional to those employed by Zhong et al.\ in their paper: here, let $N_a$ denote the number of arcs ablated from the complete tournament RPSLS dominance network; let $n_s(t)$ denote the number of surviving (non-extinct) species in the system at time $t$; let $t_\text{max}$ denote the duration of a simulation experiment, i.e.\ how many {\sc mcs} the experiment runs for before terminating, and note that in all of Zhong et al.'s experiments,  $t_\text{max}$$=$$10^5$; finally, let $N_\text{{\sc iid}}$ denote the number of independent and identically distributed ({\sc iid}) repetitions of a specific simulation experiment with a given set of (hyper-)parameter values -- in all of the experiments reported in Zhong et al.\ (2022), $N_\text{\sc iid}$$=$$500$.   

In Zhong et al.\ (2022) the primary mode of analysis and summary of experiment outcomes aggregated over some suitably large number of {\sc iid} simulation runs was plots of what Zhong et al.\ referred to as $F$, the frequency of outcome of number of species surviving at the end of the simulation, as a function of $M$, the mobility value used in the simulation: in my notation, this is more specifically denoted by $F(n_s(t_\text{max})$$=$$c)$ vs.\ $M$, for $c$$\in$$ \{1,\ldots,5\}$, 
but for brevity I'll refer to these simply as ``$FvsM$'' plots.

\clearpage
\newpage

\subsection{The {\sc Escg} algorithm}
\label{sec:Escg_algo}

Let $l$ be the 2D lattice of cells such that an individual cell with coordinates $(x,y)$ in the lattice is denoted by $l(x,y)$, and where the individual cell is the {\em position} of an individual agent $i$ in the model, which I'll denote by $\vec{p}_i$. 
Let  ${\cal U}$$\left[n_{lo},n_{hi}\right]$ denote a new random draw from a uniform distribution over the range $\left[n_{lo},n_{hi}\right] \subset {\mathbb R}$ and similarly let ${\cal U}\{ m_0, m_1, \ldots\}$ represent a new uniform (equiprobable) random choice of member from the set $\{ m_0, m_1, \ldots\}$.

Assume here the existence of a function {\sc RndNeighbour}$(l, \vec{p}_i, {\cal N}, {\cal B})$ which returns the coordinate pair $\vec{p}_n=(x_n, y_n)$ for a randomly chosen member from the neighbourhood of $\vec{p}_i=(x_i,y_i)$ where ${\cal N}$ specifies the neighborhood function to use  (e.g. von Neumann or Moore, etc), and with boundary conditions specified by ${\cal B}$ as either periodic or no-flux. 

Assume also that the three possible actions introduced above in Equations~\ref{eq:RPSLS0} to~\ref{eq:move} are encoded as three simple functions, each of which take as arguments: the lattice $l$; the lattice position $\vec{p}_i$ of the randomly chosen individual cell $c_i$; and the lattice coordinates $\vec{p}_n$ of $c_i$'s randomly chosen neighboring cell, denoted $c_n$. Additionally, {\sc Compete} requires a specification of the dominance network, denoted by ${\cal D}$:
\begin{itemize}

\item {\sc Compete}$(l,\vec{p}_i,\vec{p}_n, {\cal D})$ implements Equations~\ref{eq:RPSLS0} to~\ref{eq:RPSLS4}.
\item {\sc Reproduce}$(l,\vec{p}_i,\vec{p}_n)$ implements Equation~\ref{eq:reproduction}.

\item {\sc Move}$(l,\vec{p}_i,\vec{p}_n)$ implements Equation~\ref{eq:move}.
\end{itemize}

\noindent
Each of these functions returns the updated lattice $l$. 
These three functions will be called from within the procedure for a single elementary step, {\sc ElStep}, which takes as arguments the lattice, the position vectors $\vec{p}_i$ and $\vec{p}_n$ of $c_i$ and $c_n$ respectively, the three hyperparameter rates $\mu, \sigma,$ and $\epsilon$, and the dominance network ${\cal D}$, and returns the updated lattice. The entire algorithm for an instance of the evolutionary spatial cyclic game ({\sc Escg}) on a square 2D lattice is as shown in Algorithm~\ref{alg:game}, which calls the {\sc PopulateLattice} algorithm listed in  Algorithm~\ref{alg:poplat}, and the Original Elementary Step (OES) algorithm {\sc ElStep},  listed in Algorithm~\ref{alg:OES}.
 
As is discussed in more detail in \cite{cliff_2024_noops}, the {\sc ElStep} algorithm as shown in Algorithm~\ref{alg:OES} is inefficient in space and in time, because the contents of the randomly-chosen cells at $\vec{p}_i$ and $\vec{p}_n$ may not be compatible with the randomly chosen action selected within  {\sc ElStep}. For instance, if either $l(\vec{p}_i)$$=$$\emptyset$ or $l(\vec{p}_n)$$=$$\emptyset$ and then in {\sc ElStep} the randomly-chosen action is {\sc Compete}, no change to the lattice will occur on that elementary step, and so that step becomes a ``no-op'', i.e.\ an operation that does nothing.\footnote{Many central processor unit (CPU) chips include a no-op instruction in their assembly language: typically the no-op will consume exactly one unit of clock time while leaving everything else unchanged in the CPU and its associated memory. No-ops can be useful in parallel real-time systems where it is important to establish or maintain temporal synchronisation between the constituent sub-systems.} The elementary step also becomes a no-op if both $l(\vec{p}_i)$$\neq$$\emptyset$ and $l(\vec{p}_n)$$\neq$$\emptyset$ and {\sc ElStep} randomly selects {\sc Reproduce} as the action for that step; and also if $l(\vec{p}_i)$$=$$\emptyset$ and $l(\vec{p}_n)$$=$$\emptyset$ then {\sc ElStep} is again a no-op, for all three possible actions.


\begin{algorithm}[h]
\caption{Evolutionary Spatial Cyclic Game ({\sc Escg}) -- 2D Square}
\label{alg:game}
\begin{algorithmic}[1]
\Require $L \geq 1 \in {\mathbb N}$ \Comment{Side-length of square lattice $l$}
\Require $N_s \geq 3 \in {\mathbb N}$ \Comment{Number of species}
\Require $ P_{\text{empty}} \in [0.0,1.0] \subset {\mathbb R}$ \Comment{Probability of empty cell}
\Require $M \in (0.0, 1.0) \subset {\mathbb R}$ \Comment{Mobility}
\Require $\mu \in [0.0, 1.0] \subset {\mathbb R}$ \Comment{Competition rate}
\Require $\sigma \in [0.0, 1.0] \subset {\mathbb R}$ \Comment{Reproduction rate}
\Require $t_{\text max} \geq 1 \in {\mathbb N}$ \Comment{Experiment duration (max.\ \#{\sc mcs})}
\Require $e_{\text max} \geq 1 \in {\mathbb N}$ \Comment{\#elementary-steps per {\sc mcs}}
\Require ${\cal N} \in \{ \text{`vonNeumann', `Moore'} \} $ \Comment{Neighborhood function}
\Require ${\cal B} \in \{\text{`periodic'}, \text{`noflux'}\} $ \Comment{Boundary condition}
\Require ${\cal D} $ \Comment{Dominance network ($N_s$$\times$$N_s$ adjacency matrix)}
\Ensure $N_s = 2j + 1 ; j \in {\mathbb Z}$ \Comment{$N_s$ must be odd}
\State $N \gets L^2$ \Comment{Total number of cells in lattice}
\State $\epsilon \gets 2MN$ \Comment{Movement rate}
\State$l \gets \text{\sc PopulateLattice}(L, N_s, P_{\text{empty}})$ \Comment{Initial state of lattice}
\State $t \gets 0$ \Comment{$t$ is current timestep, in units of {\sc mcs}}
\While{$t < t_{\text max}$}\Comment{Outer {\sc mcs} loop}
\State $e \gets 0$\Comment{$e$ is current elementary step}
\While{$e < e_{\text max}$}
	\Comment{Core inner ES loop}
	\State$\vec{p}_i \gets ({\cal U}\{0,\ldots, L-1\} ,  {\cal U}\{0,\ldots, L-1\} )$	\Comment{Randomly chosen cell} 
	\State$\vec{p}_n \gets {\text{\sc RndNeighbor}}(l, \vec{p}_i, {\cal N}, {\cal B}) $ \Comment{Randomly chosen nbr cell}
	\State$l \gets$ {\sc ElStep}$(l, \vec{p}_i, \vec{p}_n, \mu, \sigma, \epsilon, {\cal D})$ \Comment{Elementary Step}
\State $e \gets e+1$
\EndWhile 
\State $t \gets t+1$
\EndWhile
\end{algorithmic}
\end{algorithm}

\clearpage

\begin{algorithm}
\caption{Populating the square lattice}
\label{alg:poplat}
\begin{algorithmic}[1]
\Require $L \geq 1 \in {\mathbb N}$ \Comment{Side-length of square lattice $l$}
\Require $N_s \geq 3 \in {\mathbb N}$ \Comment{Number of species}
\Require $ P_{\text{empty}} \in [0.0,1.0] \subset {\mathbb R}$ \Comment{Probability of empty cell}
\Ensure $  l \in {\cal M}^L_L({\mathbb Z}) $  \Comment{Lattice $l$ is a $L$$\times$$L$ matrix}
\Procedure{PopulateLattice}{$L, N_s, P_{\text{empty}}$} 
\State$x \gets 0$
\While{$x < L$}\Comment{Populate lattice with species}
	\State$y \gets 0$
	\While{$y<L$}	
		\If {${\cal U}\left[0.0,1.0\right] < P_{\text{empty}}$}
			\State $l(x,y) \gets \emptyset$ \Comment{Empty cell}
		\Else
			\State $l(x,y) \gets {\cal U}\{0, \ldots, N_s-1\}$ \Comment{All species equiprobable}
		\EndIf
		\State$y\gets y +1  $
	\EndWhile
	\State$x\gets x +1 $
\EndWhile
	\State \textbf{return} {$l$}
\EndProcedure
\end{algorithmic}
\end{algorithm}

\begin{algorithm}
\caption{Original Elementary Step (OES)}
\label{alg:OES}
\begin{algorithmic}[1]
\Require $\mu \in [0.0, 1.0] \subset {\mathbb R}$ \Comment{Competition rate}
\Require $\sigma \in [0.0, 1.0] \subset {\mathbb R}$ \Comment{Reproduction rate}
\Require $\epsilon$ \Comment{Movement rate}
\Require ${\cal D} \in {\cal M}^{N_s}_{N_s}({\mathbb N})$ \Comment{Dominance network ($N_s$$\times$$N_s$ adjacency matrix)}
\Require $l \in {\cal M}^{L}_{L}({\mathbb Z}) $ \Comment{Current state of lattice $l$}
\Require $\vec{p}_i \in {\mathbb N}^2 $ \Comment{Lattice coords of cell $c_i$}
\Require $\vec{p}_n \in {\mathbb N}^2 $ \Comment{Lattice coords of cell $c_n$}
\Procedure{ElStep}{$l, \vec{p}_i, \vec{p}_n, \mu, \sigma, \epsilon, {\cal D}$} 
	\State$P_{\mu} \gets {\mu}/(\mu + \sigma + \epsilon)$\Comment{Normalized probability of competition}
	\State$P_{\sigma} \gets {\sigma}/(\mu + \sigma + \epsilon)$\Comment{Normalized probability of reproduction}	
	\State$a \gets {\cal U}\left[0.0,1.0\right]$
	\If{$0 \leq a < P_{\mu} $}
    			\State $l \gets ${\sc Compete}$(l,\vec{p}_i,\vec{p}_n, {\cal D})$
    	\ElsIf{$P_{\mu} \leq a < (P_{\mu} + P_{\sigma})$}
		    \State $l \gets ${\sc Reproduce}$(l,\vec{p}_i,\vec{p}_n)$
	\Else
    			\State $l \gets ${\sc Move}$(l,\vec{p}_i,\vec{p}_n)$
	\EndIf
	\State \textbf{return} {$l$}
\EndProcedure
\end{algorithmic}
\end{algorithm}

In each individual {\sc Escg} experiment, the density $\rho_i(t)$ of each species $S_i$ (i.e., what proportion of the lattice cells are occupied by agents of species type $S_i$ at time $t$) was recorded after each {\sc mcs}: illustrative time-series of the $\rho_i$ values from several experiments were given in \cite{cliff_2024_noops}.
For the purposes of this paper the primary variable of interest is $n_s(t)$ the number of non-extinct species at time $t$
 recorded over a set of $N_\text{\sc iid}$ {\sc iid} simulation experiments for any given set of hyperparameter values.

\newpage
\clearpage

\section{Replicating the results of Zhong et al.\ (2022)}
\label{sec:zhong_replication}

\begin{figure}
\begin{center}
\includegraphics[trim=0cm 0cm 0cm 1.2cm, clip=true, scale=0.28]{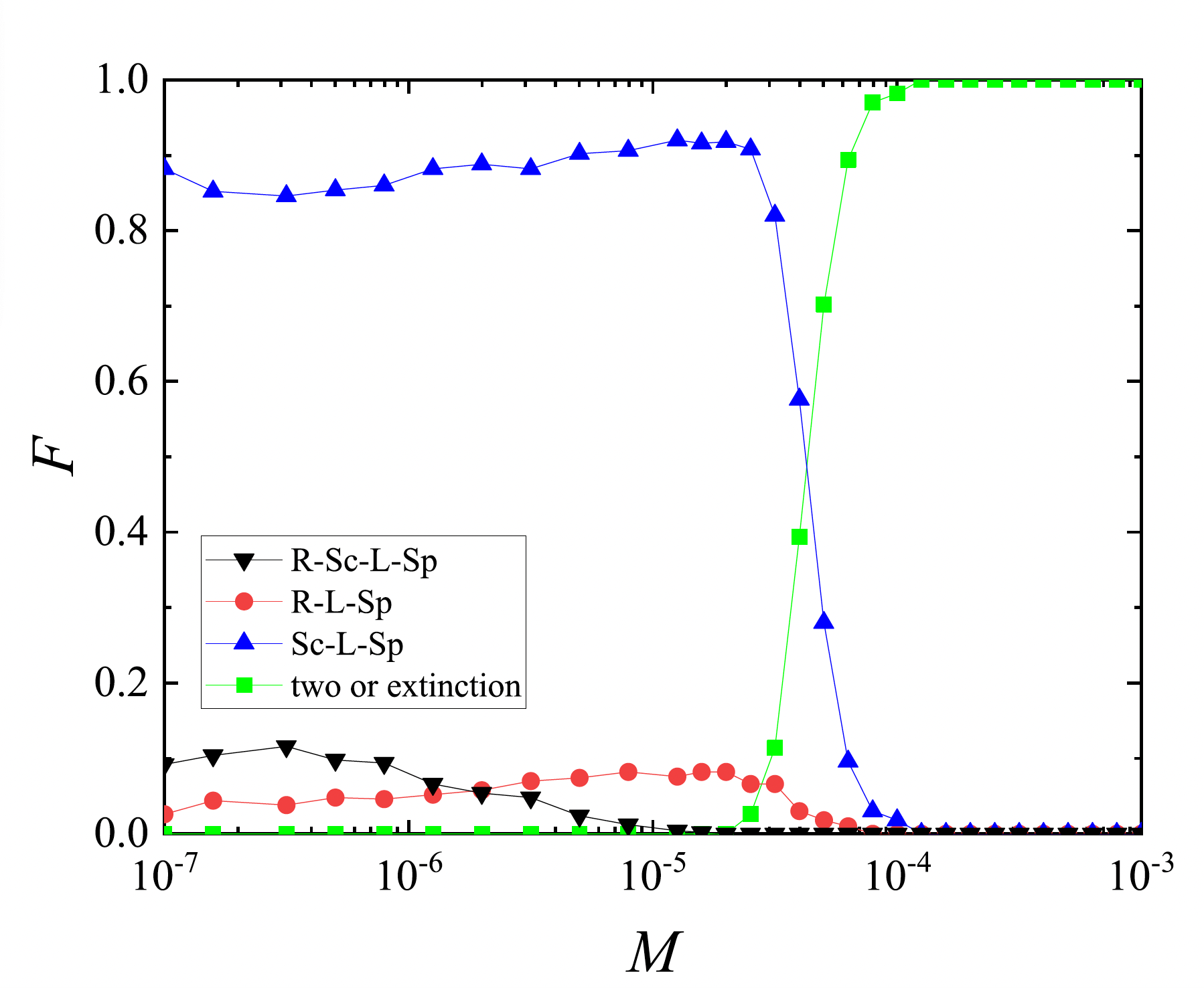}
\includegraphics[trim=0cm 0cm 0cm 0cm, clip=true, scale=0.40]{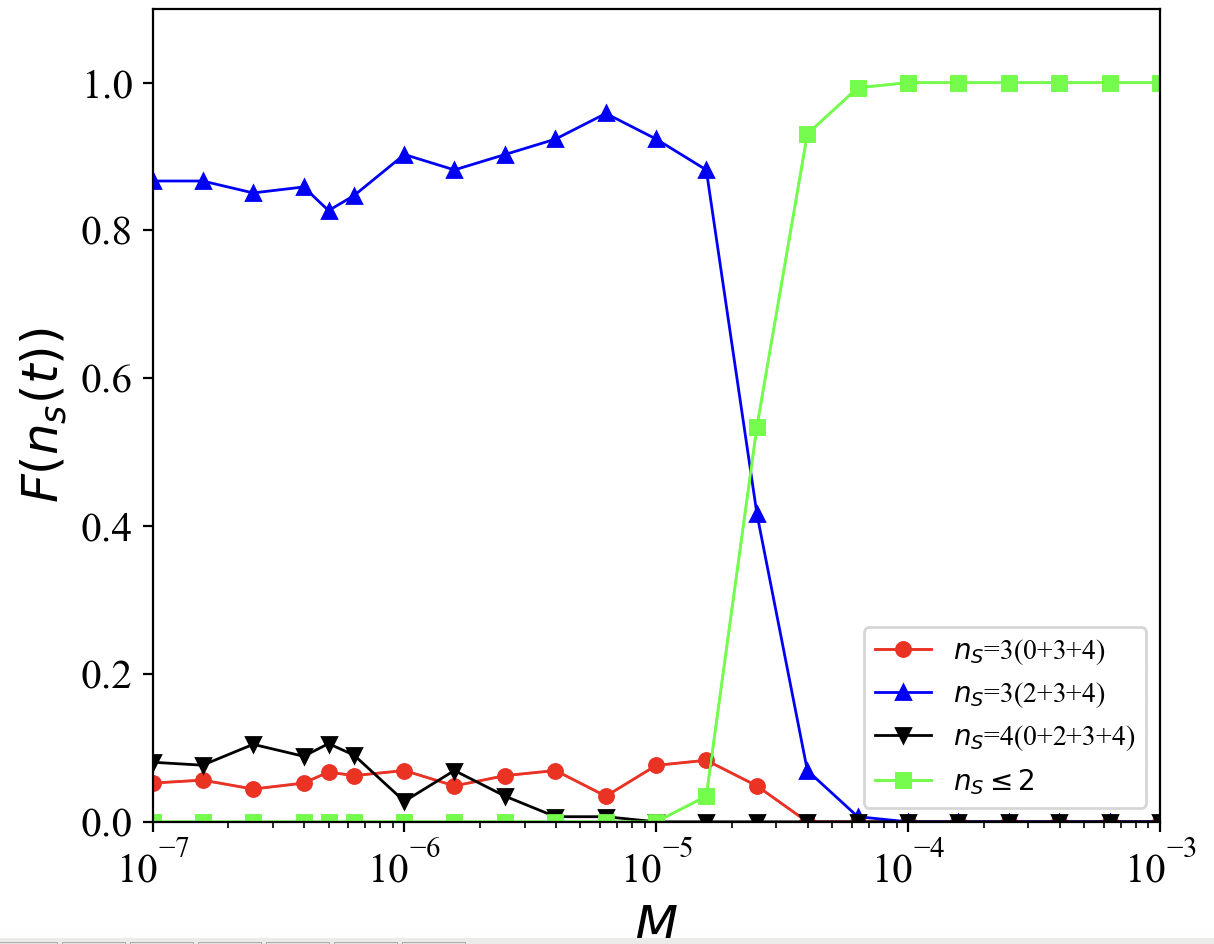}
\end{center}
\caption{
Graphs of $F(n_s(t_\text{max})$$=$$c)$ for $c \in \{1, \dots, 5\}$ (i.e., frequency of outcome of number of species surviving at the end of the experiment) vs.\ $M$ (mobility) -- referred to as $FvsM$ plots -- in my replication of the $N_a$$=$$1$ (dominance network Z1) experiments reported by Zhong et al.: upper graph is Zhong et al.'s Figure~3(a) $FvsM$ results; lower
graph shows corresponding $FvsM$ results  from my replication of the same experiments. The legend in Zhong et al.'s Figure~3(a) uses the abbreviations $R$ for `Rock', $Sc$ for `Scissors', $L$ for `Lizard', and $Sp$ for `Spock'. Note that Zhong et al.\ combine the results for $F(n_s(t_\text{max})$$=$$2)$ and $F(n_s(t_\text{max})$$=$$1)$ into a single class  of outcome labelled ``two or extinction". The legend in the lower graph of outcomes from my replication uses the numeric node-labels introduced in Figure~\ref{fig:RPSLSDomNets}: each row of the legend shows the color of the line and marker for the given value of $n_s$ followed by, for $F(n_s(t_\text{max})$$>$$2)$, in parentheses and separated by `+' symbols, the node-numbers of that set of $n_s$ surviving species. }
\label{fig:ZhongFig3}
\end{figure}

\begin{figure}
\begin{center}
\includegraphics[trim=0cm 0cm 0cm 0cm, clip=true, scale=0.45]{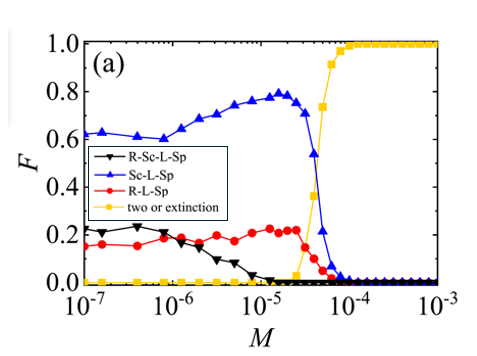}
\includegraphics[trim=0cm 0cm 0cm 0cm, clip=true, scale=0.4]{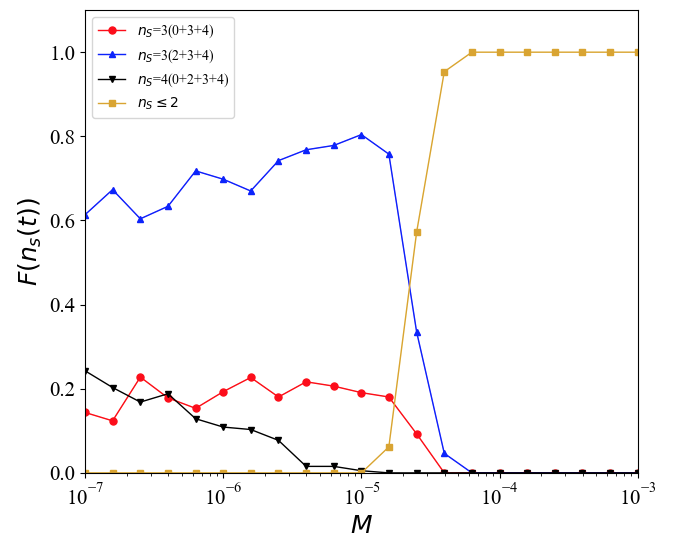}
\end{center}
\caption{
Replication of one of the two $N_a$$=$$2$ (dominance network Z2a) experiments results reported by Zhong et al.:
Upper graph is Zhong et al.'s Figure~5a (edited to include the legend); $N_\text{\sc iid}$$=$$500$; lower
graph shows corresponding results  from my replication of the same experiments; $N_\text{\sc iid}$$=$$200$. 
Format and legend labels as for Figure~\ref{fig:ZhongFig3}.
}
\label{fig:ZhongFig5}
\end{figure}

\begin{figure}
\begin{center}
\includegraphics[trim=0cm 0cm 0cm 0cm, clip=true, scale=0.4]{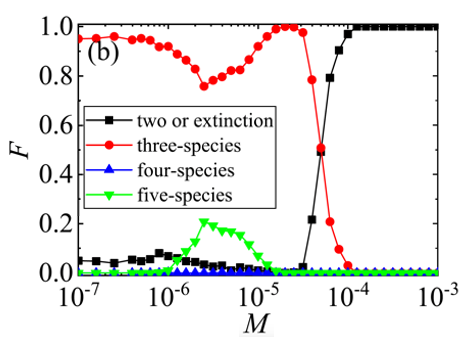}
\includegraphics[trim=0cm 0cm 0cm 0cm, clip=true, scale=0.35]{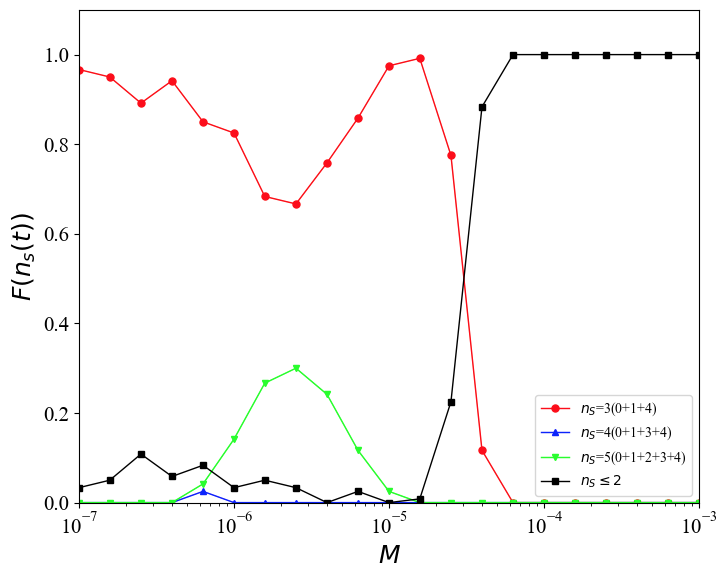}
\end{center}
\caption{
Replication of one of the two $N_a$$=$$3$ (dominance network Z3b) experiments results reported by Zhong et al.: 
Upper graph is Zhong et al.'s Figure~6b (edited to include the legend); $N_\text{\sc iid}$$=$$500$; lower
graph shows corresponding results  from my replication of the same experiments; $N_\text{\sc iid}$$=$$200$.
In Zhong et al's Figure~6 and subsequent Figure~7 (shown here in Figure~\ref{fig:ZhongFig7}) the results are no longer explicitly presented for each distinct $n_s(t_\text{max})$$=$$c$ outcome such that, for example, the possible multiple different $n_s(t_\text{max})$$=$$3$ outcomes illustrated in Figures~\ref{fig:ZhongFig3} and~\ref{fig:ZhongFig5} are instead presented in Zhong et al.'s Figures~6 and~7 as a single aggregate class of $n_s(t_\text{max})$$=$$3$ outcome: for ease of comparison, I have followed that style of presentation here. 
}
\label{fig:ZhongFig6}
\end{figure}

\begin{figure}
\begin{center}
\includegraphics[trim=0cm 0cm 0cm 0cm, clip=true, scale=0.3]{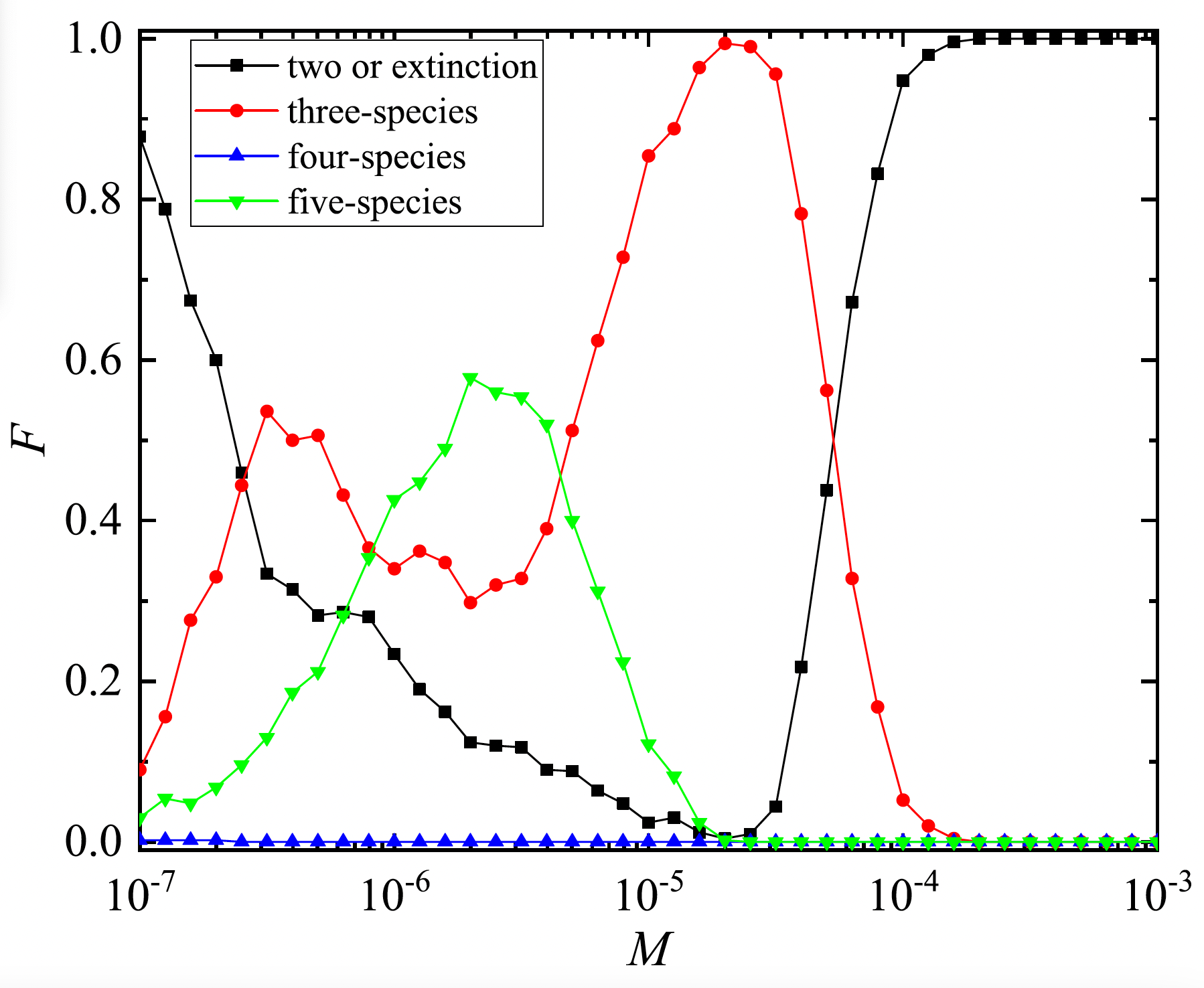}
\includegraphics[trim=0cm 0cm 0cm 0cm, clip=true, scale=0.375]{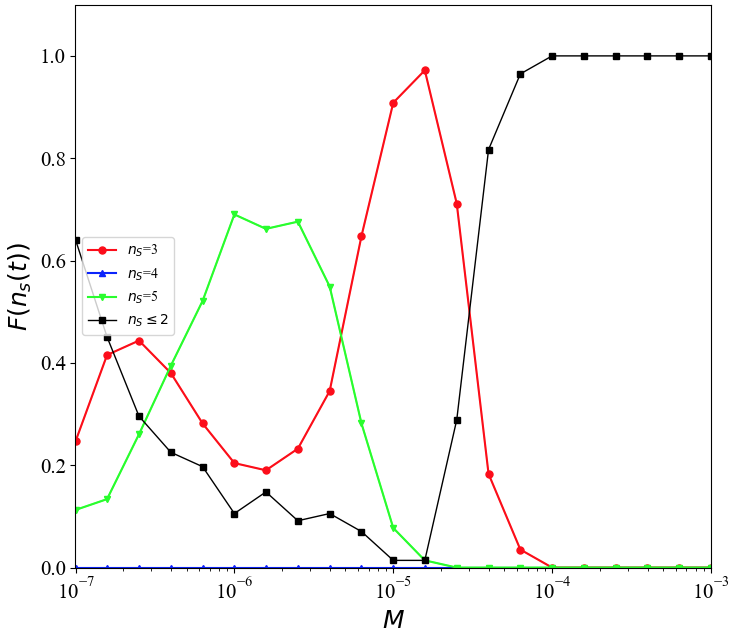}
\end{center}
\caption{
Replication of the $N_a$$=$$4$ (dominance network Z4) experiments results reported by Zhong et al.:
upper graph is Zhong et al.'s Figure~7, $N_\text{\sc iid}$$=$$500$; lower
graph shows corresponding results  from my replication of the same experiments; $N_\text{\sc iid}$$=$$200$.
Format and legend labels as for Figure~\ref{fig:ZhongFig3}.
}
\label{fig:ZhongFig7}
\end{figure}

The upper graph in each of Figures~\ref{fig:ZhongFig3}, \ref{fig:ZhongFig5}, \ref{fig:ZhongFig6}, and~\ref{fig:ZhongFig7} re-prints
the $FvsM$ results from the $N_a$$=$$1$, $N_a$$=$$2$, $N_a$$=$$3$, and $N_a$$=$$4$ experiments presented in Figures 3, 5, 6, and 7 of Zhong et al.\ (2022), respectively, and the lower graph in each of  Figures~\ref{fig:ZhongFig3}, \ref{fig:ZhongFig5}, \ref{fig:ZhongFig6}, and~\ref{fig:ZhongFig7} shows corresponding $FvsM$ results from my replication of each of those respective sets of experiments. For ease of comparison, my graphs use the same data-point markers and line-colors as Zhong et al., which (for reasons not stated by Zhong et al.) vary from graph to graph.

As can be seen, qualitatively there is extremely good agreement between the original results and my replication, but my results are not in exact quantitative agreement with the originals. Specifically, for all four of my $FvsM$ graphs, I found that using $t_\text{max}$$=$$1.7$$\times$$10^{5}$ (in contrast to the $t_\text{max}$$=$$10^{5}$ used by Zhong et al.) gave the best alignment between the two sets of results; and even with that adjustment to $t_\text{max}$, the $M$-value at the crossover point where $F(n_s(t_\text{max})>2)=F(n_s(t_\text{max})\leq2)$ is slightly -- but noticeably -- lower in my results than in Zhong et al.'s: for example, in Figure~\ref{fig:ZhongFig3} the crossover point in Zhong et al.'s $FvsM$ graph is at $M$$\approx$$4.1$$\times$$10^{-5}$ whereas in my replication the crossover is at $M$$\approx$$2.2$$\times$$10^{-5}$. 

These minor points of quantitative difference indicate that, in all likelihood, Zhong et al.'s actual implementation of their {\sc Rpsls} {\sc Escg}, their program code, does not exactly implement the {\sc Escg} laid out in Algorithm~\ref{alg:game}.  For instance, if  Algorithm~\ref{alg:game} is altered so that at line~9, where originally a cell is chosen at random from anywhere in the lattice, instead a random {\em non-empty} cell is chosen, per the amended algorithm snippet shown in Algorithm~\ref{alg:nonemptycell}, then {\em every} call to {\sc ElStep} will offer the possibility of either a competition, a reproduction, or a movement occurring, because we now guarantee that at least one of the two cells selected on each ES contains an individual, and hence the chances of a no-op on each ES are significantly reduced. This reduction in the number of no-ops per {\sc mcs} would mean that the dynamics of the {\sc Escg} would unfold at a quicker pace, because a greater number of substantive changes to the lattice would occur on each {\sc mcs}, and so the value of $t_\text{max}$ needed to get a good match with Zhong et al.'s results could be significantly reduced from the $t_\text{max}$$=$$1.7$$\times$$10^{5}$ used here.

\begin{algorithm}[h]
\caption{Amended {\sc Escg}: start with non-empty cell}
\label{alg:nonemptycell}
\begin{algorithmic}[1]
\State $N \gets L^2$ \Comment{Total number of cells in lattice}
\State $\epsilon \gets 2MN$ \Comment{Pr(move)}
\State$l \gets \text{\sc PopulateLattice}(L, N_s, P_{\text{empty}})$ \Comment{Initial state of lattice}
\State $t \gets 0$ \Comment{$t$ is current timestep, in units of {\sc mcs}}
\While{$t < t_{\text max}$}\Comment{Outer {\sc mcs} loop}
\State $e \gets 0$\Comment{$e$ is current elementary step}
\While{$e < e_{\text max}$}
	\Comment{Core inner ES loop}
	\State$\vec{p}_i \gets ({\cal U}\{0,\ldots, L-1\} ,  {\cal U}\{0,\ldots, L-1\} )$	\Comment{Randomly chosen cell}
	\While{$l(\vec{p}_i) = \emptyset$} \Comment{Empty cell?}
		\State$\vec{p}_i \gets ({\cal U}\{0,\ldots, L-1\} ,  {\cal U}\{0,\ldots, L-1\} )$	\Comment{Choose again}
	\EndWhile 
	\State$\vec{p}_n \gets {\text{\sc RndNeighbor}}(l, \vec{p}_i, {\cal N}, {\cal B}) $ \Comment{Randomly chosen nbr cell}
	\State$l \gets$ {\sc ElStep}$(l, \vec{p}_i, \vec{p}_n, \mu, \sigma, \epsilon, {\cal D})$ \Comment{Elementary Step}
\State $e \gets e+1$
\EndWhile 
\State $t \gets t+1$
\EndWhile
\end{algorithmic}
\end{algorithm}

In principle, I could make a series of alterations to my implementation of Algorithm~\ref{alg:game} such as the one just discussed, and then via a trial-and-error process eventually arrive at a better quantitative match between my replication results and those of Zhong et al., but to do so would not be a productive use of time because the arguments that follow in Sections~\ref{sec:howlong} and~\ref{sec:NA0} would be neither strengthened nor weakened by having the replication results fitting quantitatively closer to the original results: that is, the replication results as presented here are already sufficiently close to the originals to support the critique that I offer in the next two sections of this paper.  

\newpage
\clearpage

\section{Asymptote or Transient?}
\label{sec:howlong}

In the abstract of their paper, in the captions to their figures 2 to 7, and repeatedly in the text of their paper, Zhong et al.\ state that they are presenting results showing asymptotic behaviors of the system, as they explain thus:
\begin{quotation}
``To get an overview of the asymptotic behaviors against the mobility, we explore the occurrence of different asymptotic behaviors. For this aim, we run 500 realizations independently with equally and randomly distributed initial conditions for $10^5$ {\sc mcs}s and calculate the abundance of different asymptotic behaviors for each mobility.'' \cite[Section 3.1]{zhong_etal_2022_ablatedRPSLS} 

\end{quotation}
\noindent

In common technical usage, ``asymptotic behavior'' refers to how some response or dependent variable behaves as the input or independent variable tends to infinity. Here, the independent variable is time (measured in {\sc mcs}) and the dependent variable is the frequency of occurrence of the number of species coexisting in the modified {\sc Rpsls} {\sc Escg} system: for example, Zhong et al.'s caption to their Figure~2 states that two possible asymptotic behaviors are four-species coexistence and three-species coexistence, in experiments that ran for $10^5${\sc mcs}. Using the terminology introduced above in 
Section~\ref{sec:escg_defn}, Zhong et al.\ had $t_{\text{max}}$$=$$10^5$, and for each of their individual {\sc Escg} simulation experiments the frequencies of observed outcome of each possible number of coexisting species at time $t_{\text{max}}$, (denoted here by $n_s(10^5)$) was claimed to be the asymptotic behavior of that experiment. Implicit in this is the assumption that there is no need to run the simulations for longer than $10^5${\sc mcs} because there would be no further changes; that is, Zhong et al.\ assumed $10^5${\sc mcs} to be such a very large number of {\sc mcs} as to be, for practical purposes, infinite. In this section I show that this assumption is incorrect, and that it can be necessary to run the simulations for $t_\text{max}=10^7$ or longer before the true asymptotic behaviors of the system can be convincingly identified. A consequence of this is that some of the features in their $FvsM$ graphs that were highlighted for discussion by Zhong et al.\ are in fact only artefacts of slowly decaying transients in the system's dynamics: when the {\sc Escg} is run for sufficiently long, the features highlighted by Zhong et al.\ as asymptotic and worthy of discussion simply decay to nothing, and the diversity of observed outcomes reduces to near-homogeneity.

Figure~\ref{fig:Zhong3_FvsM_compare} shows an exemplar set of results that illustrate this point: it shows FvsM plots for the Z2a ablated {\sc Rpsls} dominance network at four different values of $t_\text{max}$: 170k{\sc mcs} (which gives a close match to Zhong et al.'s results for $t_\text{max}$$=$$10^5$ as was shown previously in Figure~\ref{fig:ZhongFig5}), 350k{\sc mcs}, 500k{\sc mcs}, and $10^6${\sc mcs}. Looking for instance at the $F$ values at $M$$=$$10^{-7}$, as the system is run for longer, the frequency of occurrence of four-species coexistence (i.e., $F(n_s(t_\text{max})$$=$$4)$) falls from roughly 20\% at $t_\text{max}$$=$$170$k to absolute zero at $t_\text{max}$$=$$10^6$, and the $t_\text{max}$$=$$10^6$ results show conclusively that four-species co-existence is in fact not an asymptotic behavior of this system at all: the presence of four-species outcomes in Zhong et al.'s Z2a results is merely an artefact of their experiments having not been run for long enough. 

\def\ScalePNG{0.23}

\begin{figure}[h]
\centering
\subfloat[LoF text][$FvsM$ at $t_{\text{max}}$$=$$170$k{\sc mcs}.]{
\includegraphics[trim=0cm 0cm 0cm 0cm, clip=true, scale=\ScalePNG]{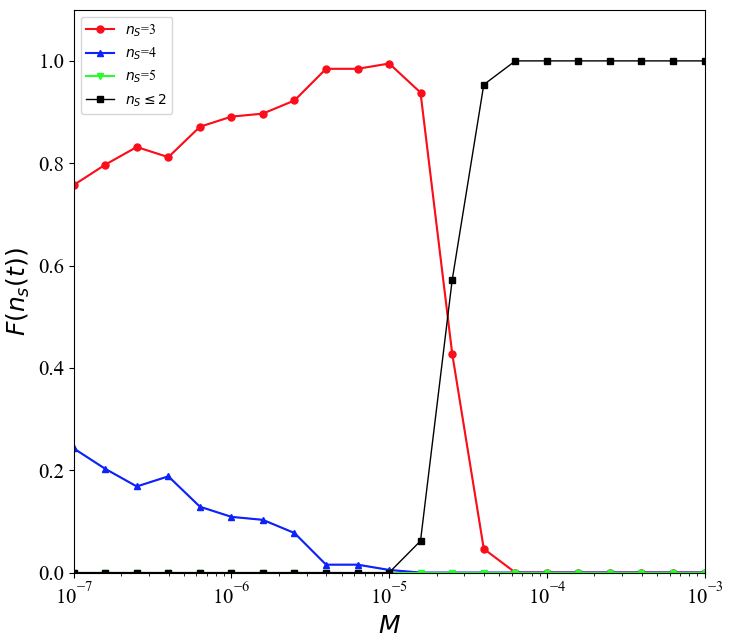}
\label{fig:Zhong3_FvsM_170k}}
\qquad
\subfloat[LoF text][$FvsM$ at $t_{\text{max}}$$=$$350$k{\sc mcs}]{
\includegraphics[trim=0cm 0cm 0cm 0cm, clip=true, scale=\ScalePNG]{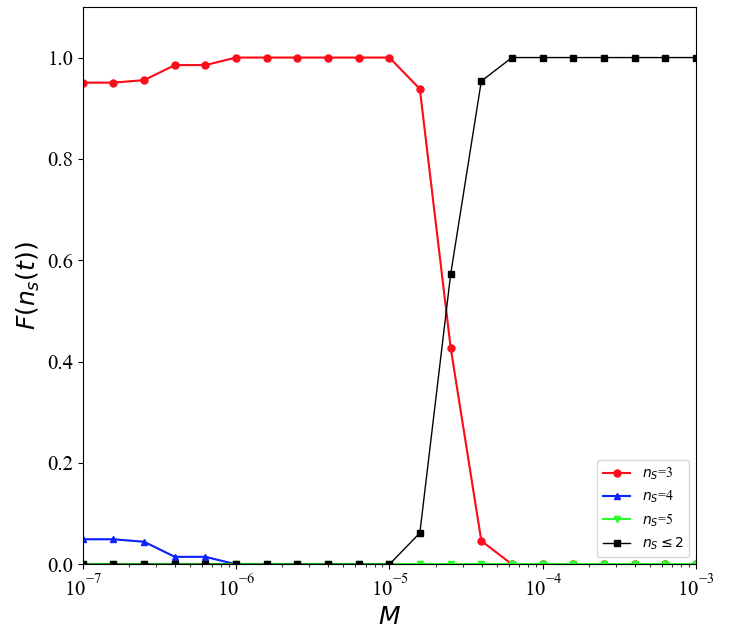}
\label{fig:Zhong3_FvsM_350k}}
\qquad
\subfloat[LoF text][$FvsM$ at $t_{\text{max}}$$=$$500$k{\sc mcs}]{
\includegraphics[trim=0cm 0cm 0cm 0cm, clip=true, scale=\ScalePNG]{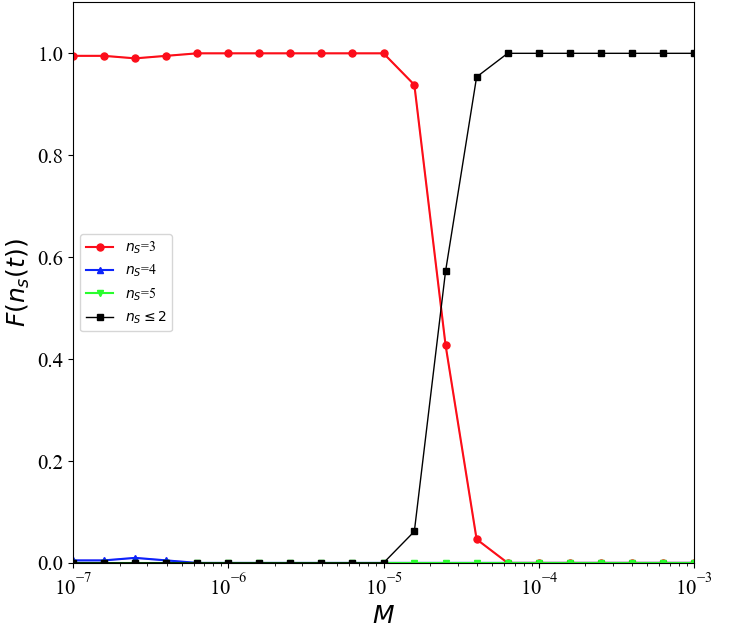}
\label{fig:Zhong3_FvsM_500k}}
\qquad
\subfloat[LoF text][$FvsM$ at $t_{\text{max}}$$=$$10^6${\sc mcs}]{
\includegraphics[trim=0cm 0cm 0cm 0cm, clip=true, scale=\ScalePNG]{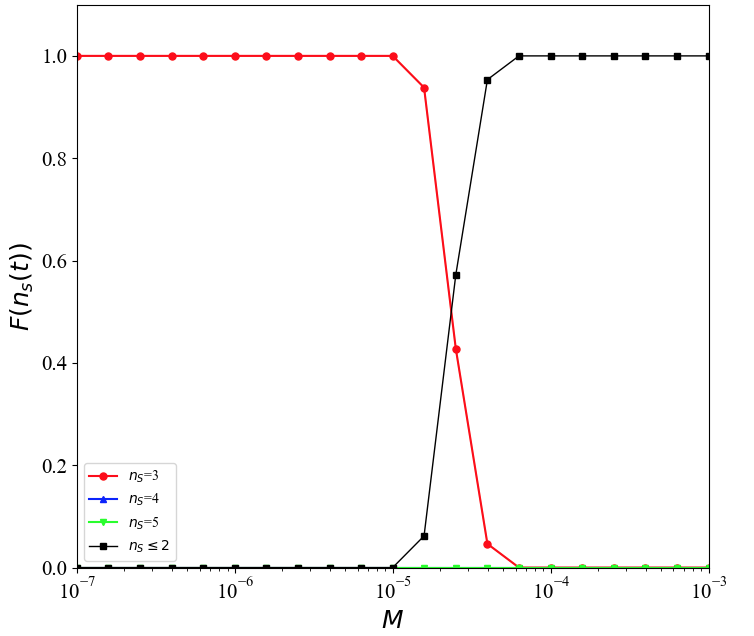}
\label{fig:Zhong3_FvsM_1000k}}
\caption{
Plots of $FvsM$ for the $N_a$$=$$2$ {\sc Rpsls} dominance network Z2a,  
$N$$=$$200$$\times$$200$, results from which were shown in Zhong et al.'s Figure~5a; for four different experiment durations: 
(a)
$t_{\text{max}}$$=$$1.7$$\times$$10^5${\sc mcs} 
(which gives the best match to Zhong et al.'s Figure~5a);
(b)
$t_{\text{max}}$$=$$3.5$$\times$$10^5${\sc mcs};
(c)
$t_{\text{max}}$$=$$5.0$$\times$$10^5${\sc mcs};
(d)
$t_{\text{max}}$$=$$10^6${\sc mcs}. 
Format and legend labels as for Figure~\ref{fig:ZhongFig6}.
In each graph, at each value of $M$, results from $N_\text{\sc iid}$$=$$200$ simulations are shown. 
As can be seen, as $t_\text{max}$ is increased, F($n_s(t_\text{max})$$=$$4)$$\rightarrow$$0$, 
and the true asymptotic frequency distribution of outcomes is shown in  
(d): for $M$ below some threshold value
of $\approx$$2.5$$\times$$10^{-4}$ 
the outcome is always $n_s(t)$$=$$3$, and once $M$ is above the threshold the outcome is always $n_s(t)$$\leq$$2$.
}
\label{fig:Zhong3_FvsM_compare}
\end{figure}

\newpage
\clearpage

\begin{figure}[h]
\begin{center}
\includegraphics[trim=0cm 0cm 0cm 0cm, clip=true, scale=0.7]{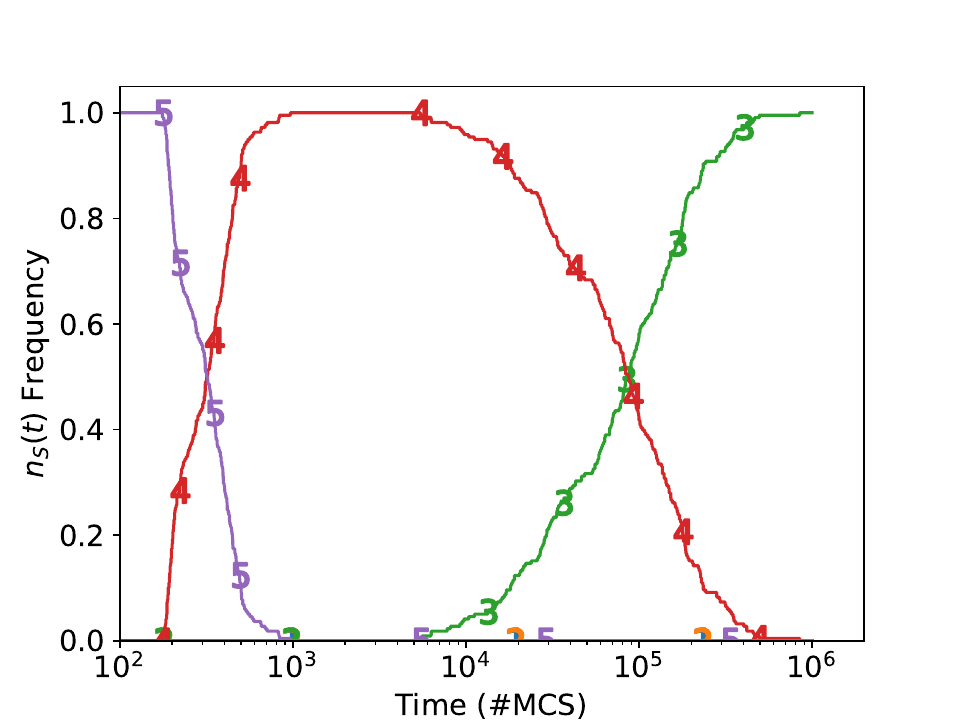}
\end{center}
\caption{
Time series of $F(n_s(t)$$=$$c)$ for $c \in \{1, \ldots, 5 \}$ (i.e., frequency of occurrence of the number of co-existing species at time $t$)
from $N_\text{\sc iid}$$=$$200$ simulation experiments using the  $N_a$$=$$2$ {\sc Rpsls} dominance network Z2a, with $t_\text{max}$$=$$10^6$ and with  $M$$=$$10^{-7}$.
For brevity, plots such as these are referred to as $FvsT$ plots. Horizontal axis is time measured in {\sc mcs}; vertical axis is $F(n_s(t)$$=$$c)$.  Marker digits along each line show the value of $c$ for that line.
}
\label{fig:Z2a_NST_M1e-07}
\end{figure}

To get a better sense of the nature of these transients,  data such as that shown in the four graphs in Figure~\ref{fig:Zhong3_FvsM_compare} (which plotted $F(n_s(t_\text{max}))$ vs.\ $M$ at different specific values of $t_\text{max}$) can be visualised instead along the orthogonal axis of projection, i.e.\ plotting  $F(n_s(t))$ vs.\ $t$ at different specific values of $M$: for brevity, I will refer to these as $FvsT$ plots. 
Figure~\ref{fig:Z2a_NST_M1e-07} shows one such graph of $FvsT$ from $N_\text{\sc iid}$$=$$500$ simulation experiments with the  $N_a$$=$$2$ {\sc Rpsls} dominance network Z2a, for $t_\text{max}$$=$$10^6$ and with $M$$=$$10^{-7}$. As would be expected, initially all experiments have five species co-existing, so $F(n_s(t)$$=$$5)$$=$$1.0$ and $F(n_s(t)$$=$$c)$$=$$0.0$ for all other values of $c$; then at $t$$\approx$$150${\sc mcs}, the simulations start to show their first extinction, and $F(n_s(t)$$=$$5)$ drops rapidly to zero, matched by the consequent rise in $F(n_s(t)$$=$$4)$ which reaches $1.0$ at $t$$\approx$$1$k{\sc mcs}; then, commencing at $t$$\approx$$3$k{\sc mcs}, we see a growing number of experiments undergoing a second extinction, leaving three co-existing species, and hence  $F(n_s(t)$$=$$3)$ starts to rise toward $1.0$ while $F(n_s(t)$$=$$4)$ attenuates, eventually falling to zero at $t$$\approx$$700$k{\sc mcs}.

\newpage
\clearpage

The fact that the true asymptotic state for this system is reached at $t$$\approx$$700$k{\sc mcs} demonstrates that Zhong et al.'s assertion that running simulations of this system for $t_\text{max}$$=$$10^5${\sc mcs} is sufficient to identify its asymptotic behaviors is simply incorrect, and simulations substantially longer than $100$k{\sc mcs} are necessary.

\begin{figure}
\begin{center}
\includegraphics[trim=0cm 0cm 0cm 0cm, clip=true, scale=0.4]{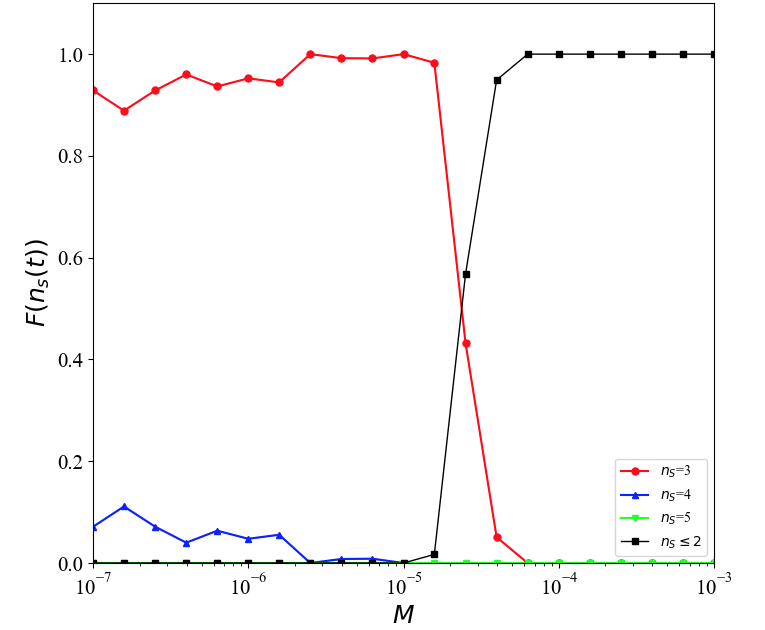}
\includegraphics[trim=0cm 0cm 0cm 0cm, clip=true, scale=0.4]{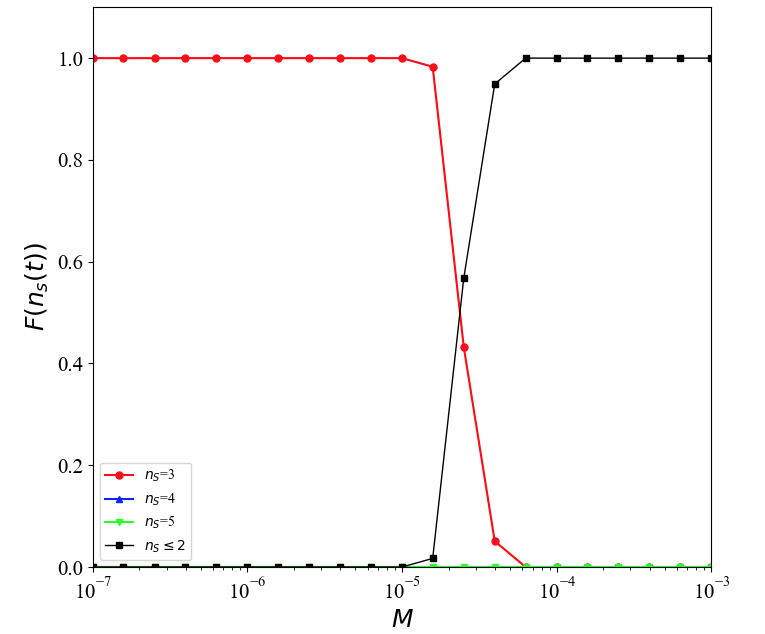}
\end{center}
\caption{
Plots of $FvsM$ for the Z1 ablated dominance network at (upper graph) 
$t_\text{max}$$=$$170k${\sc mcs} and (lower graph) $t_\text{max}$$=$$10^6${\sc mcs}.
Format and legend labels as for Figure~\ref{fig:ZhongFig6}.
}
\label{fig:Z1_FvsM_compare}
\end{figure}

\begin{figure}
\begin{center}
\includegraphics[trim=0cm 0cm 0cm 0cm, clip=true, scale=0.34]{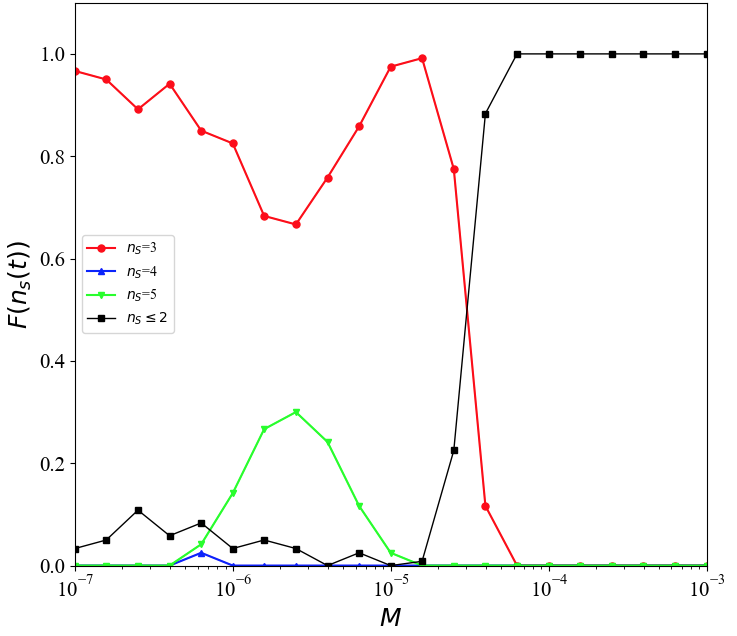}
\includegraphics[trim=0cm 0cm 0cm 0cm, clip=true, scale=0.34]{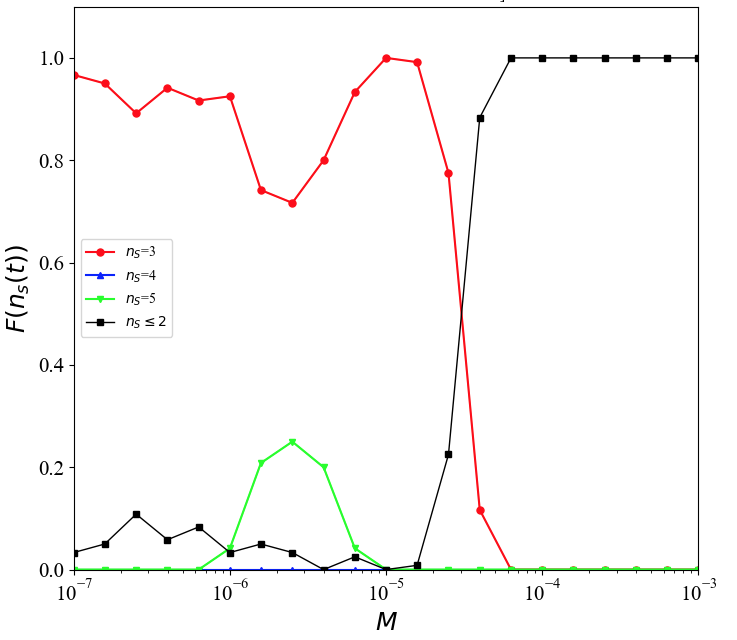}
\end{center}
\caption{
Plots of $FvsM$ for the $N_a$$=$$3$ {\sc Rpsls} dominance network Z3b, results from which were shown in Zhong et al.'s Figure~6b;  $N$$=$$200$$\times$$200$, for two different durations of experiment. Upper graph is from experiment duration $t_{\text{max}}$$=$$1.7$$\times$$10^{5}${\sc mcs}, which gives the best match to the Z3b results published by Zhong et al.\ (2022); lower graph is from experiment duration $t_{\text{max}}$$=$$10^{6}${\sc mcs}.
Format and legend labels as for Figure~\ref{fig:ZhongFig3}.
}
\label{fig:Z3b_FvsM_compare}
\end{figure}

\begin{figure}
\begin{center}
\includegraphics[trim=0cm 0cm 0cm 0cm, clip=true, scale=0.7]{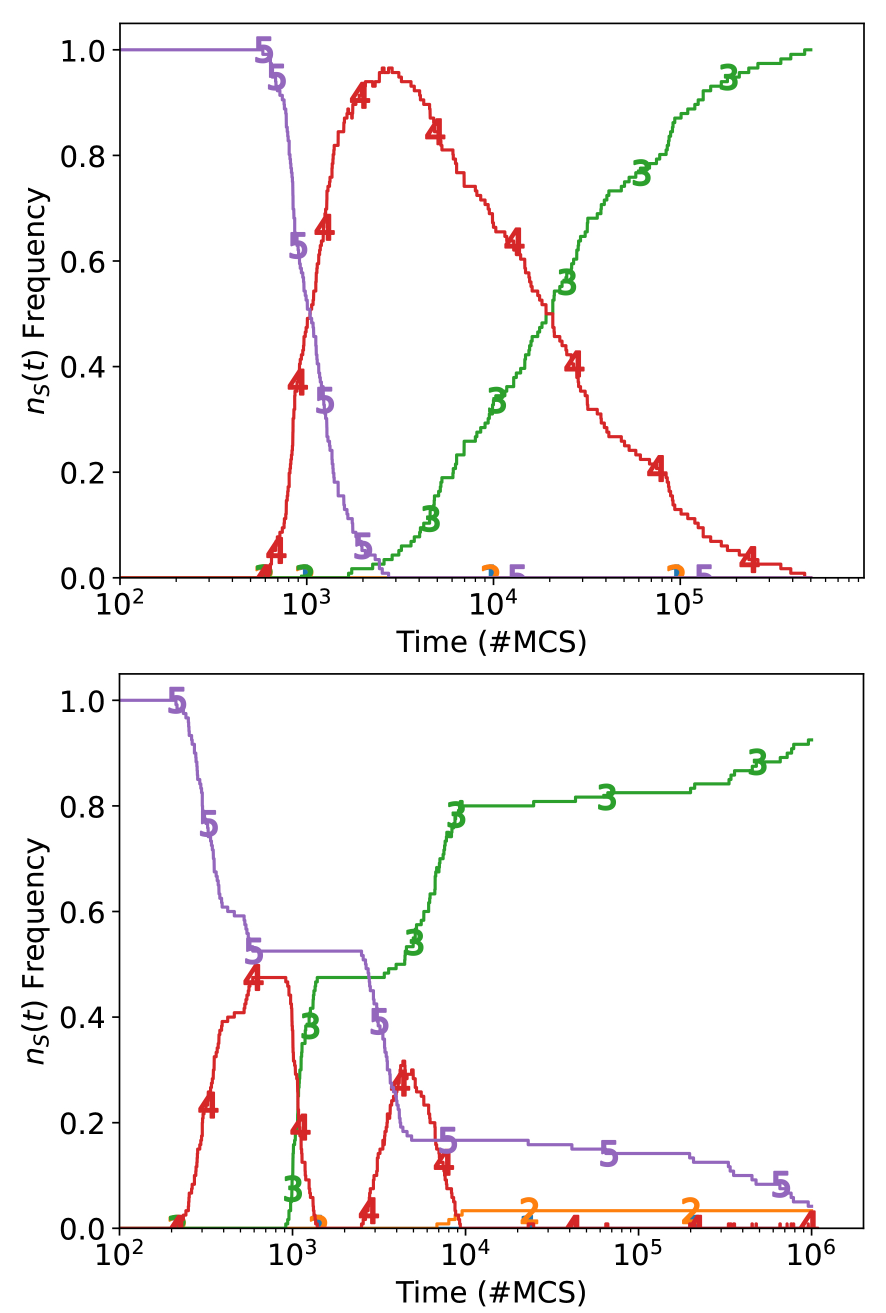}
\end{center}
\caption{
Illustrative $FvsT$ plots at $t_\text{max}$$=$$10^6$ for the Z1 ablated dominance network at $M$$=$$10^{-7}$ (upper graph) and  for the Z3b ablated dominance network at $M$$=$$10^{-6}$ (lower graph); format as for Figure~\ref{fig:Z2a_NST_M1e-07}. For Z1, the longest-persisting transient decays to zero by  $t$$\approx$$5$$\times$$10^5${\sc mcs}. 
For Z3b, even at $t_\text{max}$$=$$10^6$, the $F(n_s(t)$$=$$5)$ transient is clearly decaying but has not yet reached zero; further $FvsT$ plots for Z3b are presented in~\ref{sec:app_NA3_NST}.
}
\label{fig:Z1_Z3b_FvsT}
\end{figure}

Thus far the discussion here has concentrated on results from the Z2a ablated dominance network, but the same issue with slowly decaying transients being mis-identified as asymptotic outcomes affects the results presented by Zhong et al.\ for networks Z1 and Z3b: Figure~\ref{fig:Z1_FvsM_compare} shows for comparison the $FvsM$ plots for Z1 at $t_\text{max}=170k$ and $t_\text{max}$$=$$10^6$; Figure~\ref{fig:Z3b_FvsM_compare} shows $FvsM$ for Z3b at the same two values of $t_\text{max}$; and Figure~\ref{fig:Z1_Z3b_FvsT} shows example $FvsT$ plots for networks Z1 and Z3b, illustrating the slowly decaying transients. 

\newpage
\clearpage

The $FvsM$ plots for ablated dominance network Z1 in  Figure~\ref{fig:Z1_FvsM_compare} and the $FvsT$ plot for Z1 in Figure~\ref{fig:Z1_Z3b_FvsT}  demonstrate conclusively that four-species coexistence is not an asymptotic behavior for this system. 
However the $FvsM$ plots for ablated dominance network Z3b in Figure~\ref{fig:Z3b_FvsM_compare} and the $FvsT$ plot for Z3b in Figure~\ref{fig:Z1_Z3b_FvsT} are not so immediately conclusive. Writing about their Z3b $FvsM$ plot, Zhong et al. observe:

\begin{quotation}
``At low mobility, the state of Scissors-Lizard-Spock dominates
the population, its occurrence stays at high value except
for a dip at around 
$M$$=$$3$$\times$$10^{-6}$. 
Different from other interaction
structures in modified {\sc Rpsls}, it may allow for five-species coexistence
since the two three-species-cyclic interactions only share one
species, Lizard. Once the coexistence of Rock-Lizard-Spock becomes
possible, the coexistence of five species becomes possible. Actually,
Fig.~6(b) [{\em reproduced here as the upper graph in Figure~\ref{fig:ZhongFig6}}] shows that the occurrence of asymptotic five-species state
displays a bump at around $M$$=$$3$$\times$$10^{-6}$.'' \cite[Section 3.3]{zhong_etal_2022_ablatedRPSLS}
\end{quotation}

In Figure~\ref{fig:Z3b_FvsM_compare}  the `bump' in frequency of five-species coexistence visible in the low-$t_\text{max}$ $FvsM$ data for
 $M \in [ \approx$$4$$\times$$10^{-7}, \approx$$2$$\times$$10^{-5} ]$ 
 is still present in the high-$t_\text{max}$ $FvsM$ data, although the longer run-time has resulted in minor reductions in the bump's height and width. 
Clearly some further simulation work is required to make a convincing case for what are the true asymptotic behaviors of the system with the Z3b dominance network. 

\begin{figure}
\begin{center}
\includegraphics[trim=0cm 0cm 0cm 0cm, clip=true, scale=0.425]{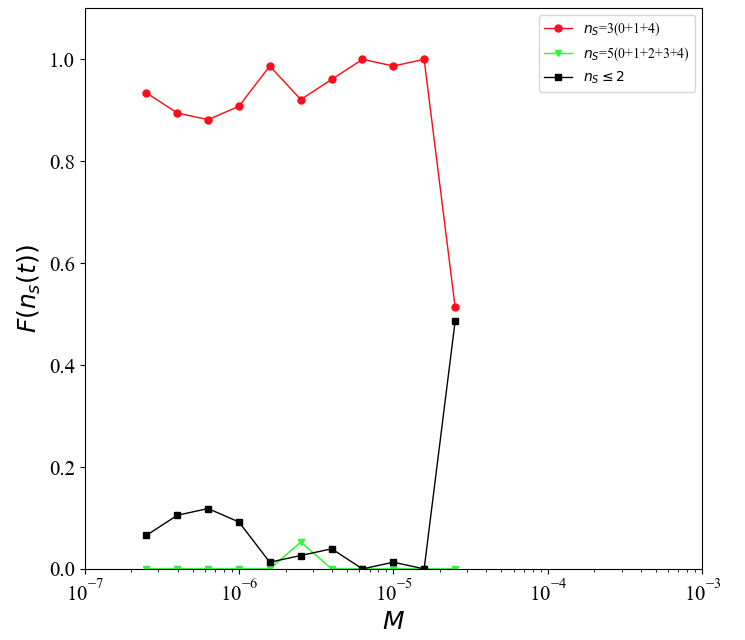}
\end{center}
\caption{
Plot of $FvsM$ for the Z3b ablated dominance network run to $t_\text{max}$$=$$10^7$, one hundred times longer than that used by Zhong et al., data generated only over the range $M$$\in$$[2.51$$\times$$10^{-7},2.51$$\times$$10^{-5},]$, the values for which the $FvsM$ plots from shorter-running experiments showed a `bump' of five-species co-existence.  As can be seen, the bump has almost entirely disappeared as a consequence of increasing the run-time to  $t_\text{max}$$=$$10^7$.
Format and legend labels as for Figure~\ref{fig:ZhongFig3}.
}
\label{fig:Z3b_FvsM_1e7MCS}
\end{figure}

\begin{figure}
\begin{center}
\includegraphics[trim=0cm 0cm 0cm 0cm, clip=true, scale=0.7]{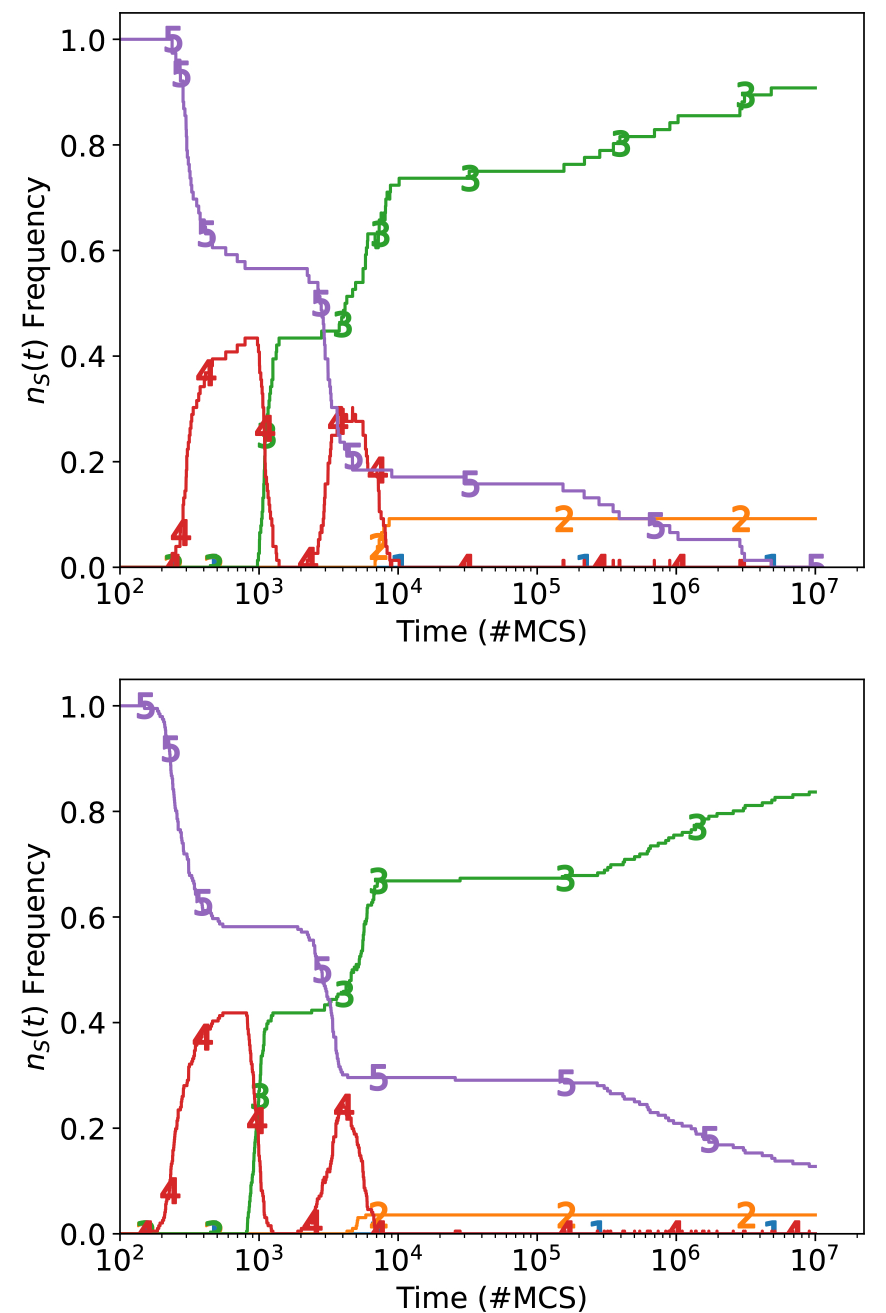}
\end{center}
\caption{
Plots of $FvsT$ for the Z3b ablated dominance network run to $t_\text{max}$$=$$10^7$, one hundred times longer than that used by Zhong et al.: upper graph is for $M$$=$$1.58$$\times$$10^{-6}$; lower graph is for $M$$=$$2.51$$\times$$10^{-6}$. In both graphs, at $t$$=$$10^5${\sc mcs} (the value of $t_\text{max}$ used by Zhong et al.\ to determine the ``asymptotic behavior'' of the system) the frequency of five-species coexistence $F(n_s(t)$$=$$5)$$>$$20\%$ but when the experiment is continued to $t_\text{max}$$=$$10^7$,  the value of $F(n_s(t)$$=$$5)$ steadily decays toward zero, and it seems reasonable to conjecture that zero is its true asymptotic value. 
Format as for Figure~\ref{fig:Z2a_NST_M1e-07}. 
}
\label{fig:Z3b_FvsT_1e7MCS}
\end{figure}

\begin{figure}[h]
\begin{center}
\includegraphics[trim=0cm 0cm 0cm 0cm, clip=true, scale=0.35]{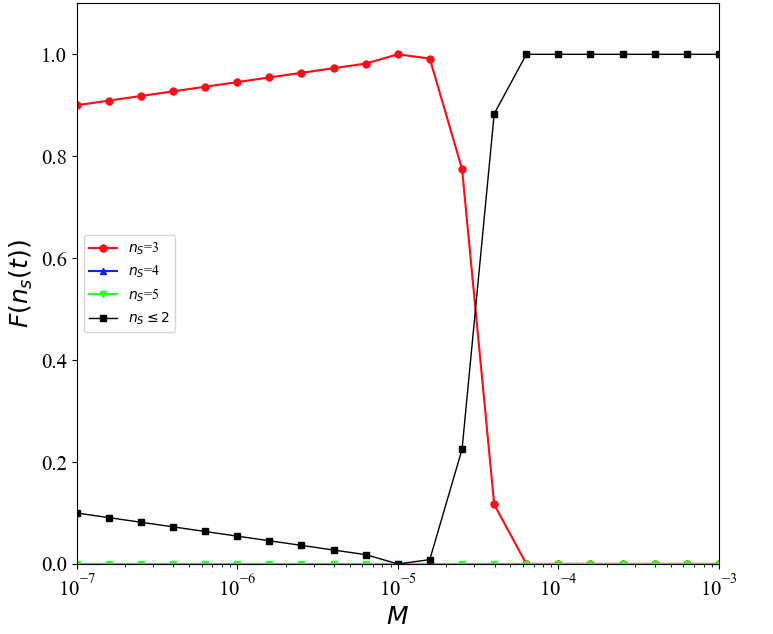}
\end{center}
\caption{
Plot of predicted/projected $FvsM$ for the true asymptotic behavior of the Z3b ablated dominance network, based on the data previously shown in 
Figures~\ref{fig:Z3b_FvsM_compare} and 
\ref{fig:Z3b_FvsM_1e7MCS}, and with $F(n_s(t)$$=$$5)$ predicted to be zero at $M$$=$$2.51$$\times$$10^{-6}$ for the reasons given in the text.
The $F$ values shown for $M$$\geq$$10^{-5}$ 
are results from simulations lasting $t_\text{max}$$=$$10^6$
(from Figure~\ref{fig:Z3b_FvsM_compare});
the $F$ values for $10^{-7}$$\leq$$M$$<$$10^{-5}$  are synthetic data showing outcomes predicted on the basis of Figure~\ref{fig:Z3b_FvsM_1e7MCS} and the arguments given in the text. 
}
\label{fig:Z3b_composite_FvsM}
\end{figure}

To that end, Figure~\ref{fig:Z3b_FvsM_1e7MCS} shows an $FvsM$ plot of outcomes from a set of Z3b experiments where $t_\text{max}$$=$$10^7$, ten times longer than the $t_\text{max}$$=$$10^6$ Z3b results shown in 
Figure~\ref{fig:Z3b_FvsM_compare},
and 100 times longer than the $t_\text{max}$$=$$10^5$ used by Zhong et al.\ in their experiments. As can be seen, when run to $t_\text{max}$$=$$10^7$, the `bump' of five-species coexistence is almost completely flattened to zero. The slowly decaying transient of five-species coexistence, which causes that bump in the $FvsM$ plot when the experiments are not run for long enough, is further illustrated in the  $FvsT$ plots of Figure~\ref{fig:Z3b_FvsT_1e7MCS}. However, for the bump to be completely eliminated, and the true asymptotic behavior of the system to be visible in the $FvsM$ plot, clearly the simulation would need to be run even longer than $t_\text{max}$$=$$10^7$: looking at the rate of decay for $F(n_s(t)$$=$$5)$ in the lower graph of Figure~\ref{fig:Z3b_FvsT_1e7MCS}, it seems likely to hit zero by $t_\text{max}$$\approx$$10^8$. Given that the Z3b $F(n_s(t)$$=$$5)$ decays to zero for all other values of $M$, I think it intuitively reasonable to conjecture -- to make a projection or prediction -- that the true asymptotic behavior Z3b at $M$$\approx$$2.5$$\times$$10^{-6}$ is  $F(n_s(t)$$=$$5)$$=$$0$, which gives the $FvsM$ plot for the true asymptotic behavior of the Z3b system that is shown in Figure~\ref{fig:Z3b_composite_FvsM}.

\newpage
\clearpage

To summarise the critique thus far: the true asymptotic $FvsM$ plots for
Z1 (shown in Figure~\ref{fig:Z1_FvsM_compare}),
Z2a (Figure~\ref{fig:Zhong3_FvsM_1000k}),
and Z3b (Figure~\ref{fig:Z3b_composite_FvsM}),
are all visualizations of essentially the same outcome, and are also equally similar to the $FvsM$ results for ablated dominance networks Z2b and Z3a shown in Zhong et al.'s Figures~5b and~6a.  That is, for each of the five ablated {\sc Rpsls} systems Z1 to Z3b studied by Zhong et al., when the system is run with a sufficiently large value of $t_\text{max}$, the true asymptotic behavior is essentially identical across all five cases. The only thing that varies is how many time-steps the simulations of the differently ablated networks have to be run for, for the system to converge on its asymptotic behavior.

In the next section, I go on to show that the true asymptotic behavior of the original {\sc Rpsls} system with the {\em unablated}\/ dominance network Z0 is {\em also} essentially identical to the asymptotic behaviors of the five ablated {\sc Rpsls} systems Z1 to Z3b.

\newpage
\clearpage


\section{Baseline: dynamics of unablated {\sc Rpsls} at $N$$=$$200$$\times$$200$}
\label{sec:NA0}

\begin{figure}
\begin{center}
\includegraphics[trim=0cm 0cm 0cm 0cm, clip=true, scale=0.375]{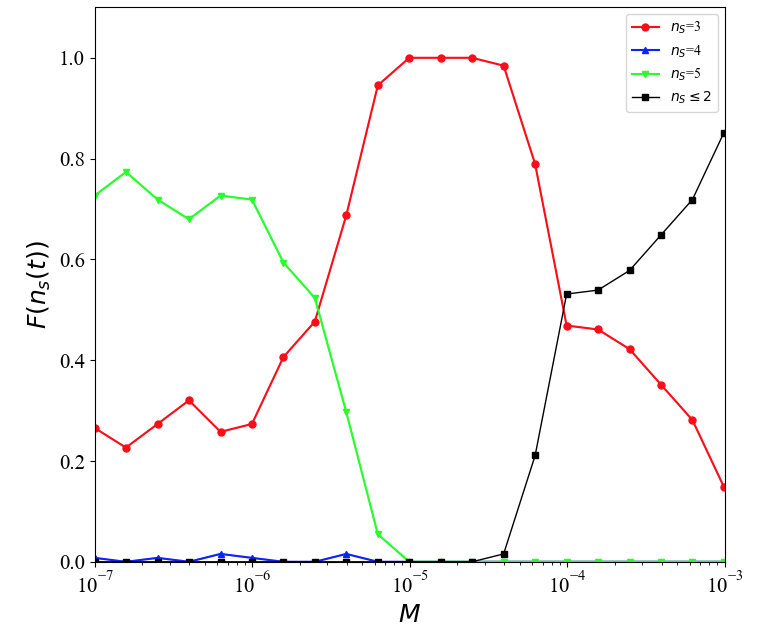}
\includegraphics[trim=0cm 0cm 0cm 0cm, clip=true, scale=0.375]{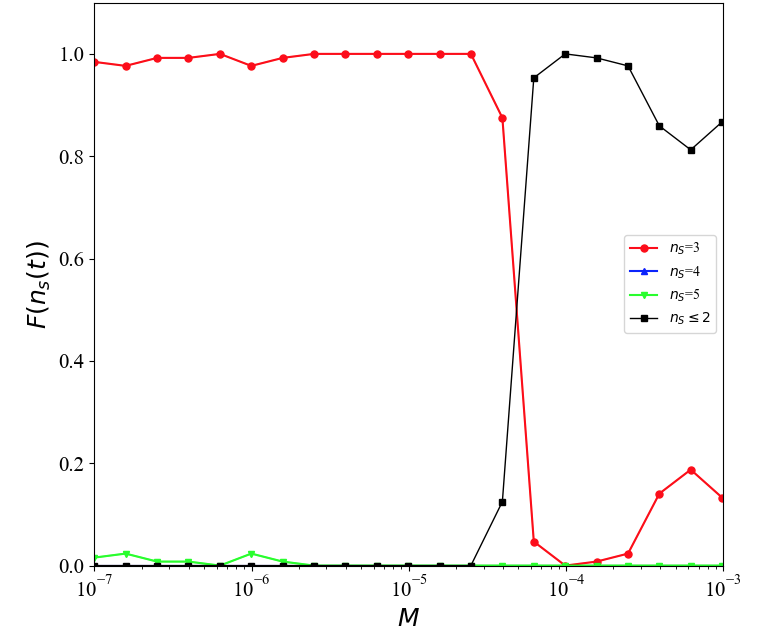}
\end{center}
\caption{
Plots of $FvsM$ for the unablated ($N_a$$=$$0$) {\sc Rpsls} dominance network Z0 (illustrated in Figure~\ref{fig:RPSLSDomNets}) with $N$$=$$200$$\times$$200$, for two different durations of experiment. Upper graph is from experiment duration $t_{\text{max}}$$=$$10^{5}${\sc mcs} as used by Zhong et al. (2022); lower graph is from experiment duration $t_{\text{max}}$$=$$10^{6}${\sc mcs}.
Format and legend labels as for Figure~\ref{fig:ZhongFig3}.
}
\label{fig:OES_NS05_NA0_L200_FvsM_1e6MCS}
\end{figure}

Figure~\ref{fig:OES_NS05_NA0_L200_FvsM_1e6MCS} shows $FvsM$ plots for the unablated ($N_a$$=$$0$) original {\sc Rpsls} system with dominance network Z0 (illustrated in Figure~\ref{fig:RPSLSDomNets}) at $t_\text{max}$$=$$10^5$ and $t_\text{max}$$=$$10^6$. As can be seen, the results from  $t_\text{max}$$=$$10^5$ are far from asymptotic, given the degree of change brought about by extending the run-time to $t_\text{max}$$=$$10^6$. The Z0 $FvsM$ plot for $t_\text{max}$$=$$10^6$ is similar to those for Z1, Z2a, and Z3 (shown in Figures~\ref{fig:Z1_FvsM_compare}, \ref{fig:Zhong3_FvsM_1000k}, and~\ref{fig:Z3b_composite_FvsM}, respectively),  
except that in the range $M$$\in$$[10^{-4},10^{-3}]$ there is another `bump'', this time in the $F(n_s(t)$$=$$3)$ values, similar to the bump in $F(n_s(t)$$=$$5)$ values highlighted by Zhong et al.\ in their  Z3b results, as discussed in the previous section. And, as with the Z3b results, the presence of this bump in the Z0 $FvsM$ plot is an indication that running experiments to  $t_\text{max}$$=$$10^6$ is too short for the true asymptotic behavior of this system to be shown. 

\begin{figure}
\begin{center}
\includegraphics[trim=0cm 0cm 0cm 0cm, clip=true, scale=0.4]{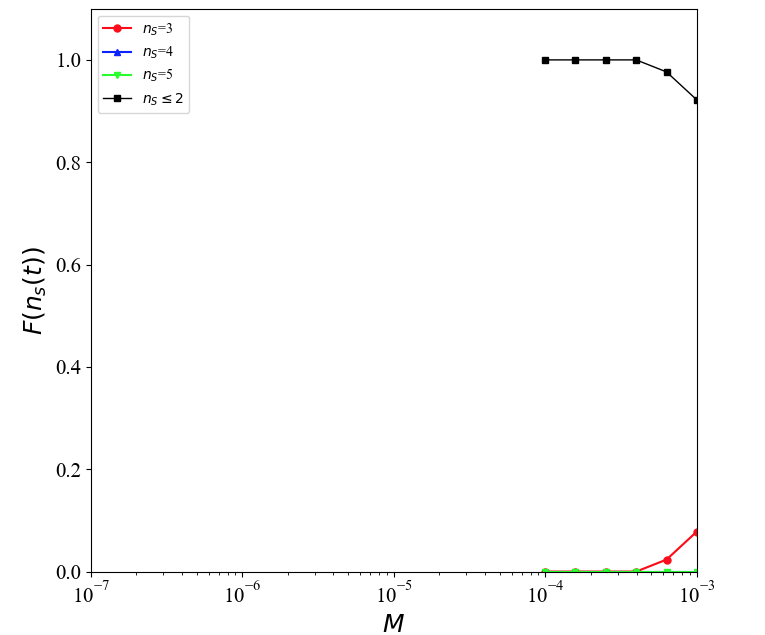}
\end{center}
\caption{
Plot of $FvsM$  at  $t_{\text{max}}$$=$$10^{7}${\sc mcs} for the unablated ($N_a$$=$$0$) {\sc Rpsls} dominance network Z0 
(illustrated in Figure~\ref{fig:RPSLSDomNets}) with $N$$=$$200$$\times$$200$, 
over the range of $M$ values where the `bump' in $F(n_s(t)$$=$$3$$)$ 
occurred in the $FvsM$ plot for  $t_{\text{max}}$$=$$10^{6}${\sc mcs} shown in Figure~\ref{fig:OES_NS05_NA0_L200_FvsM_1e6MCS}: 
here, with a tenfold longer run-time, the bump in $F(n_s(t)$$=$$3$$)$ has almost entirely disappeared, remaining nonzero in this plot only at $M$$>$$\approx$$6$$\times$$10^{-4}$. Format and legend labels as for Figure~\ref{fig:ZhongFig3}.
}
\label{fig:OES_NS05_NA0_L200_FvsM_1e7MCS}
\end{figure}

\newpage
\clearpage

\begin{figure}[h]
\begin{center}
\includegraphics[trim=0cm 0cm 0cm 0cm, clip=true, scale=0.7]{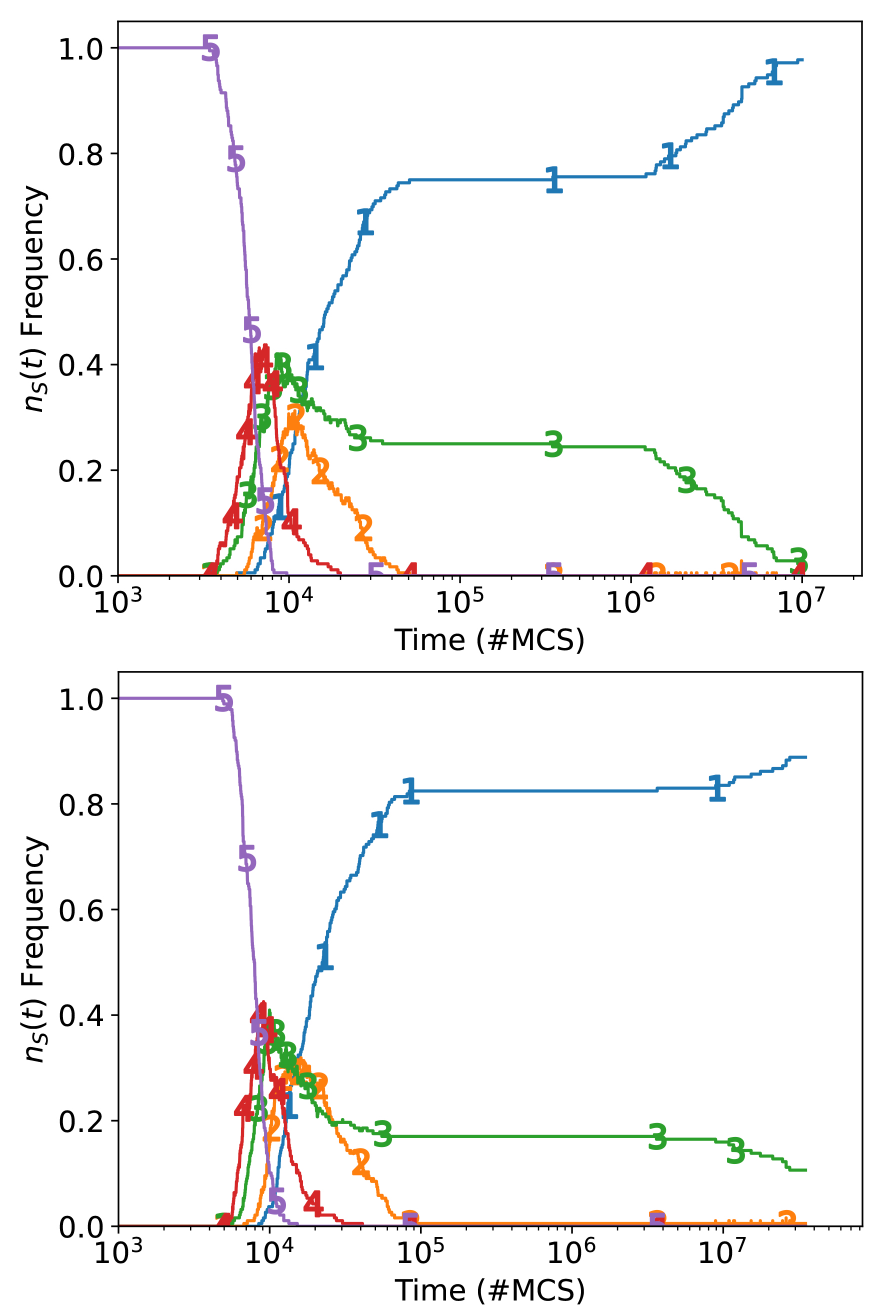}
\end{center}
\caption{
Plots of $FvsT$ to $t_\text{max}$$\geq$$10^7$ for the unablated ($N_a$$=$$0$) original {\sc Rpsls} dominance network Z0 
(illustrated in Figure~\ref{fig:RPSLSDomNets}) with $N$$=$$200$$\times$$200$: upper graph shows results 
to $t$$=$$10^7$
for $M$$=$$6.3$$\times$$10^{-4}$ 
from $N_\text{\sc iid}$$=$$175$ simulations, where $F(n_s(t)$$=$$3)$ holds almost constant at $\approx$$25\%$ for 
$4$$\times$$10^4$$\leq$$t$$\leq$$10^6$, but which then falls steadily toward zero; 
lower graph shows results 
to $t$$=$$3.5$$\times$$10^7$
for $M$$=$$10^{-3}$ from $N_\text{\sc iid}$$=$$400$ simulations, where the rate of decay of
$F(n_s(t)$$=$$3)$ is shallow 
over $10^5$$\leq$$t$$\leq$$8$$\times$$10^6$ but then steepens for 
$t$$>$$8$$\times$$10^6$. Further illustrative $FvsT$ plots for Z0 are given in \ref{sec:app_Z0_NST}.
}
\label{fig:OES_NS05_NA0_L200_NST_1e7MCS}
\end{figure}

\begin{figure}[h]
\begin{center}
\includegraphics[trim=0cm 0cm 0cm 0cm, clip=true, scale=0.475]{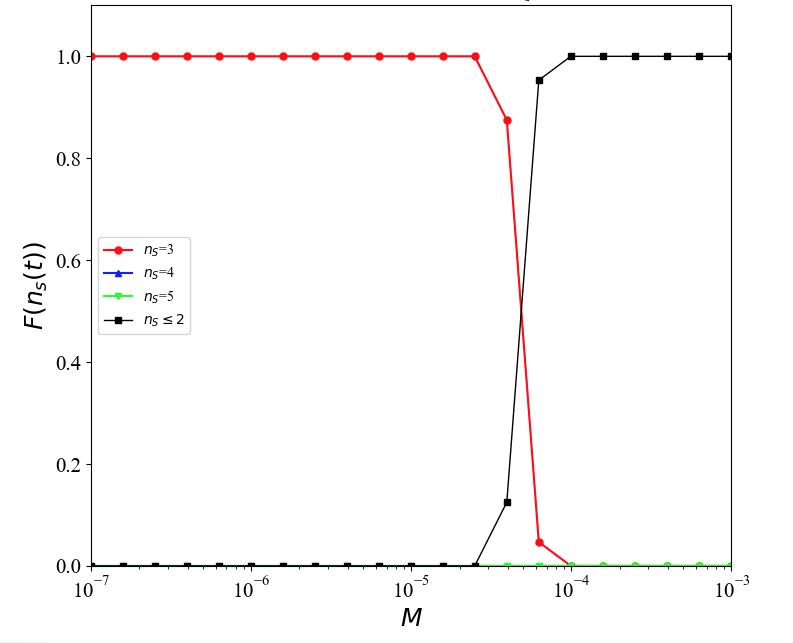}
\end{center}
\caption{
Plot of $FvsM$ for the true asymptotic behavior of the Z0 unablated dominance network (i.e., the original {\sc Rpsls} system) predicted/projected from the data previously shown in Figures~\ref{fig:OES_NS05_NA0_L200_FvsM_1e6MCS}, 
\ref{fig:OES_NS05_NA0_L200_FvsM_1e7MCS}, 
and~\ref{fig:OES_NS05_NA0_L200_NST_1e7MCS}.
The $F$ values for $M$$<$$6.31$$\times$$10^{-4}$ 
are projected results for  $10^6$$<$$t_\text{max}$$\leq$$10^7$;
and
the $F$ values for $M$$\geq$$6.31$$\times$$10^{-4}$  
are projections from the results for $10^7$$<$$t_\text{max}$$\leq$$3$$\times$$10^7$ shown in Figure~\ref{fig:OES_NS05_NA0_L200_NST_1e7MCS}.
Format and legend labels as for Figure~\ref{fig:ZhongFig3}.
}
\label{fig:Z0_composite_FvsM}
\end{figure}

To address that, the Z0 experiments were extended to  $t_\text{max}$$=$$10^7$ for the range of $M$ values where the bump in $F(n_s(t)$$=$$3))$ values is visible in the $t_\text{max}$$=$$10^6$ plot of $FvsM$ plot, and the results from these longer experiments are shown in Figure~\ref{fig:OES_NS05_NA0_L200_FvsM_1e7MCS}: with the longer run-time, the $FvsM$ bump in the $F(n_s(t)$$=3)$ data over range $M$$\in$$[10^{-4},10^{-3}]$ has almost entirely disappeared, remaining visible here only  for $M$$\geq$$6.3$$\times$$10^{-4}$.  
Plots of  $FvsT$ at $t$$\geq$$10^7$ for $M$$=$$6.3$$\times$$10^{-4}$ 
and for $M$$=$$10^{-3}$ are shown in Figure~\ref{fig:OES_NS05_NA0_L200_NST_1e7MCS}: given that for lower values of 
$M$ the $F(n_s(t)$$=$$3)$ values decay to zero by $t$$=$$10^7$, it seems intuitively obvious that these two long-lasting transients in $F(n_s(t)$$=$$3)$
are both monotonically approaching zero once  $t$$>$$10^5$ (with the consequence that $F(n_s(t)$$=$$1)$$\rightarrow$$1.0$) and would eventually reach zero if the durations of the experiments were sufficiently extended. That is, a reasonable projection to make here is that the true asymptotic behavior for the Z0 system with $M$$\geq$$6.3$ is $F(n_s(t)$$=$$3))$$=$$0$ and $F(n_s(t)$$=$$1))$$=$$1$: this projection is illustrated in the summary Z0 asymptotic $FvsM$ plot of Figure~\ref{fig:Z0_composite_FvsM}.

\newpage
\clearpage

To summarise the critique thus far, for ease of comparison Figure~\ref{fig:Z0Z1Z2Z3_FvsM_summary} shows again the true asymptotic $FvsM$ plots for each of networks Z0, Z1, Z2a, and Z3b. 
These are all essentially the same, and so it seems that the introduction of one or more ablations to the {\sc Rpsls} network does not qualitatively alter the asymptotic outcomes of the system, but instead merely changes the number of {\sc mcs} time-steps needed to be simulated for the system to fully converge onto its true asymptotic state.

\def\ScalePNG{0.22}

\begin{figure}[h]
\centering
\subfloat[LoF text][True asymptotic $FvsM$ for Z0]{
\includegraphics[trim=0cm 0cm 0cm 0cm, clip=true, scale=\ScalePNG]{./Figures/Z0_FvsM_projected.png}
\label{fig:Z0_true_FvsM}}
\qquad
\subfloat[LoF text][True asymptotic $FvsM$ for Z1]{
\includegraphics[trim=0cm 0cm 0cm 0cm, clip=true, scale=\ScalePNG]{./Figures/NST_OES_NS05_NA1_L200_Z1_500kMCS}
\label{fig:Z1_true_FvsM}}
\qquad
\subfloat[LoF text][True asymptotic $FvsM$ for Z2a]{
\includegraphics[trim=0cm 0cm 0cm 0cm, clip=true, scale=\ScalePNG]{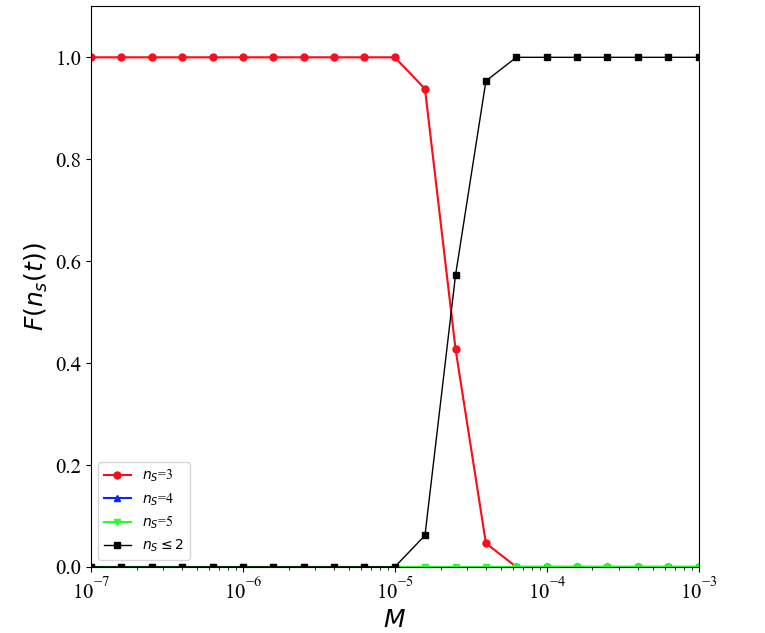}
\label{fig:Z2a_true_FvsM}}
\qquad
\subfloat[LoF text][True asymptotic $FvsM$ for Z3b]{
\includegraphics[trim=0cm 0cm 0cm 0cm, clip=true, scale=\ScalePNG]{./Figures/Z3b_FvsM_projected.png}
\label{fig:Z3b_true_FvsM}}
\caption{
Summary $FvsM$ plots of true asymptotic behaviors of the {\sc Rpsls} system with ablated dominance networks: 
(a) Z0 (shown previously in Figure~\ref{fig:Z0_composite_FvsM});
(b) Z1 (at $t_\text{max}$$=$$5$$\times$$10^5${\sc mcs}, shown previously in Figure~\ref{fig:Z1_FvsM_compare});
(c) Z2a (at $t_\text{max}$$=$$10^6${\sc mcs}, shown previously in Figure~\ref{fig:Zhong3_FvsM_1000k});
(d) Z3b (shown previously in Figure~\ref{fig:Z3b_composite_FvsM}).
As can be seen, in all four graphs, for $M$ below some threshold value $M_\theta$$\approx$$10^{-5}$ the outcome is always $n_s(t)$$=$$3$ for Z0, Z1 and Z2a, and  $n_s(t)$$=$$3$ in $\geq$$90\%$ of the Z3b outcomes; and then once $M$$\geq$$M_\theta$ there is a sharp transition to outcomes always being $n_s(t)$$\leq$$2$ for all four systems.
Format and legend labels as for Figure~\ref{fig:ZhongFig6}.
}
\label{fig:Z0Z1Z2Z3_FvsM_summary}
\end{figure}


\newpage
\clearpage

\section{What about Z4?} 
\label{sec:Z4}

Understanding why Z4 gives the results shown in Figure~\ref{fig:ZhongFig7} remains something of a work in progress, and for that reason Z4 is dealt with separately here.

Zhong et al.\ in \cite{zhong_etal_2022_ablatedRPSLS} offered no discussion of potential causal explanations for why the Z4 results are as they are, opting instead to offer only a verbal description of features manifestly evident from visual inspection of the $FvsM$ plot. Discussing their Z4 results, the entirety of Zhong et al.'s narrative on the $FvsM$ plot is as follows:

\begin{quotation}
``In this case, only one prey-predator interaction remains in the alternative competition. 
We consider its representative example shown in Fig.1(e) 
[{\em reprinted here as Z4 in Figure~\ref{fig:ZhongDomNets}}].
The results are presented in Fig.7 
[{\em reprinted here as the upper plot in Figure~\ref{fig:ZhongFig7}}]. The possible asymptotic behaviors include five-species state, three-species state (Rock-Lizard-Paper), two-species states and single-species states. The latter two are the states of noninteracting species. Interestingly, the state of noninteracting species may exist at low mobility whose occurrence displays a non-mono\-tonic behavior against the mobility.
The competition between three-species state and five-species state also induces interesting behaviors. The occurrence of
the five-species state displays a maximum at around $M$$=$$2$$\times$$10^{-6}$
In contrast, the occurrence of the three-species state shows a dip
at around $M$$=$$2$$\times$$10^{-6}$ and reaches its maximum at around $M$$=$$3$$\times$$10^{-5}$ 
before yielding to the state of non-interacting species.''
\cite[Section 3.4]{zhong_etal_2022_ablatedRPSLS}
\end{quotation}

\begin{figure}
\begin{center}
\includegraphics[trim=0cm 0cm 0cm 0cm, clip=true, scale=0.3]{./Figures/Zhong7_FvsM_1.0e+5MCS_leq2.png}
\includegraphics[trim=0cm 0cm 0cm 0cm, clip=true, scale=0.3]{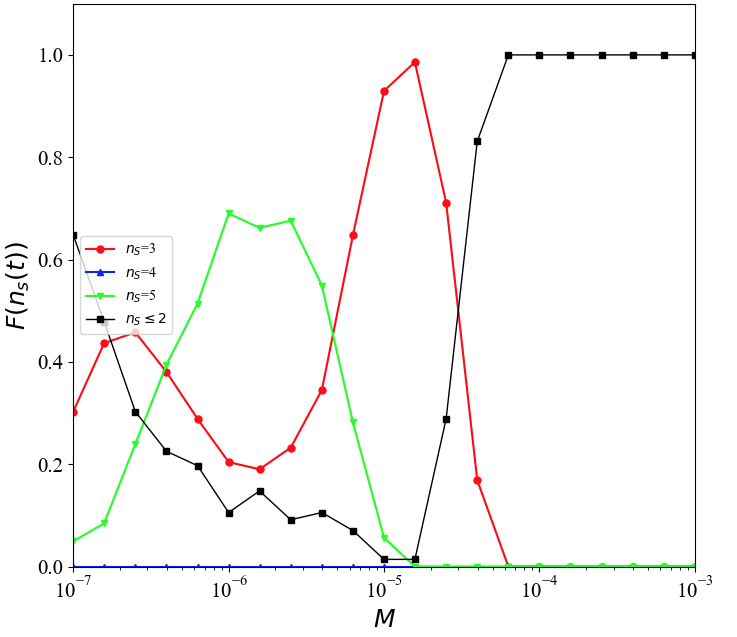}
\end{center}
\caption{
Plots of $FvsM$ for the $N_a$$=$$4$ {\sc Rpsls} dominance network Z4, results from which were shown in Zhong et al.'s Figure~7;  
$N$$=$$200$$\times$$200$, for two different durations of experiment. Upper graph is from experiment duration $t_{\text{max}}$$=$$170$k{\sc mcs}, which gives the best match to the results published by  Zhong et al.\ in \cite{zhong_etal_2022_ablatedRPSLS}; lower graph is from experiment duration $t_{\text{max}}$$=$$10^{6}${\sc mcs}. 
The graph for $t_{\text{max}}$$=$$10^{5}${\sc mcs} shows a large ``bump'' in $F(n_s(t_\text{max})$$=$$5)$
(the frequency of occurrence of the experiment ending with five-species coexistence) centred at $M$$\approx$$2$$\times$$10^{-6}$.
The longer run-time brings some small reductions in $F(n_s(t_\text{max})$$=$$5)$ at $M$$=$$10^{-7}$ and at $M$$=$$10^{-5}$,  but the bulk of the `bump' is otherwise essentially unchanged.
Format and legend labels as for Figure~\ref{fig:ZhongFig3}.
}
\label{fig:Z4_FvsM_compare}
\end{figure}

Figure~\ref{fig:Z4_FvsM_compare} shows plots of $FvsM$ for outcomes from Z4 simulations, results from which were shown in Zhong et al.'s Figure~7, for two different durations of experiment: the upper graph is from my experiments with $t_{\text{max}}$$=$$170$k{\sc mcs}, which gives the best match to Zhong et al.'s results; the lower graph is from my experiments with $t_{\text{max}}$$=$$10^{6}${\sc mcs}.

As with Zhong et al.'s Z4 $FvsM$ plot, my replication's $FvsM$ plot for $t_{\text{max}}$$=$$10^{5}${\sc mcs} shows a large ``bump'' in  $F(n_s(t_\text{max})$$=$$5)$ (i.e., the frequency of occurrence of five-species coexistence)  centered at $M$$\approx$$2$$\times$$10^{-6}$.
The longer run-time brings some small reductions in $F(n_s(t_\text{max})$$=$$5)$ at $M$$=$$10^{-7}$ and at $M$$=$$10^{-5}$,  but the bulk of the bump in  $F(n_s(t_\text{max})$$=$$5)$ is otherwise unchanged. {\em Prima facie}, this appear likely to be another case where, when the simulation is run for longer, the bump will disappear. 
Additional runs were executed with the Z4 system at $t_{\text{max}}$$=$$10^{7}${\sc mcs}, i.e.\ running the system for an additional nine million {\sc mcs} but this had very little impact on the shape, the gross morphology, of the bump in $F(n_s(t_\text{max})$$=$$5)$ values on the $FvsM$ graph.

\begin{figure}
\begin{center}
\includegraphics[trim=0cm 0cm 0cm 0cm, clip=true, scale=0.7]{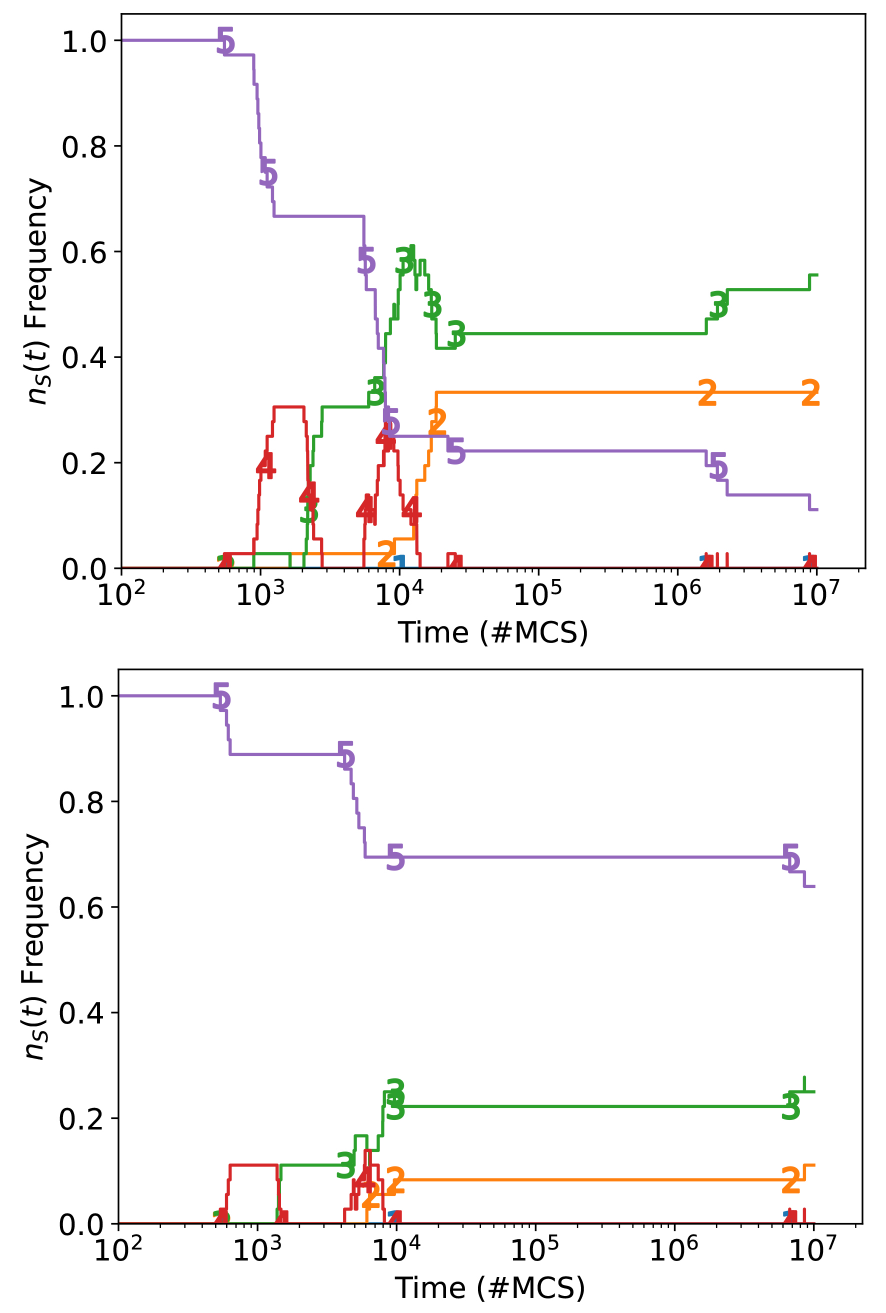}
\end{center}
\caption{
Plots of $FvsT$ from the Z4 system at $t_{\text{max}}$$=$$10^{7}${\sc mcs} for values of $M$ in the range over which the ``bump'' in $F(n_s(t)$$=$$5)$ was prominent in the $FvsM$ plots of Figure~\ref{fig:Z4_FvsM_compare}.
Upper graph shows $FvsT$ for $M$$=$$6.3$$\times$$10^{-7}$;
lower graph shows $FvsT$ for $M$$=$$6.3$$\times$$10^{-7}$. Format and legend labels for both graphs are as for Figure~\ref{fig:Z2a_NST_M1e-07}.  
In both graphs there is a long period of equilibrium over $3\times$$10^4$$<$$t$$<$$1.5$$\times$$10^6$, 
which could be  mistaken for asymptotic behavior, but then from $t$$\approx$$1.5$$\times$$10^6$ in the upper graph and 
$t$$\approx$$6$$\times$$10^6$ in the lower graph, the value of $F(n_s(t)$$=$$5)$ starts to drop, with commensurate rises in $F(n_s(t)$$=$$3)$ and $F(n_s(t)$$=$$2)$.
}
\label{fig:Z4_FvsT_1e7mcs}
\end{figure}

However, Figure~\ref{fig:Z4_FvsT_1e7mcs} shows $FvsT$ plots for two values of $M$ at $t_\text{max}$$=$$10^7$: in both plots, the system appears to have settled to an asymptotic equilibrium with $F(n_s(t)$$=$$c)$ for $c \in \{2, 3, 5\}$ flatlining over a long period of apparent stasis over $3$$\times$$10^{4}$$<$$t$$<$$10^6$, but then at $t$$\approx$$1.5$$\times$$10^6$ in the upper plot and at  $t$$>$$6$$\times$$10^6$ in the lower plot, we start to see a reduction in  $F(n_s(t)$$=$$5)$ and commensurate rises in the frequencies of two-species and three-species co-existence. Clearly this is a situation in which running the simulation to $t_\text{max}$$=$$10^8$ or $t_\text{max}$$=$$10^9$ would be necessary to identify the true asymptotic behaviors, but where again it seems reasonable to conjecture that the eventual true asymptotic behavior will be for $F(n_s(t)$$=$$5)\rightarrow 0$ as $t \rightarrow \infty$.  That would mean the `bump' of five-species outcomes disappears, but it would not explain why the frequency of $n_s(t_\text{max})$$\leq$$2$ outcomes for Z4 is $\approx$$0.6$ at $M$$=$$10^{-7}$, falling steadily to near-zero at $M$$=$$10^{-5}$, before sharply rising to unity by  $M$$=$$10^{-4}$: explaining this multi-phasic response in the $F(n_s(t_\text{max})$$\leq$$2)$ (which is also evident, albeit to a  much lesser extent, in the Z3b asymptotic behavior $FvsM$ plot of Figure~\ref{fig:Z3b_true_FvsM}) remains one topic for further work. Additional topics for further work are discussed in the next section.

\newpage
\clearpage

\section{Discussion and Further Work}
\label{sec:discussion}

The primary aims of this paper were to independently reproduce the results of Zhong et al.\ published in \cite{zhong_etal_2022_ablatedRPSLS}, to identify the true asymptotic behaviors for the ablated {\sc Rpsls} systems, and to compare those to the true asymptotic behavior of the original unablated Z0 network. The results presented thus far have achieved all of those aims. Conclusions drawn from my results are discussed in Section~\ref{sec:conclusion}.

Working on this replication has felt reminiscent of a notable replication from more than 30 years ago, when a paper by Nowak and May  \cite{nowak_may_1992} was replicated by Huberman and Glance \cite{huberman_glance_1993}. 
The Nowak and May paper, which was the front-cover feature article in {\em Nature}, explored the dynamics of another discrete-time Cartesian-lattice-based evolutionary spatial game system, based not on Rock-Paper-Scissors but instead on the iterated Prisoner's Dilemma (see e.g.\ \cite{axelrod_1984,axelrod_1997}): Nowak and May noted that visualizations of the time-varying spatial patterns of cooperators and defectors distributed over the lattice in their experiments changed chaotically and wrote:``{\em \ldots these ever-changing sequences of spatial patterns -- dynamic fractals -- can be extraordinarily beautiful, and have interesting mathematical properties}''  \cite[Abstract]{nowak_may_1992}. 

Not long later, Huberman and Glance \cite{huberman_glance_1993} published their replication, showing that the ``dynamic fractal'' features highlighted by Nowak and May were simple artefacts of Nowak and May's decision to implement their evolutionary spatial game systems using a synchronous update rule (where, as the discrete-time system transitioned from time-step $t$ to $t$$+$$1$, every cell in the lattice would -- in perfect synchrony -- update to its new state for time $t$$+$$1$ on the basis of the states it observed in its neighbours at time $t$): 
when Huberman and Glance altered  this aspect of the simulation to instead use asynchronous updating, updating one randomly-chosen cell at a time (as is used in the experiments reported in this paper), every cell in the lattice quickly evolved to be defector,  and the intricate spatiotemporal patterns identified by Nowak and May disappeared, visualizations of the lattice becoming instead a single static solid slab of one color. In concluding their paper, Huberman and Glance wrote ``{\em These results show that while computer experiments provide a versatile approach for studying complex systems, an understanding of their subtle characteristics is required in order to reach valid conclusions about real-world systems.''} 

Those words are quite definitely just as relevant here and now as they were when first written in 1993: in this paper I have demonstrated that there are subtle characteristics in the ablated {\sc Rpsls} systems introduced by Zhong et al., characteristics that only reveal themselves when a rigorously principled approach is taken to identifying the true asymptotic outcomes for these systems, and when the outcomes of those ablation treatments are compared to an appropriate control, which in this case is the result from the relavent unablated system, the Z0 network. Conclusions drawn from those results are given below in Section~\ref{sec:conclusion}.

This paper serves to correct the published record on the asymptotic outcomes of the ablated {\sc Rpsls} systems introduced by Zhong et al., but there is clearly more work that could be done to take this line of enquiry further. Current research is proceeding on two fronts. One strand of work is exploring the extent to which a seemingly minor revision to the Elementary Step (Algorithm~\ref{alg:OES}), reported in \cite{cliff_2024_noops}, changes  the population dynamics of the ablated {\sc Rpsls} systems and significantly alters the frequency distribution of observed outcome for all the ablated systems studied in \cite{zhong_etal_2022_ablatedRPSLS}, with extinctions being much rarer events: results from those studies will be published in \cite{cliff_2024_RES_ablatedRPSLS}. 
A second strand of work is exploring the frequency distributions of outcomes in {\sc Rpsls}-style systems with ablated dominance networks  but where $N_s$$>5$, such that any one arc ablation represents a smaller percentage loss of arcs from the overall network: early results from this work have been published in \cite{cliff_2024_circulants} and a more extensive study is currently being worked on, to be released as \cite{bloom_cliff_2024}.  

One additional avenue for exploration in future work is to explore and identify what gives rise to the ``punctuated equilibria'' nature of many of the $FvsT$ plots presented here, where the distribution of frequencies of species-counts over a large number of {\sc iid} simulation experiments is stable for long periods and then undergoes a (relatively) brief period of sudden change, indicating that all the {\sc iid} experiments underwent a species extinction at roughly the same time. For example: in Figure~\ref{fig:Z2a_NST_M1e-07}, for the first $\approx$$150${\sc mcs} all 500 {\sc iid} experiments hold steady at $n_s(t)=5$ but then over the period $t \in [150,1000]${\sc mcs}, every one of the simulations undergoes its first extinction such that by $t$$=$$1000, F(n_s(t)$$=$$5)$$=$$0$ and $F(n_s(t)$$=$$4)$$=$$1$; 
the upper plot $FvsT$ plot in Figure~\ref{fig:Z1_Z3b_FvsT} shows a similar sequence of events, with $F(n_s(t)$$=$$5)$ collapsing from unity to zero and $F(n_s(t)$$=$$4)$ rising rapidly from zero to approach 1.0 over the period running roughly $t \in [500, 2500]$; 
and to give a final example, in the upper $FvsT$ plot of Figure~\ref{fig:Z3b_FvsT_1e7MCS}  there is a long equilibrium running from $t$$\approx$$10^4$ to $t$$\approx$$10^5$ where $F(n_s(t)$$=$$5)$$\approx$$0.35$ and  $F(n_s(t)$$=$$3)$$\approx$$0.65$ and then, from $t$$\approx$$10^5$  onwards, $F(n_s(t)$$=$$5)$ starts to fall steadily and hence $F(n_s(t)$$=$$3)$ starts to rise. 
In all three of these examples, and in every other $FvsT$ plot that shows a similar long-standing equilibrium that is then ``punctuated'' by temporally clustered extinctions occurring across the set of {\sc iid} simulations, presumably during the period of equilibrium there is a some sequence of changes in the lattice, gradually moving the state of the system toward some kind of tipping-point which, once crossed, brings the equilibrium to an end, and this happens on roughly the same timescale for all {\sc iid} repetitions of the specific simulation. 
These observations prompt a set of related research questions: precisely {\em what} is it about the state of the lattice that is changing during the equilibrium periods; what is the best metric by which we may measure that which is changing during the equilibria; and what is the nature of the tipping-point or threshold-crossing that triggers these extinctions that are so strongly temporally concentrated across large numbers of {\sc iid} simulations? One potentially promising line of enquiry here would be to adapt and extend the work of Dong, Li, and Yang \cite{dong_li_yang_2010} who studied extinction patterns in the three-species RPS game, finding that the maximal amplitude of density fluctuation, relative to the average density, and the average Potts energy per lattice-cell are significant factors in determining extinction patterns.

\section{Conclusion}
\label{sec:conclusion}

All research in science and engineering relies on the publication of results that are replicable, and several fields of research have struggled in recent years each with their own ``replicability crisis'' (see e.g.\ \cite{maxwell_lau_howard_2015,milkowski_hensel_hohol_2018,moody_keister_ramos_2022,nosek_etal_2022,jensen_kelly_pedersen_2023}). I have demonstrated here in Section~\ref{sec:zhong_replication} that, with some digging through the prior literature to establish full details of the algorithms employed, the results presented by Zhong et al.\ in \cite{zhong_etal_2022_ablatedRPSLS} can be reproduced\footnote{Writing in \cite{milkowski_hensel_hohol_2018}, Milkowski et al.\ make the distinction  
that {\em replicability} of a computational simulation or modelling experiment involves independent researchers obtaining the same output as the original experiment, using the experiment's original code and data, whereas {\em reproducibility} of a computational model or experiment involves independent researchers being able to recreate the results {\em without} access to the original code and data. Using that distinction strictly, my work reported in this paper demonstrates the reproducibility of Zhong et al.'s results, but is not a replication of those results because I wrote all my own code from scratch.     
} very well: this is, in principle, a good thing.

Zhong et al.'s paper sets out the rationale for their simulation experiments, explains the simulated {\sc Escg} model in some detail, plots  graphs summarising the outcomes of large numbers of experiments, and then offers brief verbal descriptions of features in the plots that are immediately obvious from visual inspection of the graphs. Their narratives on the visualizations seem superficially to be offering detailed causal explanations (such as: {\em``it may allow for five-species coexistence
since the two three-species-cyclic interactions only share one
species, Lizard. Once the coexistence of Rock-Lizard-Spock becomes
possible, the coexistence of five species becomes possible \ldots''}). 

However, my results demonstrate that if Zhong et al.\ had run their experiments for longer, and/or if they had plotted $FvsT$ graphs, they would have needed to radically re-write the stories they tell about the ablated systems they studied. In light of this, Zhong et al.'s narratives run the risk of seeming to be ``just-so-stories'', filling a gap in the paper where a carefully argued and rigorously evaluated causal explanation would normally be expected \cite{smith_2016_justso}. 

Furthermore, as I discussed in more  detail in \cite{cliff_2024_noops}, the only dominance-network ablations that Zhong et al.\ chose to explore in their paper are those that led to reductions in the number of co-existing species, but other equally plausible ablated {\sc Rpsls} dominance networks show no such effect, suffer no extinctions, and their species-count remains constant at five for the entire duration of the experiment: at best,  \cite{zhong_etal_2022_ablatedRPSLS} is selective, and  tells only half the story. 

But the design of experiments in \cite{zhong_etal_2022_ablatedRPSLS} is more deeply questionable on two counts: Zhong et al.\ assert, incorrectly, and without any justification or supporting evidence, that asymptotic behaviors are observable in the ablated {\sc Rpsls} system at $t_\text{max}$$=$$10^5$; and they also chose not to show any results from the original unablated Z0 network, which would serve as a baseline (i.e., a ``pre-treatment'' evaluation or {\em control}) against which meaningful comparisons could then be made with the results from the ``post-treatment'' ablated systems. The reasoning behind these omissions is not clear to me, nor why no-one involved in the peer-review of the manuscript picked up on these issues. I have demonstrated here that when the ablated systems are simulated for longer the results they then produce become, if anything, {\em less} interesting, because the diversity of outcomes reported and narrated on by Zhong et al.\ disappears once each ablated system is given long enough to converge on its true asymptotic outcome: when analysed in $FvsM$ form, the true asymptotic outcomes for the ablated systems Z0-Z3b are very similar to each other, and are each in turn very similar to the $FvsM$ outcome for the data from the control, the unablated Z0 system. 

I have shown that the results presented in Zhong et al.'s paper, and their narrative explanations of those results, are in need of revision and are not as straightforward as Zhong et al.\ set them out to be. When the ablated {\sc Rpsls} systems are explored in more depth, for sufficient durations, and when they are then properly compared to results from the unablated Z0 as a baseline control, the actual effects of the dominance-network ablations seems to be somewhat moot, and hence the results from the ablated {\sc Rpsls} systems are probably not as interesting or relevant to studies of real-world biodiversity and species coexistence as {\sc Escg} research of this type is often claimed to be. 

I've made my source-code available so that other researchers can  replicate, explore,  and (I hope) extend the work presented here. The appendices present additional illustrative $FvsT$ plots, for reference.

\newpage
\clearpage

\section*{Funding}
Computational resources used for the simulation experiments reported here were provided in part by Syritta Ltd, who paid for AWS cloud usage and also provided multiple 8-core Apple Mac computers. Other than that, this research did not receive any specific grant from funding agencies in the public, commercial, or not-for-profit sectors.

\appendix{}
\section[pfx={Appendix\space}]{Dynamics of $N_a$$=$$0$ {{\sc Rpsls}}{Z0} with $\mu$$=$$\sigma$$=$$1.0$, $L$$=$$200$}
\label{sec:app_Z0_NST}

\begin{figure}
\centering \includegraphics[width=0.7\textwidth] {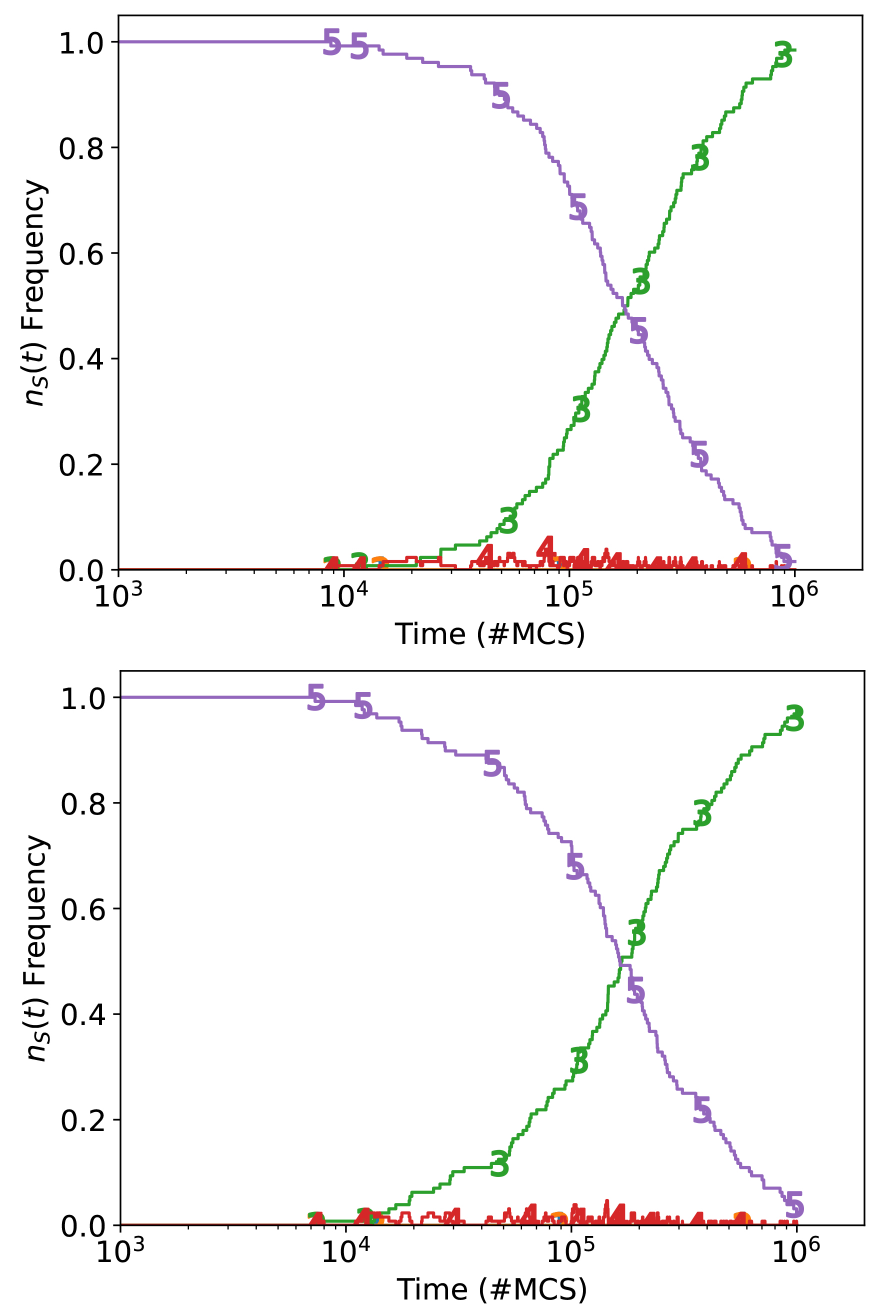}
\caption{
Plots of $FvsT$ 
for {\sc Rpsls} experiments with no ablations ($N_a
$$
=
$$
0$): $L
$$
=
$$
200$; $\mu
$$
=
$$
\sigma
$$
=
$$
1.0$; 
$M
$$
=
$$
10^{-7}$ (upper),
$M
$$
=
$$
10^{-6}$ (lower).
Format as for  Figure~\ref{fig:Z2a_NST_M1e-07}.\label{fig:OES_NS05_NA0_L200_A}}
\end{figure}

\begin{figure}
\centering \includegraphics[width=0.7\textwidth] {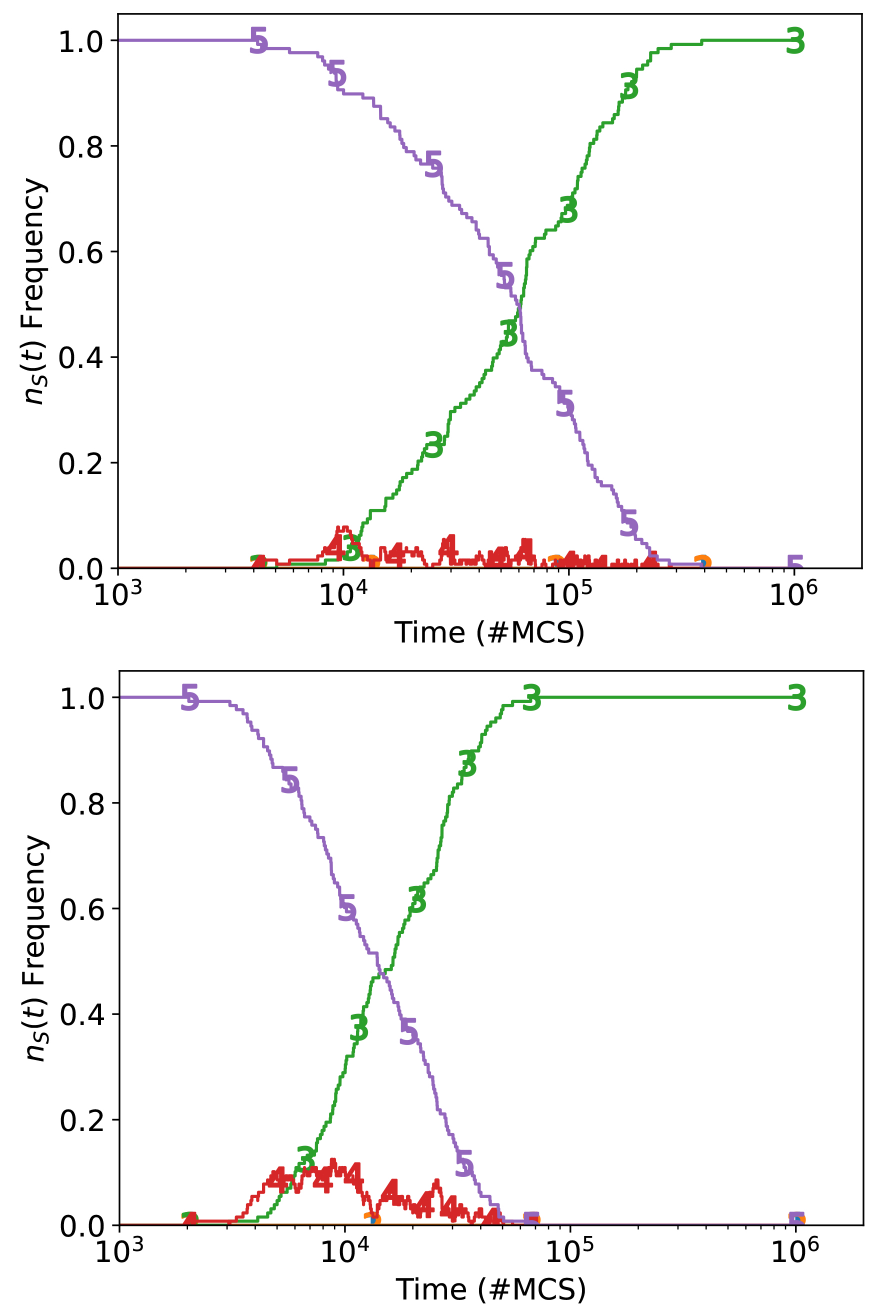}
\caption{
Plots of $FvsT$
for {\sc Rpsls} experiments with no ablations ($N_a
$$
=
$$
0$): $L
$$
=
$$
200$; $\mu
$$
=
$$
\sigma
$$
=
$$
1.0$; 
$M
$$
=
$$
3.98
$$
\times
$$
10^{-6}$ (upper),
$M
$$
=
$$
10^{-5}$ (lower).
Format as for  Figure~\ref{fig:Z2a_NST_M1e-07}.\label{fig:OES_NS05_NA0_L200_B}}
\end{figure}

\begin{figure}
\centering \includegraphics[width=0.7\textwidth] {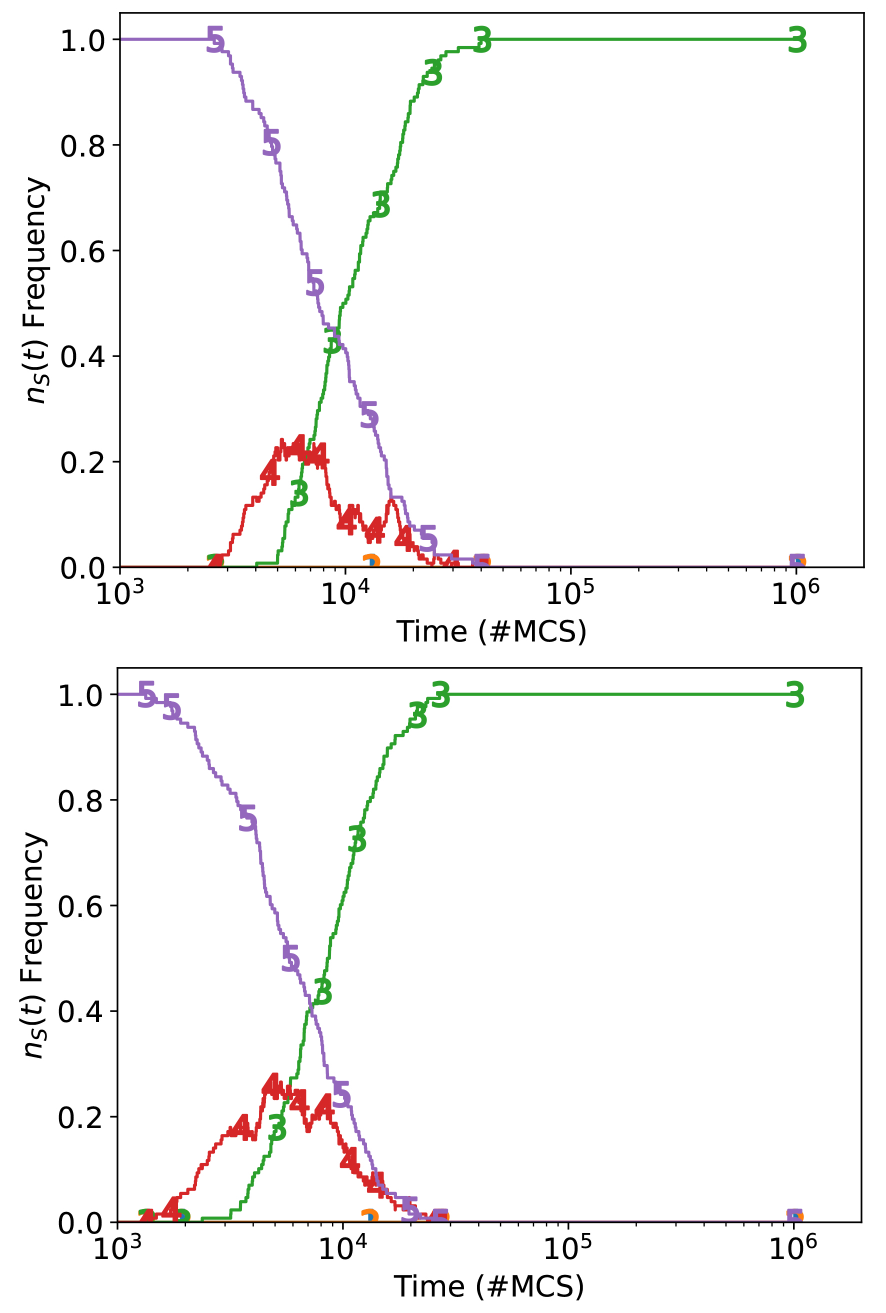}
\caption{
Plots of $FvsT$
for {\sc Rpsls} experiments with no ablations ($N_a
$$
=
$$
0$): $L
$$
=
$$
200$; $\mu
$$
=
$$
\sigma
$$
=
$$
1.0$; 
$M
$$
=
$$
1.58
$$
\times
$$
10^{-5}$ (upper),
$M
$$
=
$$
2.58
$$
\times
$$
10^{-5}$ (lower).
Format as for  Figure~\ref{fig:Z2a_NST_M1e-07}.\label{fig:OES_NS05_NA0_L200_C}}
\end{figure}

\begin{figure}
\centering \includegraphics[width=0.7\textwidth] {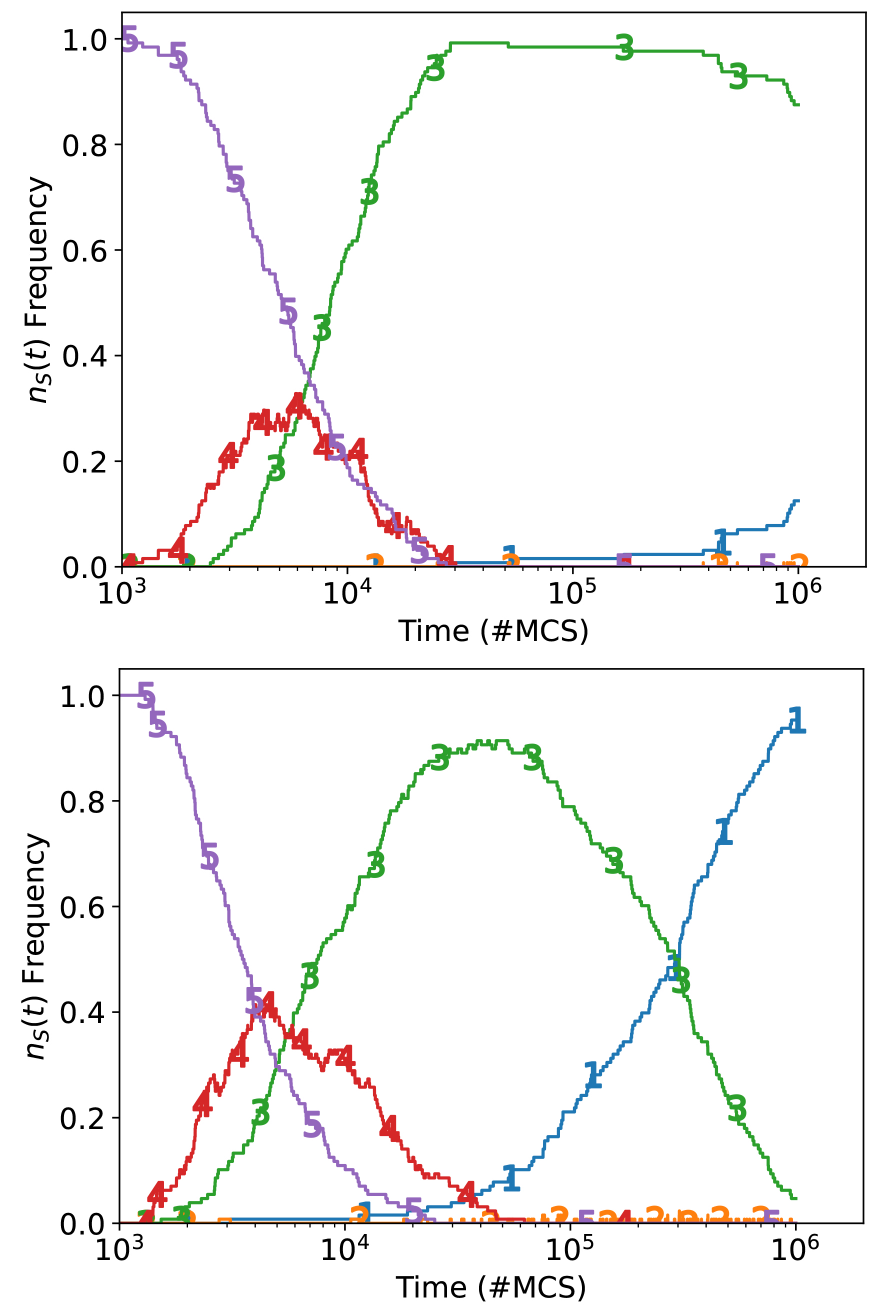}
\caption{
Plots of $FvsT$
for {\sc Rpsls} experiments with no ablations ($N_a
$$
=
$$
0$): $L
$$
=
$$
200$; $\mu
$$
=
$$
\sigma
$$
=
$$
1.0$; 
$M
$$
=
$$
3.98
$$
\times
$$
10^{-5}$ (upper),
$M
$$
=
$$
6.31
$$
\times
$$
10^{-5}$ (lower).
Format as for  Figure~\ref{fig:Z2a_NST_M1e-07}.\label{fig:OES_NS05_NA0_L200_D}}
\end{figure}

\begin{figure}
\centering \includegraphics[width=0.7\textwidth] {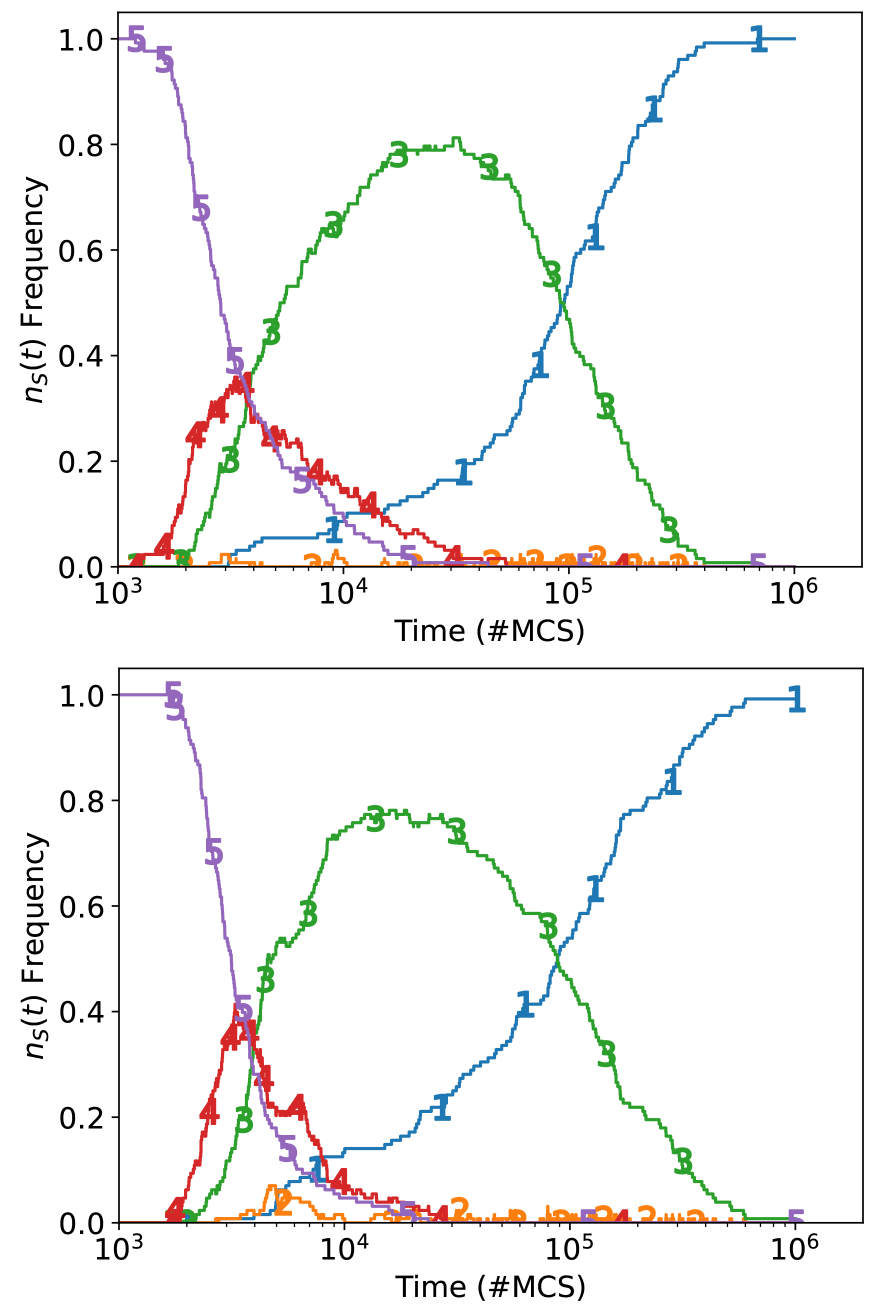}
\caption{
Plots of $FvsT$
for {\sc Rpsls} experiments with no ablations ($N_a
$$
=
$$
0$): $L
$$
=
$$
200$; $\mu
$$
=
$$
\sigma
$$
=
$$
1.0$; 
$M
$$
=
$$
10^{-4}$ (upper),
$M
$$
=
$$
1.58
$$
\times
$$
10^{-4}$ (lower).
Format as for  Figure~\ref{fig:Z2a_NST_M1e-07}.\label{fig:OES_NS05_NA0_L200_E}}
\end{figure}

\begin{figure} 
\centering \includegraphics[width=0.7\textwidth] {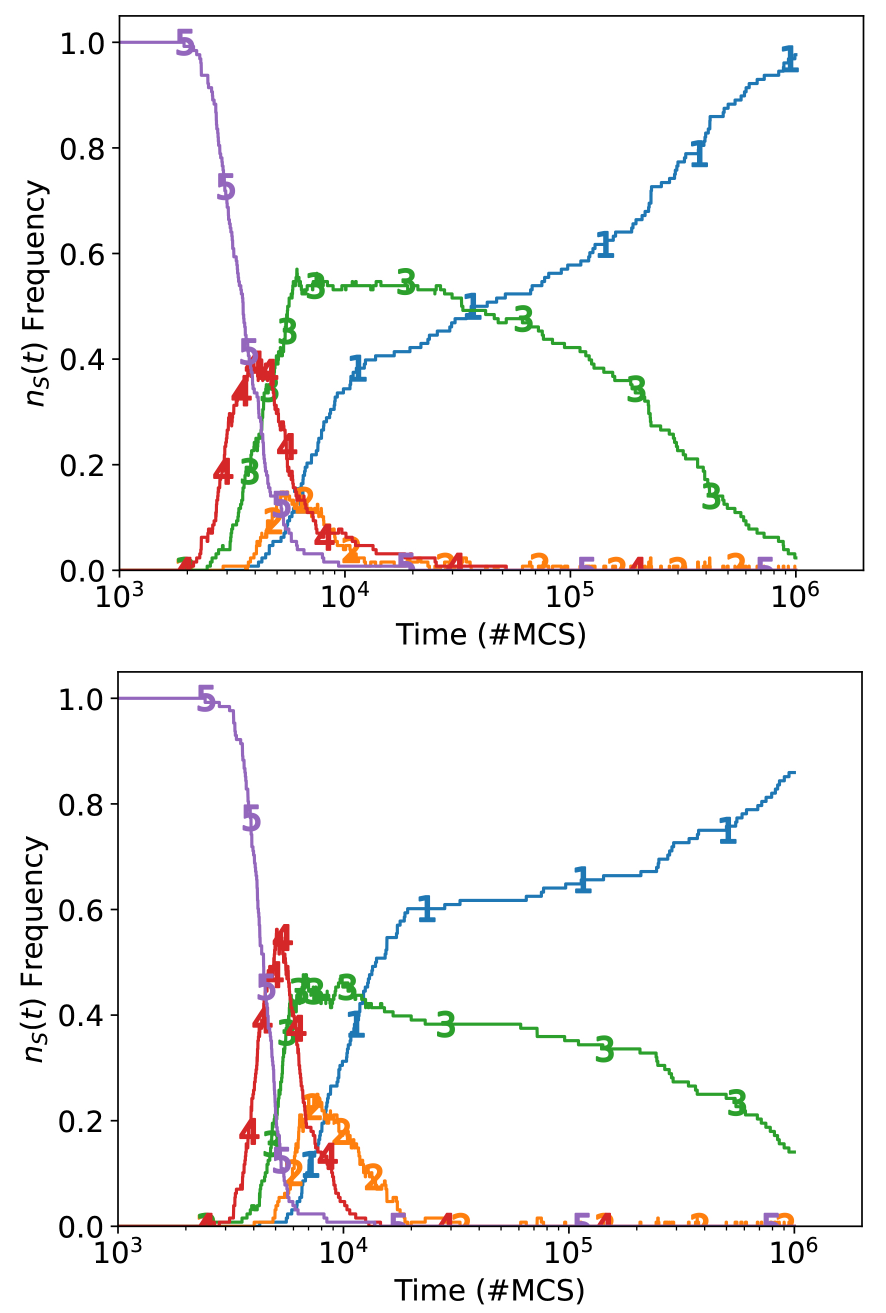}
\caption{
Plots of $FvsT$
for {\sc Rpsls} experiments with no ablations ($N_a
$$
=
$$
0$): $L
$$
=
$$
200$; $\mu
$$
=
$$
\sigma
$$
=
$$
1.0$; 
$M
$$
=
$$
2.51
$$
\times
$$
10^{-4}$ (upper),
$M
$$
=
$$
3.98
$$
\times
$$
10^{-4}$ (lower).
Format as for  Figure~\ref{fig:Z2a_NST_M1e-07}.\label{fig:OES_NS05_NA0_L200_F}}
\end{figure}

\begin{figure}
\centering \includegraphics[width=0.7\textwidth]{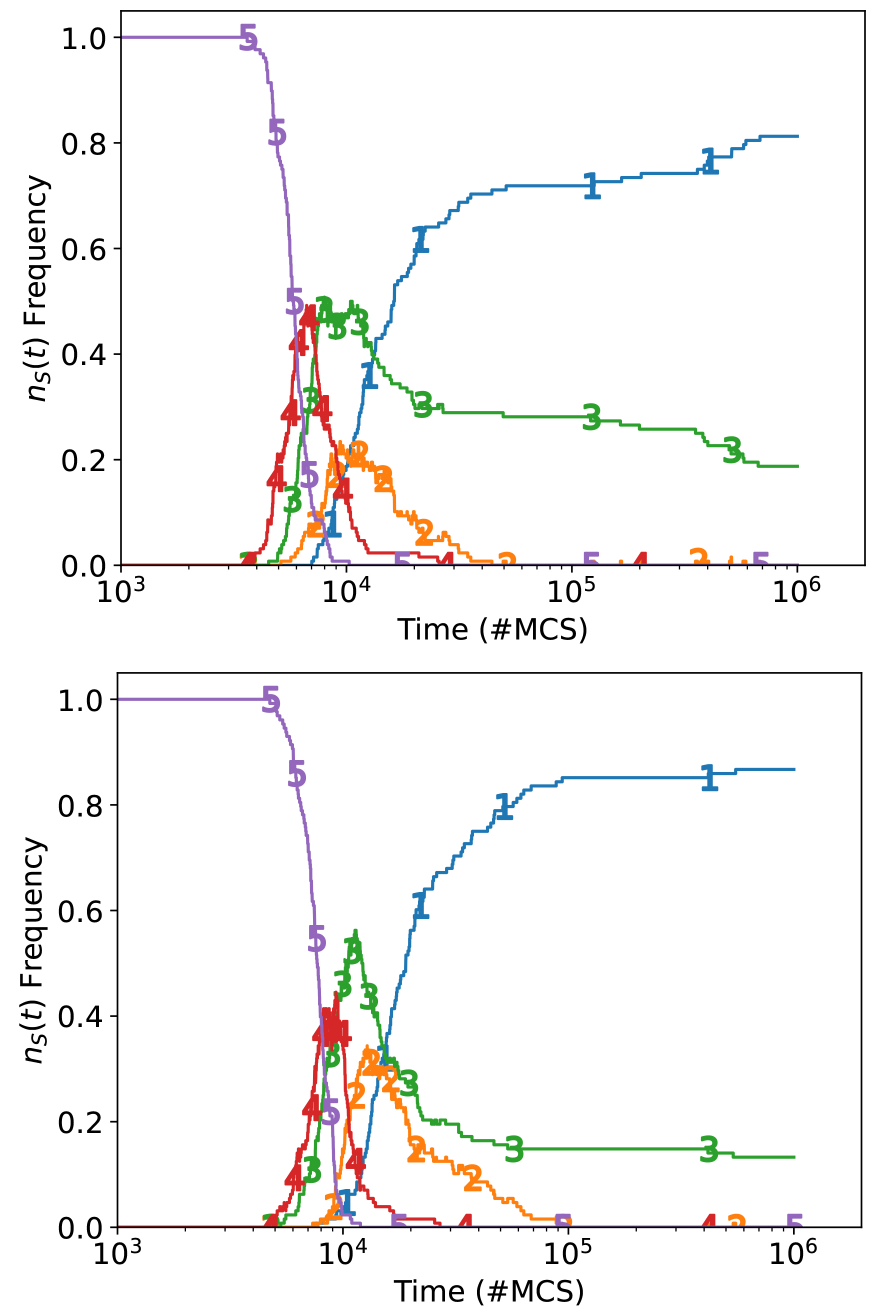}
\caption{
Plots of $FvsT$
for {\sc Rpsls} experiments with no ablations ($N_a
$$
=
$$
0$): $L
$$
=
$$
200$; $\mu
$$
=
$$
\sigma
$$
=
$$
1.0$; 
$M
$$
=
$$
6.31
$$
\times
$$
10^{-4}$ (upper),
$M
$$
=
$$
10^{-3}$ (lower).
Format as for  Figure~\ref{fig:Z2a_NST_M1e-07}.\label{fig:OES_NS05_NA0_L200_G}}
\end{figure}

Figures~\ref{fig:OES_NS05_NA0_L200_A} to~\ref{fig:OES_NS05_NA0_L200_G} show $FvsT$ plots of time-series of the relative frequencies of experiments in which the number of surviving species $n_S(t)
 $$
 =
 $$
 c$ for $c\in\{1,\dots,5\}$, and for $t_\text{max}
 $$
 =
 $$
 10^6${\sc mcs}, for the original {\sc Rpsls} {\sc Escg} with the unablated ($N_a
$$
=
$$
0$) dominance network Z0, as was illustrated in  Figure~\ref{fig:RPSLSDomNets}, with
$L
$$
=
$$
200$ and $\mu
$$
=
$$
\sigma
$$
=
$$
1.0$, and for exponentially spaced values of $M
$$
\in
$$
[ 10^{-7}, 10^{-3}]$, with 128 {\sc iid} simulations executed for each value of $M$ sampled. 

Given that $N_a
$$
=
$$
0$, the progressive reductions
in $n_s(t)$ seen in all these time-series are each underflow extinctions,
i.e.\ they are a consequence of the variation in population dynamics inherent
in the five-species {\sc Rpsls} model running up against the discretization
limit of $L
$$
=
$$
200$: see~\cite{cliff_2024_noops}
for further discussion of this point.

\mbox{}

\section[pfx={Appendix\space}]{Dynamics of $N_a
$$
=
$$
3$ {{\sc Rpsls}}
 {{Z3b}} with $\mu
$$
=
$$
\sigma
$$
=
$$
1.0$, $L
$$
=
$$
200$}
\label{sec:app_NA3_NST}

\begin{figure} 
\centering \includegraphics[width=0.7\textwidth] {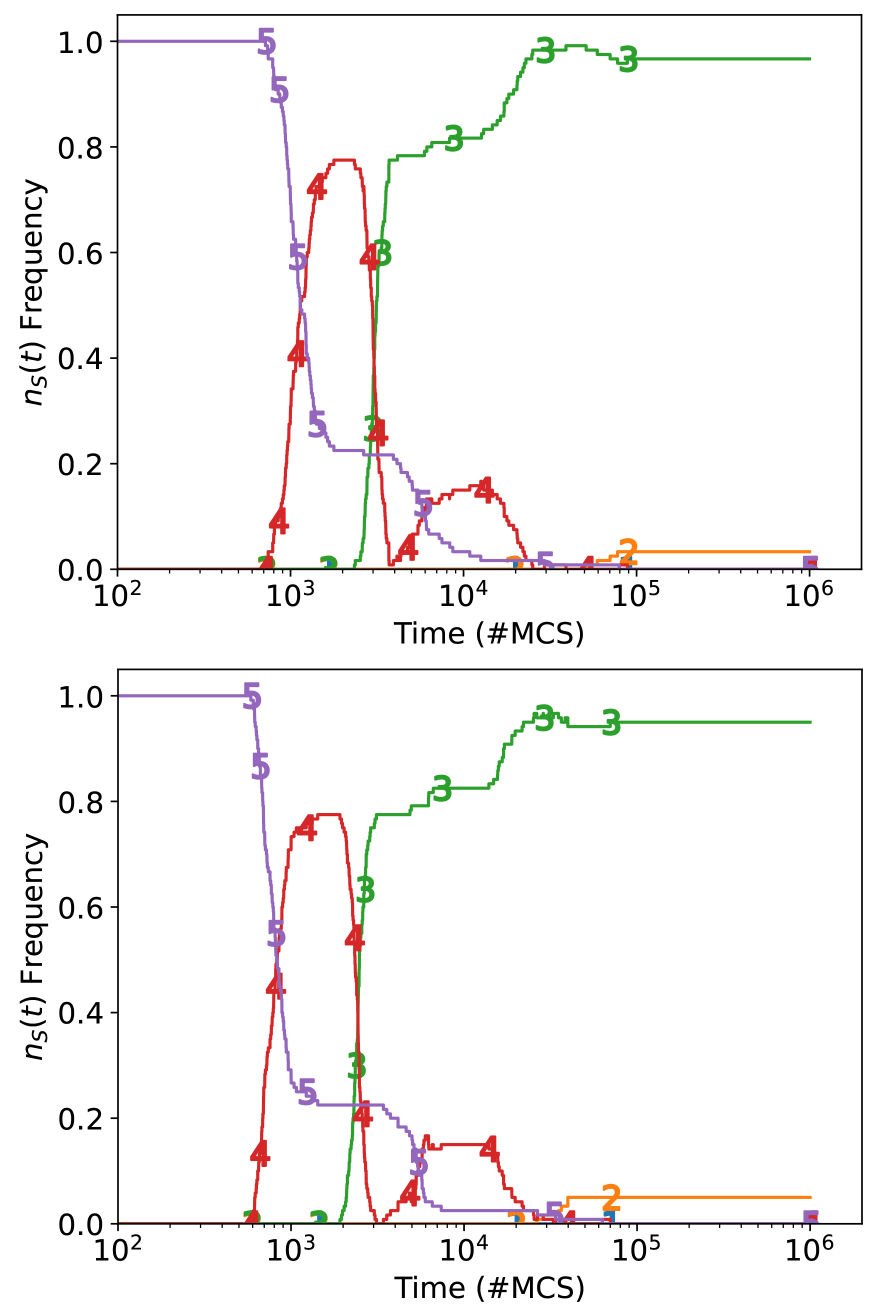}
\caption{
Plots of $FvsT$ 
for {\sc Rpsls} experiments with three ablations ($N_a
$$
=
$$
3$): $L
$$
=
$$
200$; $\mu
$$
=
$$
\sigma
$$
=
$$
1.0$; 
$M
$$
=
$$
10^{-7}$ (upper),
$M
$$
=
$$
1.58
$$
\times
$$
10^{-7}$ (lower).
Format as for  Figure~\ref{fig:Z2a_NST_M1e-07}.\label{fig:OES_NS05_NA3_L200_A}}
\end{figure}

\begin{figure} 
\centering \includegraphics[width=0.7\textwidth] {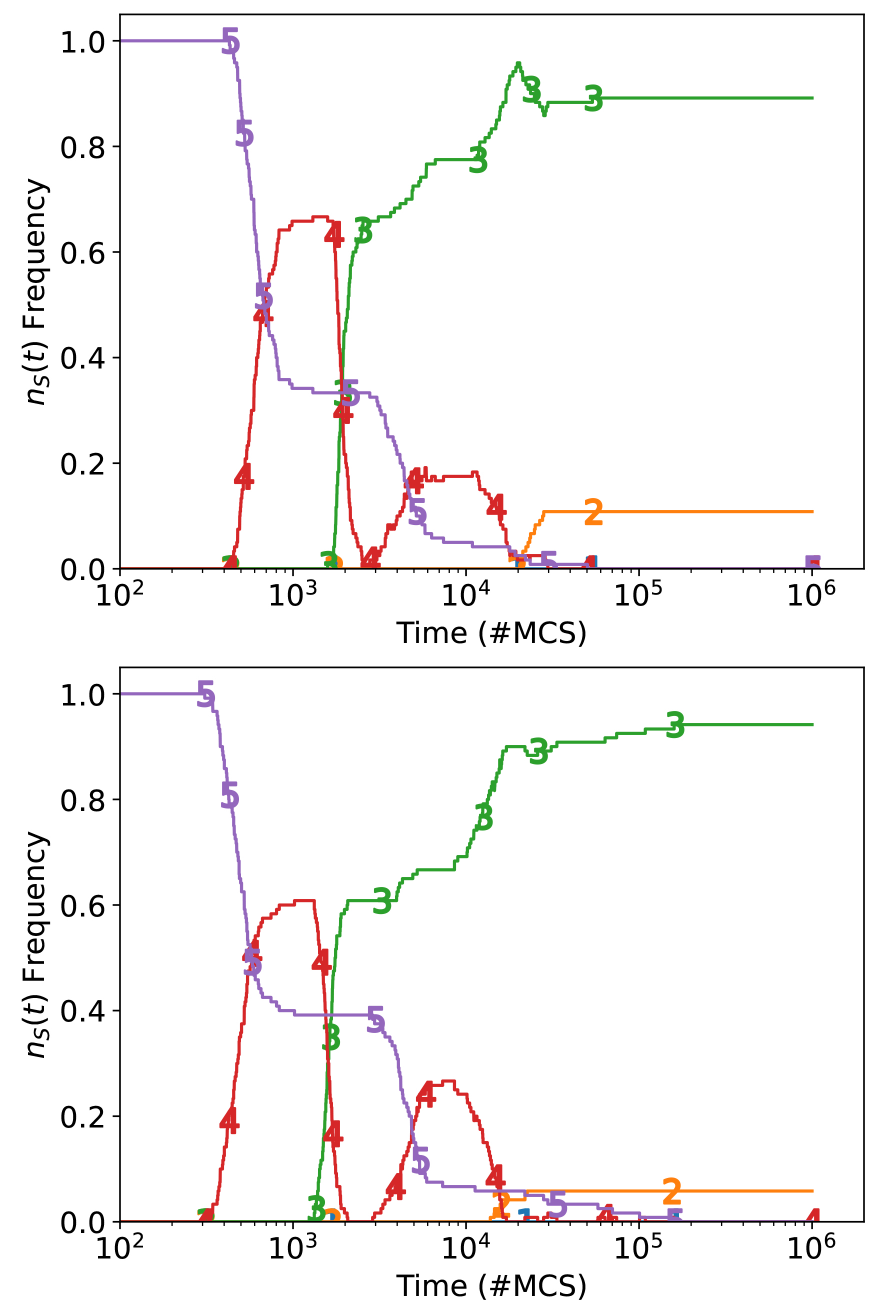}
\caption{
Plots of $FvsT$
for {\sc Rpsls} experiments with three ablations ($N_a
$$
=
$$
3$): $L
$$
=
$$
200$; $\mu
$$
=
$$
\sigma
$$
=
$$
1.0$; 
$M
$$
=
$$
2.51
$$
\times
$$
10^{-7}$ (upper,
$M
$$
=
$$
3.98
$$
\times
$$
10^{-7}$ (lower).
Format as for  Figure~\ref{fig:Z2a_NST_M1e-07}.\label{fig:OES_NS05_NA3_L200_B}}
\end{figure}

\begin{figure} 
\centering \includegraphics[width=0.7\textwidth] {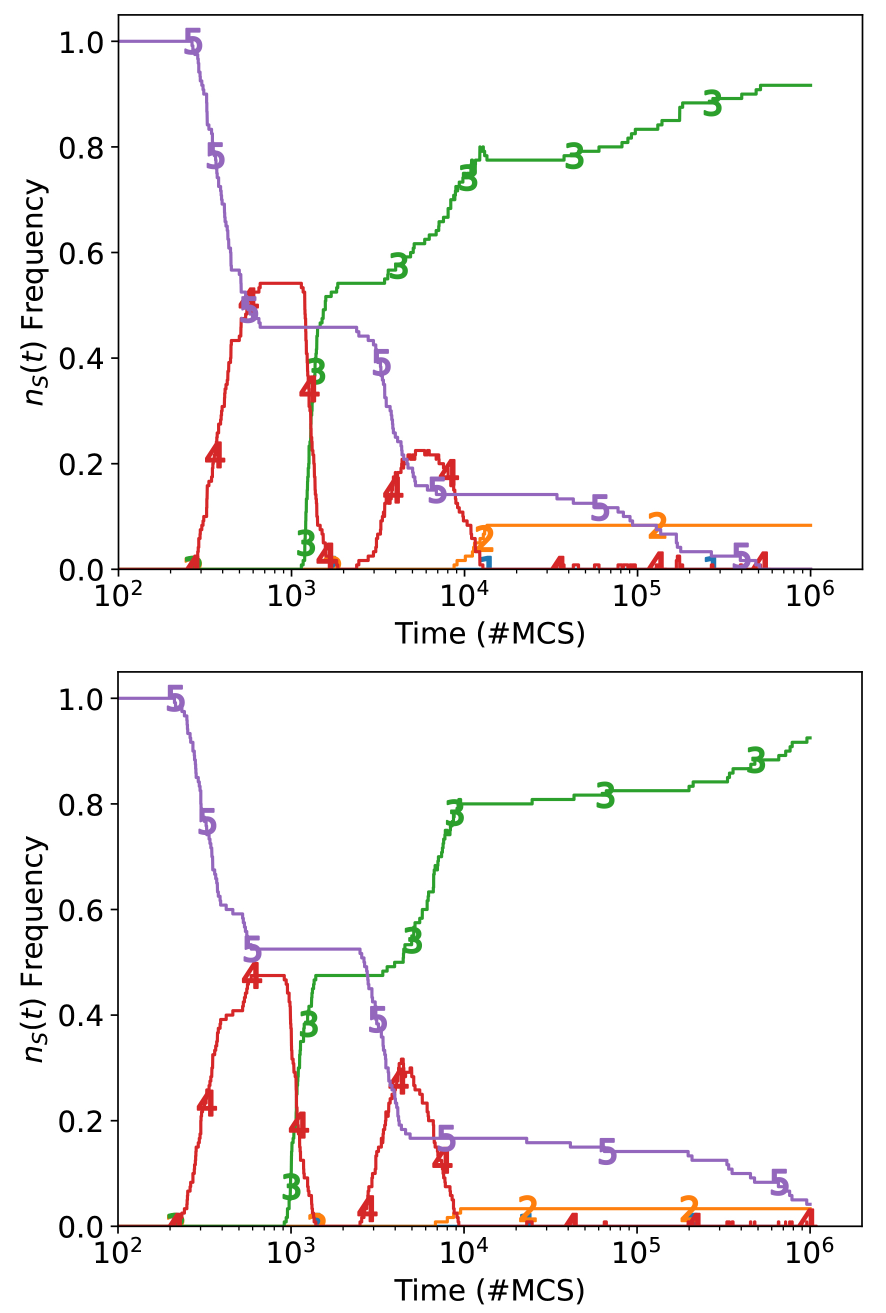}
\caption{
Plots of $FvsT$
for {\sc Rpsls} experiments with three ablations ($N_a
$$
=
$$
3$): $L
$$
=
$$
200$; $\mu
$$
=
$$
\sigma
$$
=
$$
1.0$; 
$M
$$
=
$$
6.31
$$
\times
$$
10^{-7}$ (upper),
$M
$$
=
$$
10^{-6}$ (lower).
Format as for  Figure~\ref{fig:Z2a_NST_M1e-07}.\label{fig:OES_NS05_NA3_L200_C}}
\end{figure}

\begin{figure} 
\centering \includegraphics[width=0.7\textwidth] {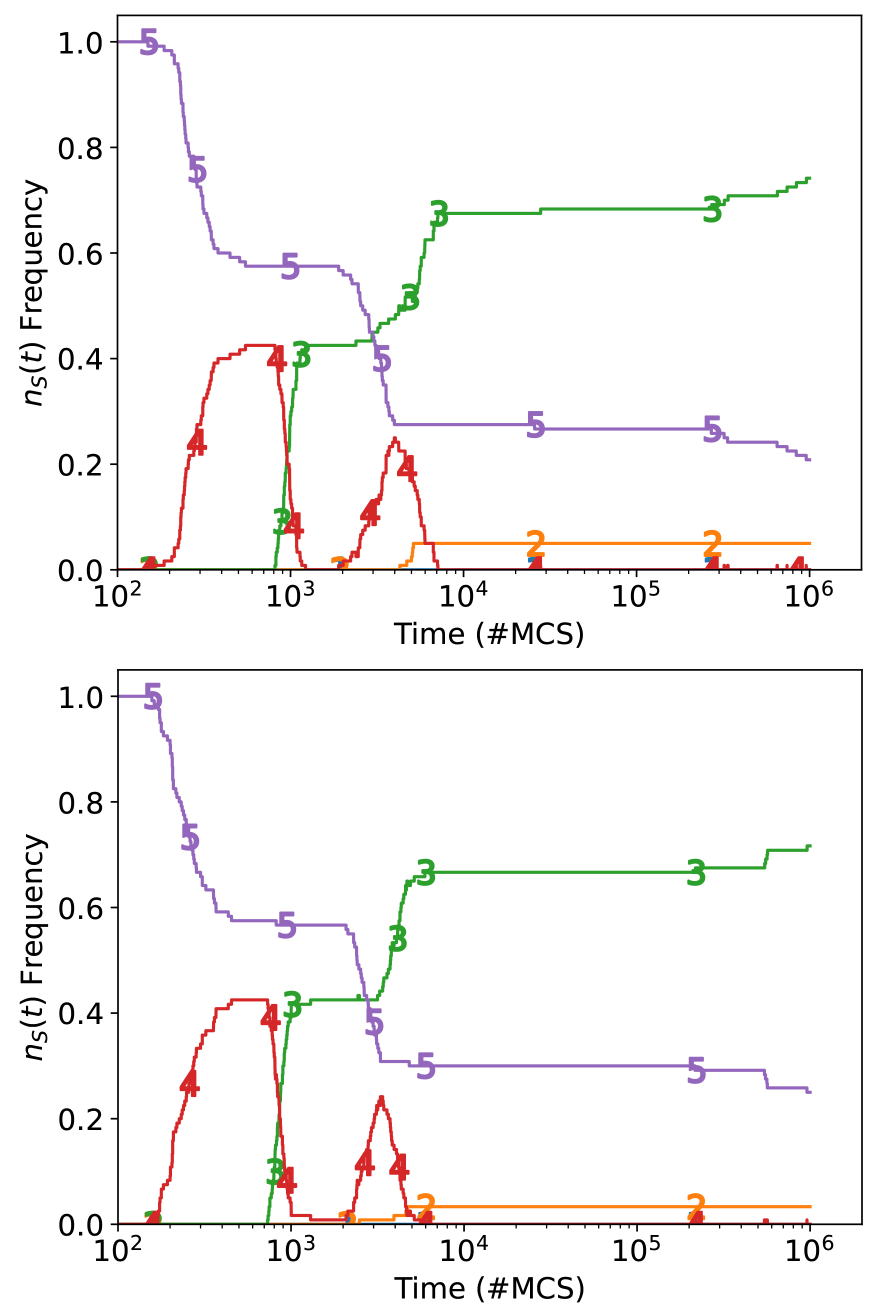}
\caption{
Plots of $FvsT$
for {\sc Rpsls} experiments with three ablations ($N_a
$$
=
$$
3$): $L
$$
=
$$
200$; $\mu
$$
=
$$
\sigma
$$
=
$$
1.0$; 
$M
$$
=
$$
1.58
$$
\times
$$
10^{-6}$ (upper),
$M
$$
=
$$
2.51
$$
\times
$$
10^{-6}$ (lower).
Format as for  Figure~\ref{fig:Z2a_NST_M1e-07}.\label{fig:OES_NS05_NA3_L200_D}}
\end{figure}

\begin{figure} 
\centering \includegraphics[width=0.7\textwidth] {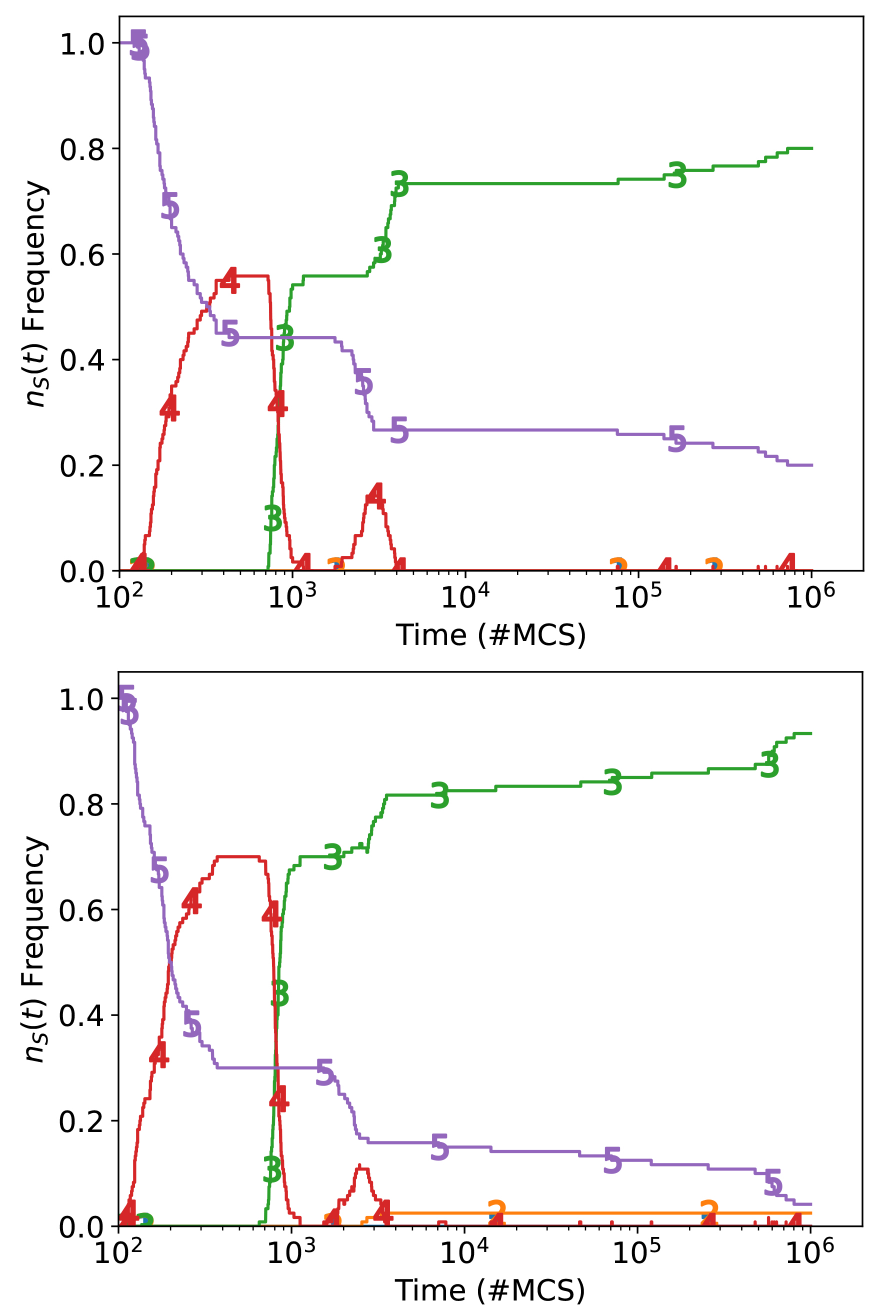}
\caption{
Plots of $FvsT$
for {\sc Rpsls} experiments with three ablations ($N_a
$$
=
$$
3$): $L
$$
=
$$
200$; $\mu
$$
=
$$
\sigma
$$
=
$$
1.0$; 
$M
$$
=
$$
3.98
$$
\times
$$
10^{-6}$ (upper),
$M
$$
=
$$
6.31
$$
\times
$$
10^{-6}$ (lower). 
Format as for  Figure~\ref{fig:Z2a_NST_M1e-07}.\label{fig:OES_NS05_NA3_L200_E}}
\end{figure}

\begin{figure} 
\centering \includegraphics[width=0.7\textwidth] {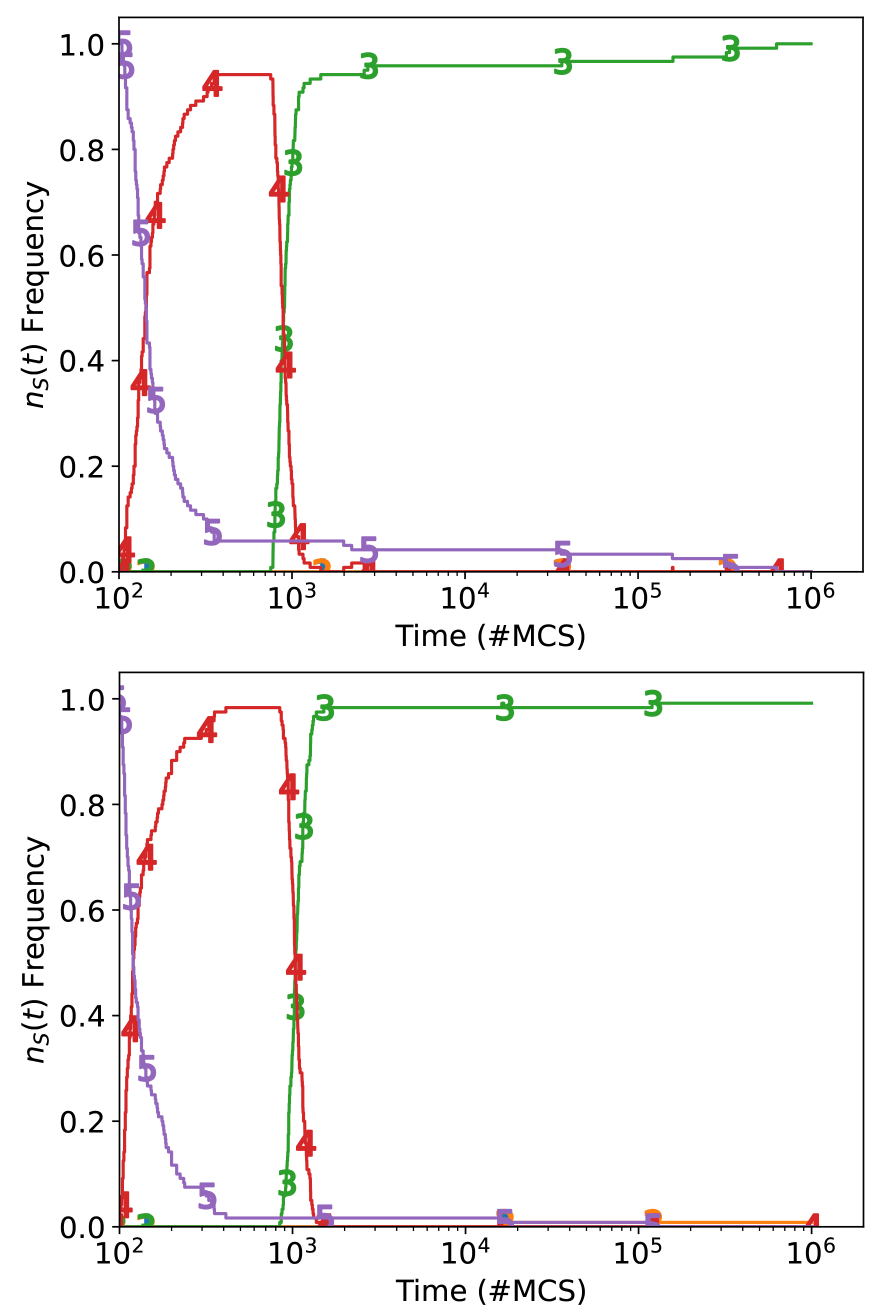}
\caption{
Plots of $FvsT$
for {\sc Rpsls} experiments with three ablations ($N_a
$$
=
$$
3$): $L
$$
=
$$
200$; $\mu
$$
=
$$
\sigma
$$
=
$$
1.0$; 
$M
$$
=
$$
10^{-5}$ (upper),
$M
$$
=
$$
1.58
$$
\times
$$
10^{-5}$ (lower).
Format as for  Figure~\ref{fig:Z2a_NST_M1e-07}.\label{fig:OES_NS05_NA3_L200_F}}
\end{figure}

\begin{figure} 
\centering \includegraphics[width=0.7\textwidth] {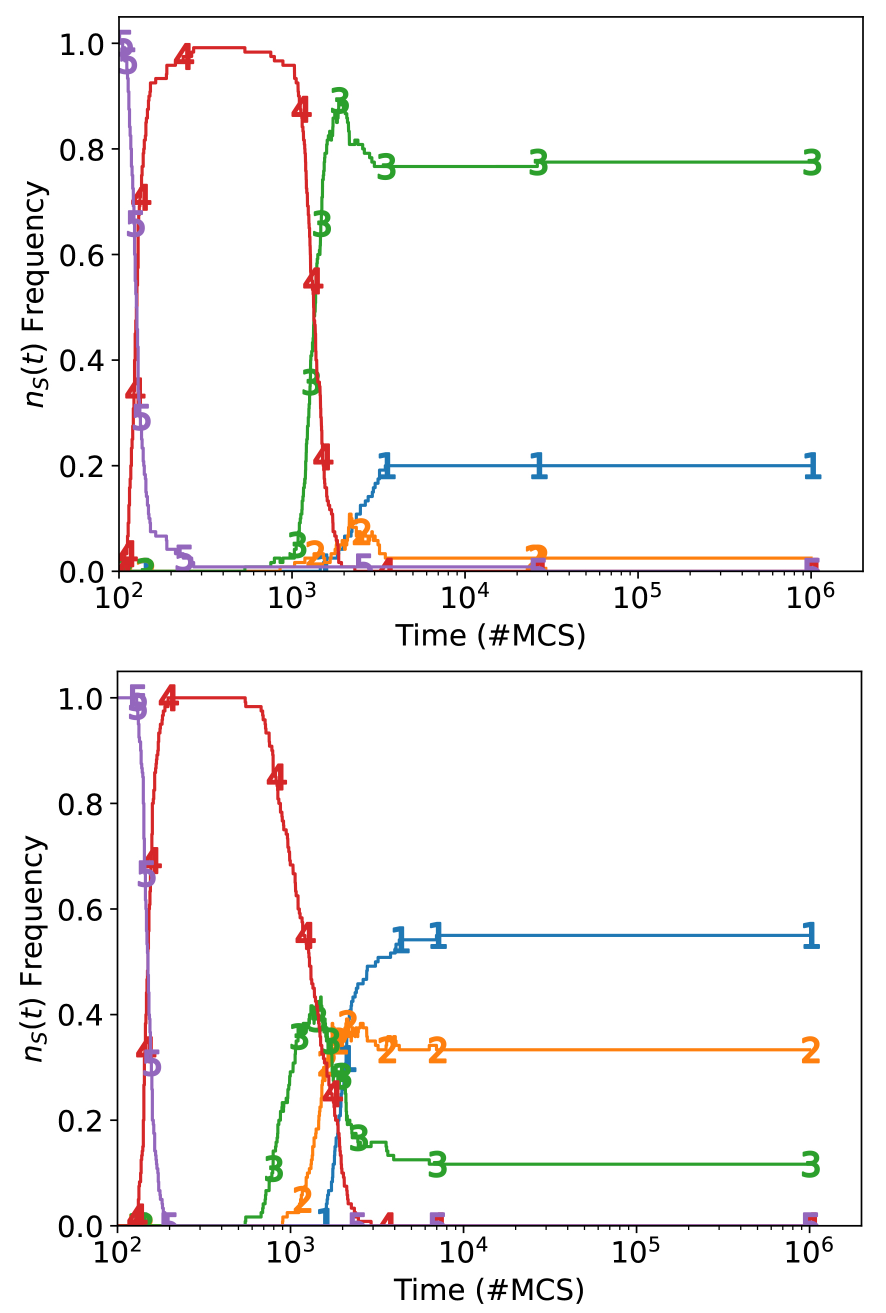}
\caption{
Plots of $FvsT$
for {\sc Rpsls} experiments with three ablations ($N_a
$$
=
$$
3$): $L
$$
=
$$
200$; $\mu
$$
=
$$
\sigma
$$
=
$$
1.0$; 
$M
$$
=
$$
2.51
$$
\times
$$
10^{-5}$ (upper),
$M
$$
=
$$
3.98
$$
\times
$$
10^{-5}$ (lower).
Format as for  Figure~\ref{fig:Z2a_NST_M1e-07}.\label{fig:OES_NS05_NA3_L200_G}}
\end{figure}

\begin{figure} 
\centering \includegraphics[width=0.7\textwidth] {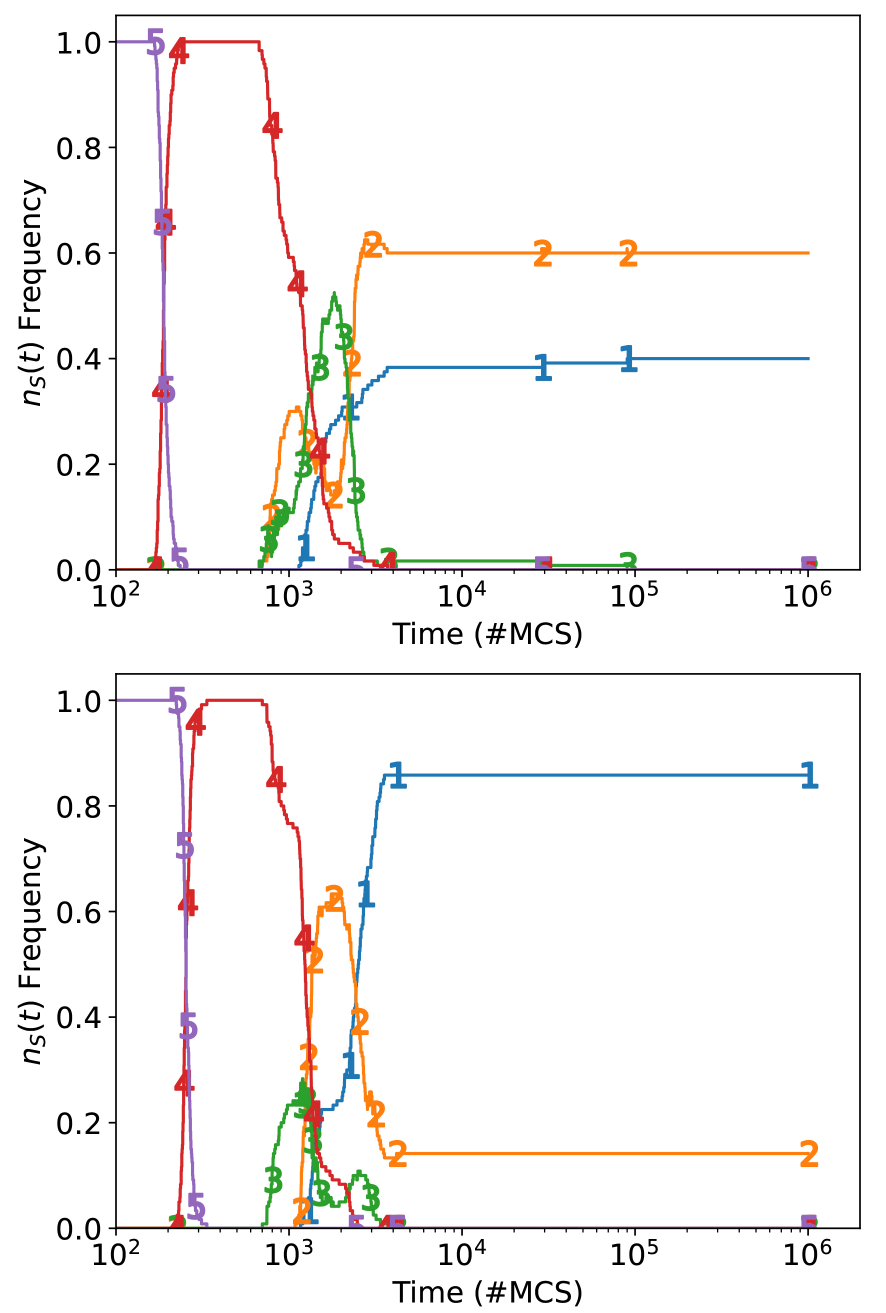}
\caption{
Plots of $FvsT$
for {\sc Rpsls} experiments with three ablations ($N_a
$$
=
$$
3$): $L
$$
=
$$
200$; $\mu
$$
=
$$
\sigma
$$
=
$$
1.0$; 
$M
$$
=
$$
6.31
$$
\times
$$
10^{-5}$ (left),
$M
$$
=
$$
10^{-4}$ (right).
Format as for  Figure~\ref{fig:Z2a_NST_M1e-07}.\label{fig:OES_NS05_NA3_L200_H}}
\end{figure}

\begin{figure} 
\centering \includegraphics[width=0.7\textwidth] {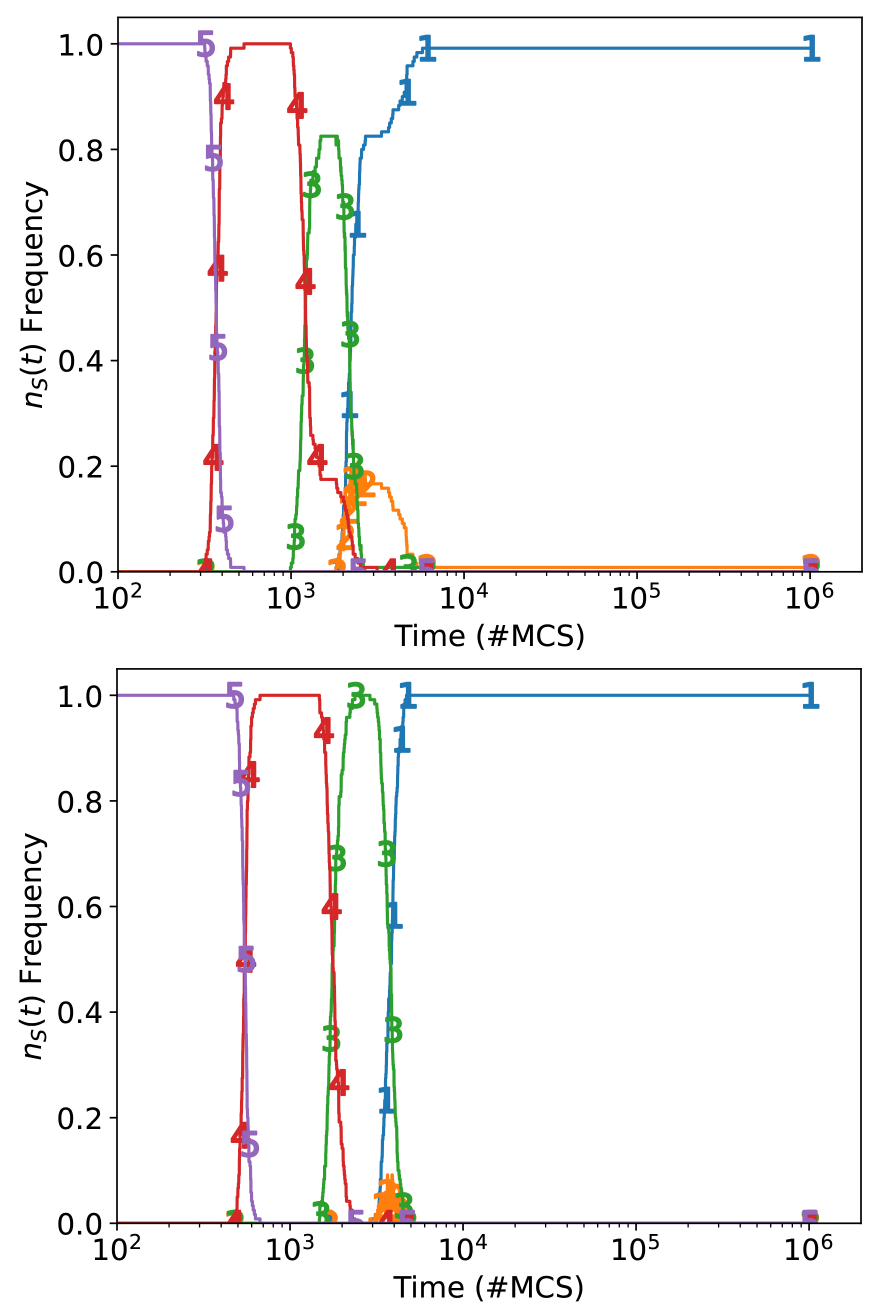}
\caption{
Plots of $FvsT$
for {\sc Rpsls} experiments with three ablations ($N_a
$$
=
$$
3$): $L
$$
=
$$
200$; $\mu
$$
=
$$
\sigma
$$
=
$$
1.0$; 
$M
$$
=
$$
1.58
$$
\times
$$
10^{-4}$ (left),
$M
$$
=
$$
2.51
$$
\times
$$
10^{-4}$ (right).
Format as for  Figure~\ref{fig:Z2a_NST_M1e-07}.\label{fig:OES_NS05_NA3_L200_I}}
\end{figure}

\begin{figure} 
\centering \includegraphics[width=0.7\textwidth] {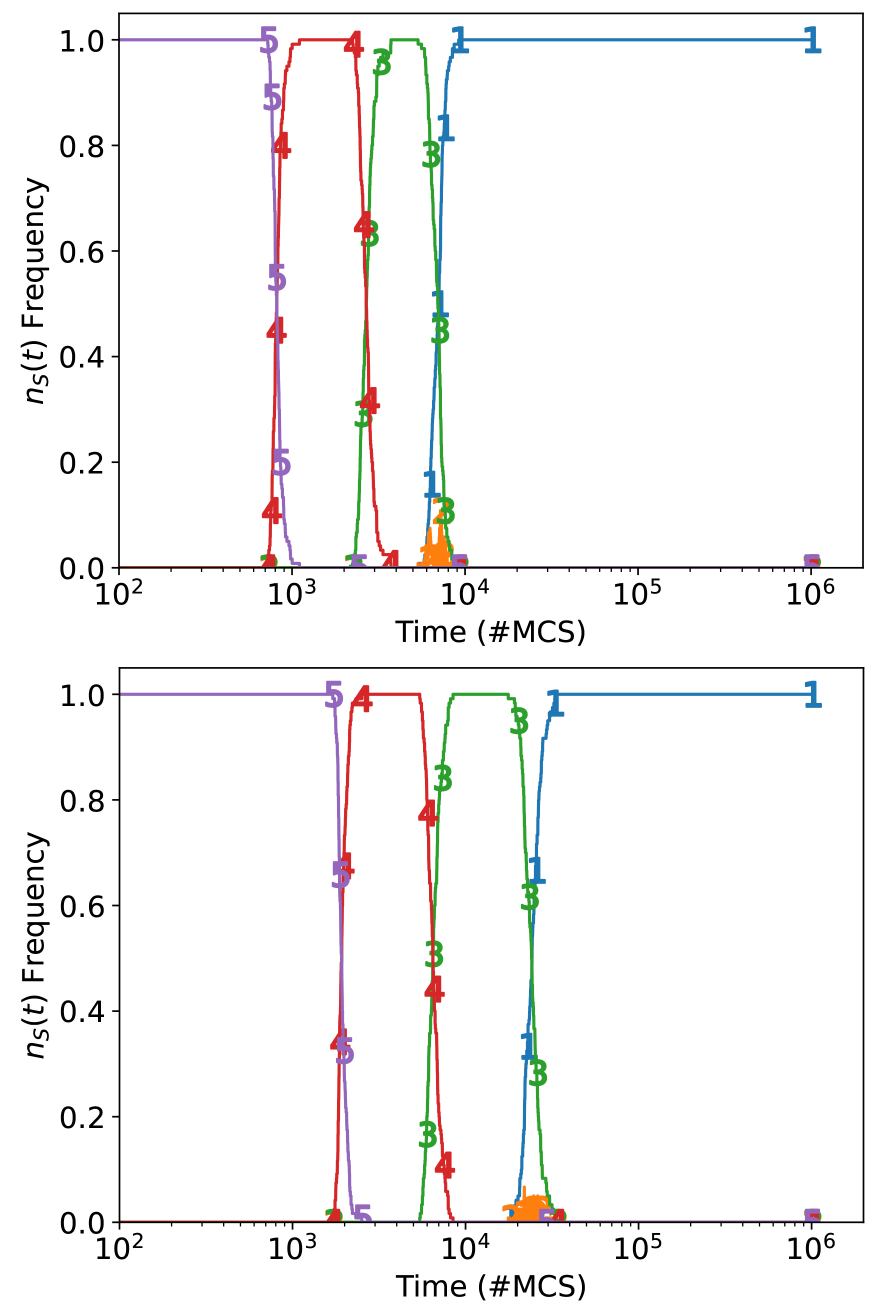}
\caption{
Plots of $FvsT$
for {\sc Rpsls} experiments with three ablations ($N_a
$$
=
$$
3$): $L
$$
=
$$
200$; $\mu
$$
=
$$
\sigma
$$
=
$$
1.0$; 
$M
$$
=
$$
3.98
$$
\times
$$
10^{-4}$ (left),
$M
$$
=
$$
10^{-3}$ (right).
Format as for  Figure~\ref{fig:Z2a_NST_M1e-07}.\label{fig:OES_NS05_NA3_L200_J}}
\end{figure}

 Figures~\ref{fig:OES_NS05_NA3_L200_A} to~\ref{fig:OES_NS05_NA3_L200_J} show $FvsT$ plots of time-series of the evolution of the number of surviving species $n_s(t)$ over $10^6${\sc mcs} for the {\sc Rpsls} {\sc Escg} with the three-ablation ($N_a
$$
=
$$
3$)  dominance network Z3b (as illustrated in  Figure~\ref{fig:ZhongDomNets}), 
$L
$$
=
$$
200$ and $\mu
$$
=
$$
\sigma
$$
=
$$
1.0$ for various values of $M
$$
\in
$$
[ 10^{-7}, 10^{-3}]$, with 120 {\sc iid} simulations executed for each value of $M$ sampled. 

\mbox{}
  
\section[pfx={Appendix\space}]{Dynamics of $N_a
$$
=
$$
4$ {{\sc Rpsls}}
 {{Z4}} with $\mu
$$
=
$$
\sigma
$$
=
$$
1.0$, $L
$$
=
$$
200$}

\label{sec:app_NA4_NST}

\begin{figure}
\centering \includegraphics[width=0.7\textwidth] {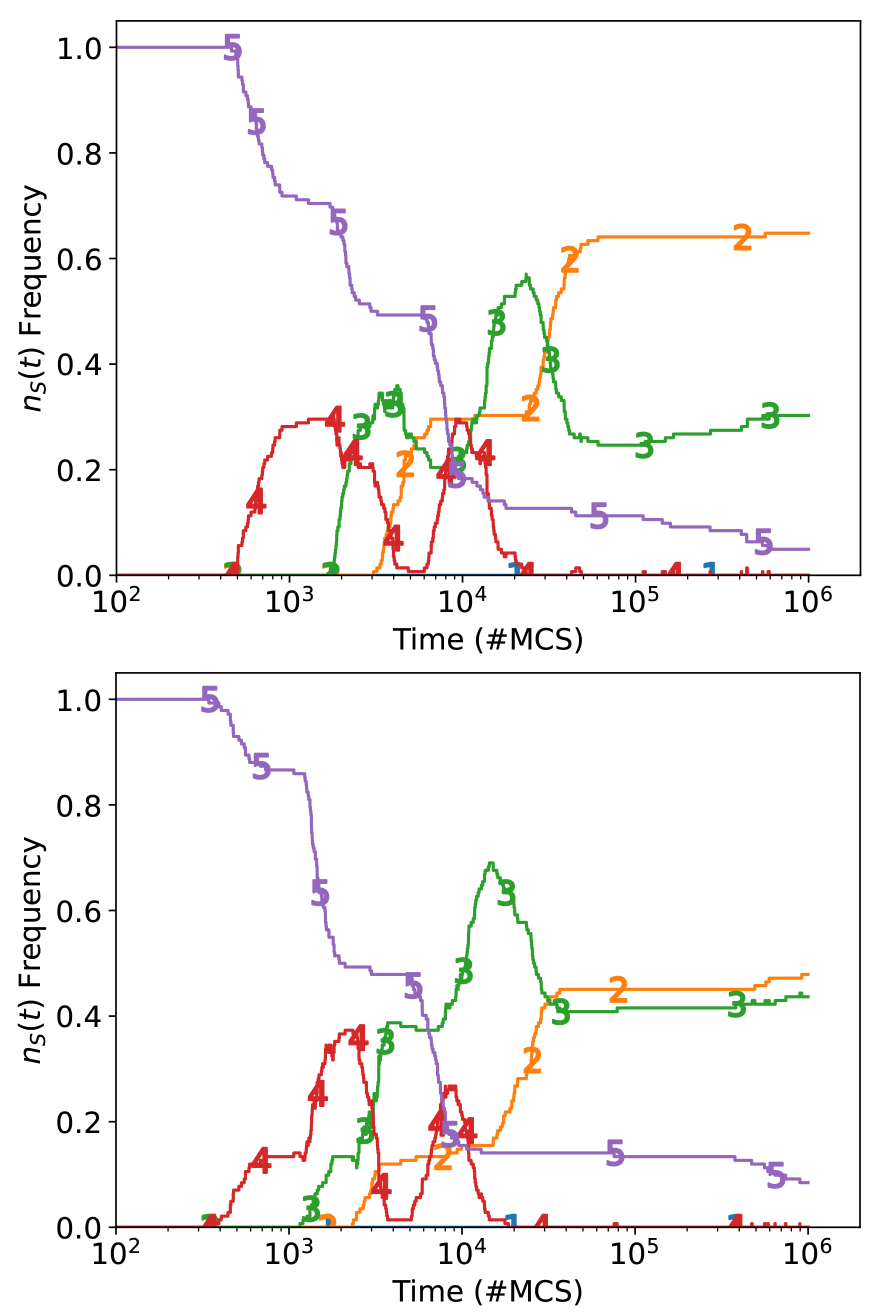}
\caption{
Plots of $FvsT$
for {\sc Rpsls} experiments with four ablations ($N_a
$$
=
$$
4$): $L
$$
=
$$
200$; $\mu
$$
=
$$
\sigma
$$
=
$$
1.0$; 
$M
$$
=
$$
10^{-7}$ (upper),
$M
$$
=
$$
1.58
$$
\times
$$
10^{-7}$ (lower).
Format as for  Figure~\ref{fig:Z2a_NST_M1e-07}.\label{fig:OES_NS05_NA4_L200_A}}
\end{figure}

\begin{figure}
\centering \includegraphics[width=0.7\textwidth] {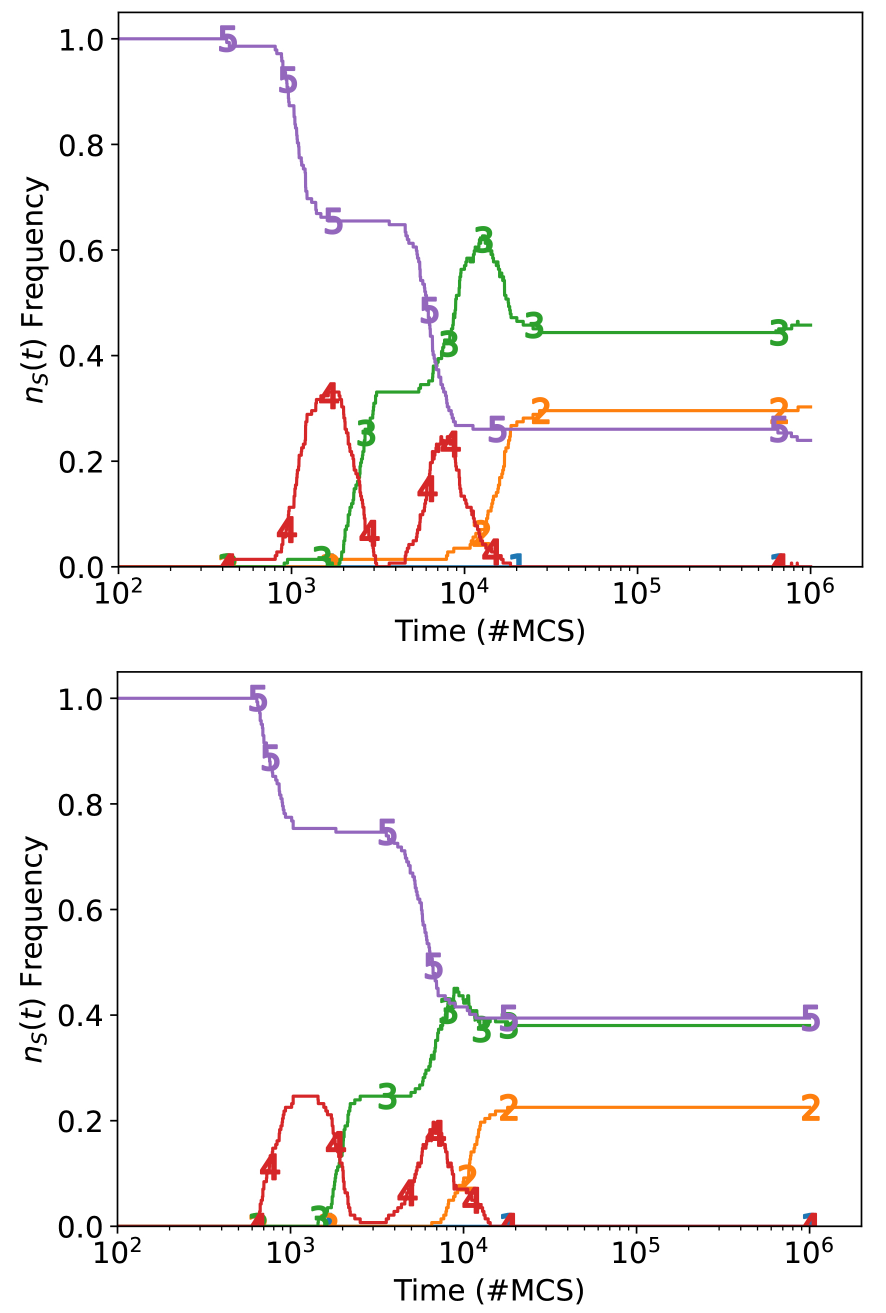}
\caption{
Plots of $FvsT$
for {\sc Rpsls} experiments with four ablations ($N_a
$$
=
$$
4$): $L
$$
=
$$
200$; $\mu
$$
=
$$
\sigma
$$
=
$$
1.0$; 
$M
$$
=
$$
2.51
$$
\times
$$
10^{-7}$ (upper),
$M
$$
=
$$
3.98
$$
\times
$$
10^{-7}$ (lower).
Format as for  Figure~\ref{fig:Z2a_NST_M1e-07}.\label{fig:OES_NS05_NA4_L200_B}}
\end{figure}

\begin{figure}
\centering \includegraphics[width=0.7\textwidth] {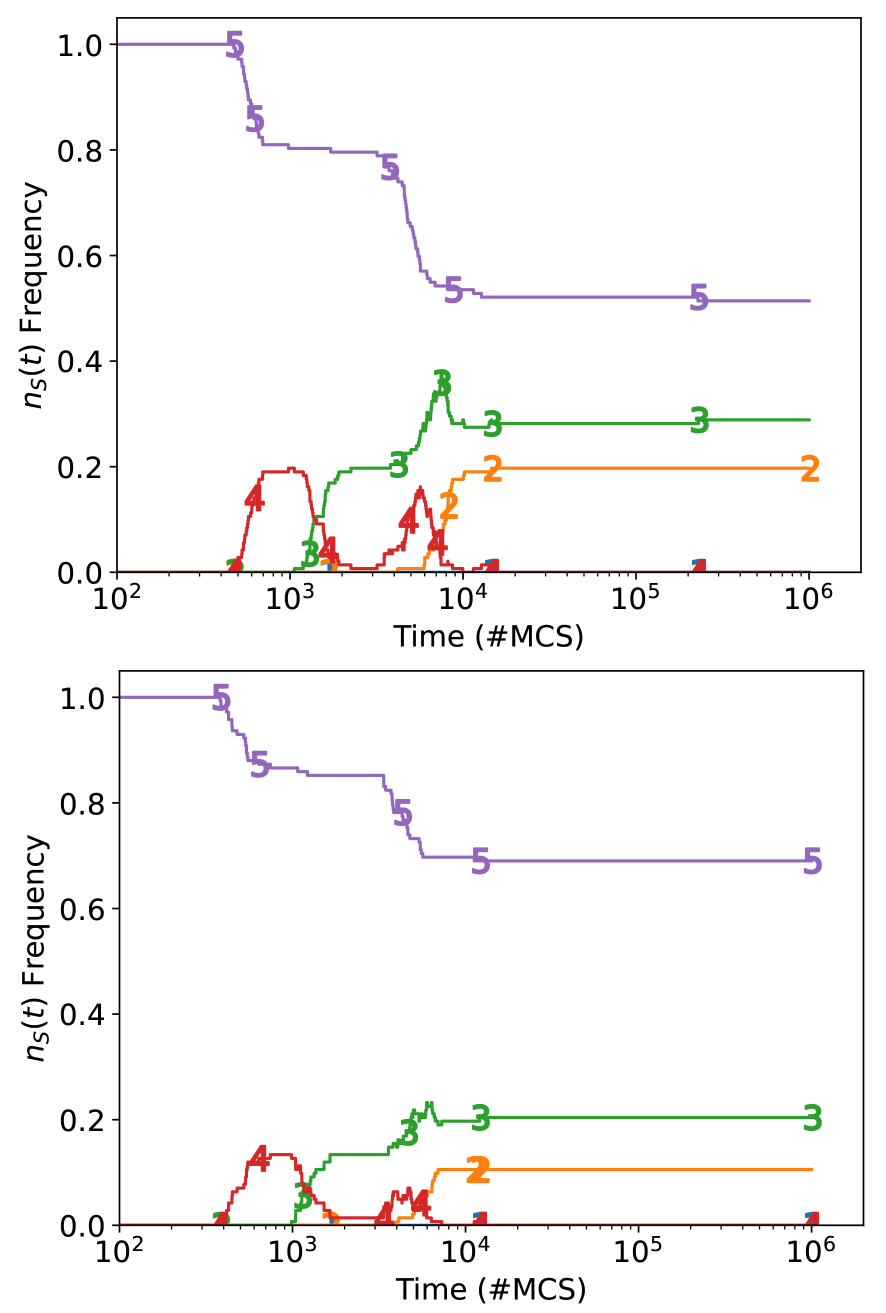}
\caption{
Plots of $FvsT$
for {\sc Rpsls} experiments with four ablations ($N_a
$$
=
$$
4$): $L
$$
=
$$
200$; $\mu
$$
=
$$
\sigma
$$
=
$$
1.0$; 
$M
$$
=
$$
6.31
$$
\times
$$
10^{-7}$ (upper),
$M
$$
=
$$
10^{-6}$ (lower).
Format as for  Figure~\ref{fig:Z2a_NST_M1e-07}.
\label{fig:OES_NS05_NA4_L200_C}}
\end{figure}

\begin{figure}
\centering \includegraphics[width=0.7\textwidth] {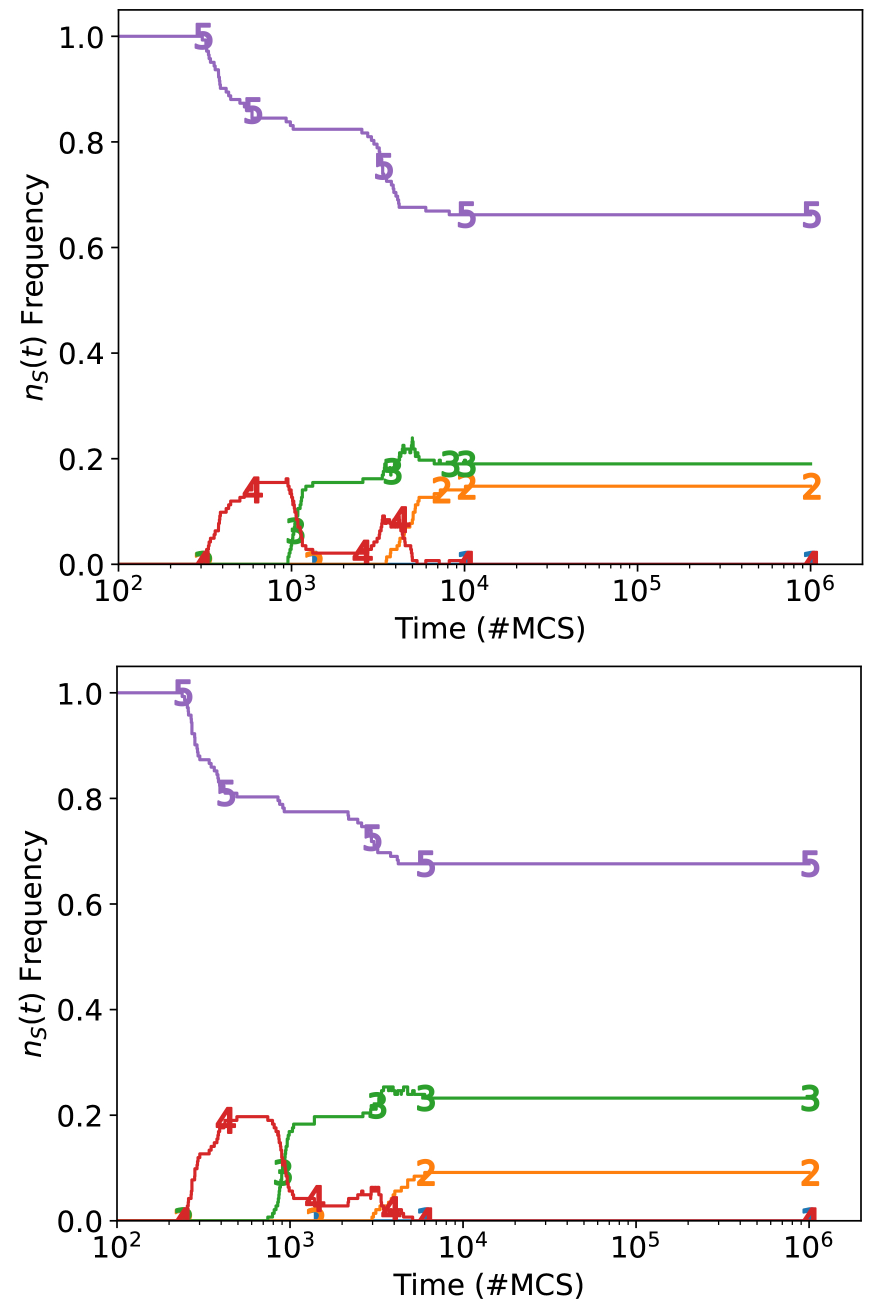}
\caption{
Plots of $FvsT$
for {\sc Rpsls} experiments with four ablations ($N_a
$$
=
$$
4$): $L
$$
=
$$
200$; $\mu
$$
=
$$
\sigma
$$
=
$$
1.0$; 
$M
$$
=
$$
1.58
$$
\times
$$
10^{-6}$ (upper),
$M
$$
=
$$
2.51
$$
\times
$$
10^{-6}$ (lower).
Format as for  Figure~\ref{fig:Z2a_NST_M1e-07}.\label{fig:OES_NS05_NA4_L200_D}}
\end{figure}

\begin{figure}
\centering \includegraphics[width=0.7\textwidth] {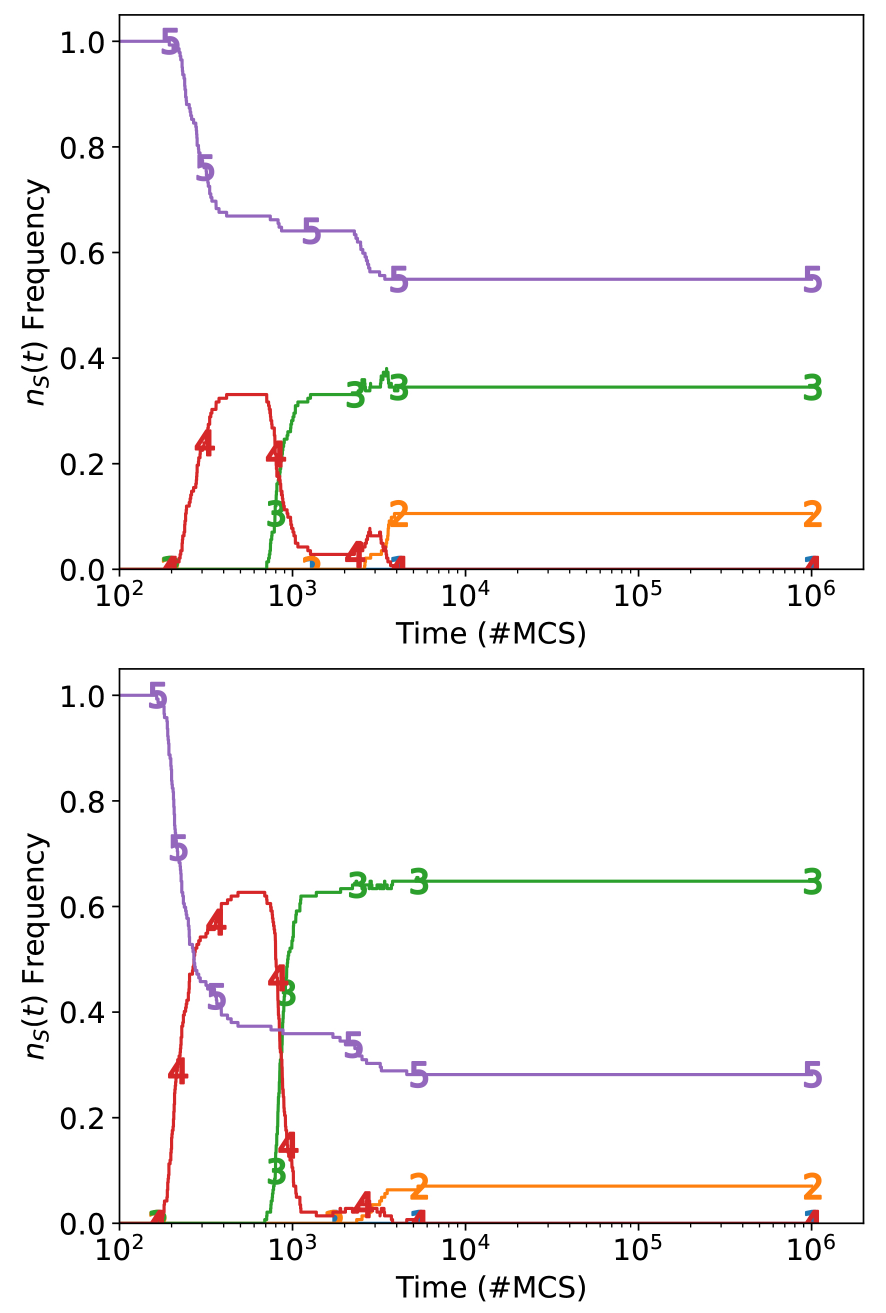}
\caption{
Plots of $FvsT$
for {\sc Rpsls} experiments with four ablations ($N_a
$$
=
$$
4$): $L
$$
=
$$
200$; $\mu
$$
=
$$
\sigma
$$
=
$$
1.0$; 
$M
$$
=
$$
3.98
$$
\times
$$10^{-6}$ (left),
$M$$=6.31$$\times$$10^{-6}$ (lower).
Format as for  Figure~\ref{fig:Z2a_NST_M1e-07}.\label{fig:OES_NS05_NA4_L200_E}}
\end{figure}

\begin{figure}
\centering \includegraphics[width=0.7\textwidth] {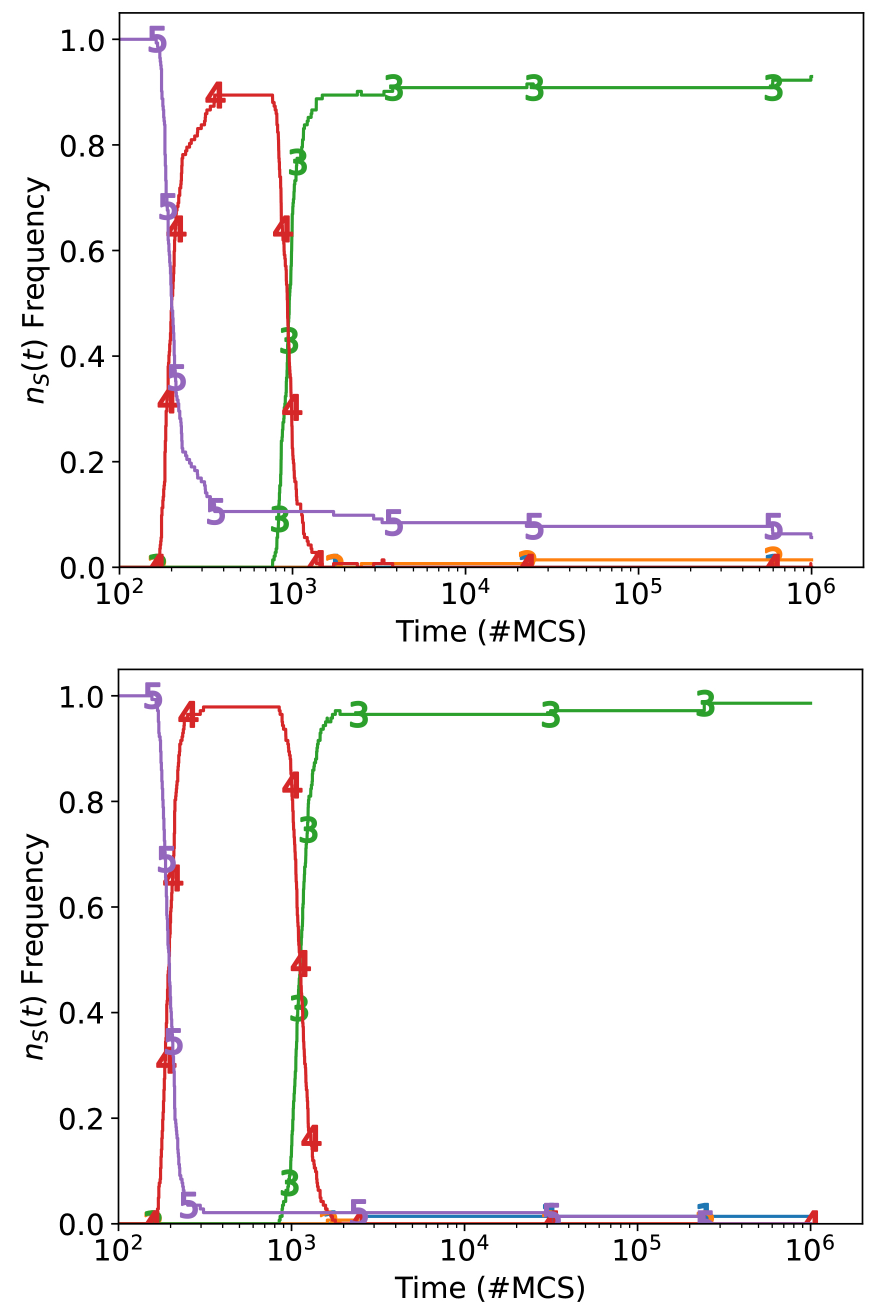}
\caption{
Plots of $FvsT$
for {\sc Rpsls} experiments with four ablations ($N_a$$=4$): $L$$=200$; $\mu$$=\sigma$$=1.0$; 
$M$$=10^{-5}$ (upper),
$M$$=1.58$$\times$$10^{-5}$ (lower).
Format as for  Figure~\ref{fig:Z2a_NST_M1e-07}.\label{fig:OES_NS05_NA4_L200_F}}
\end{figure}

\begin{figure}
\centering \includegraphics[width=0.7\textwidth] {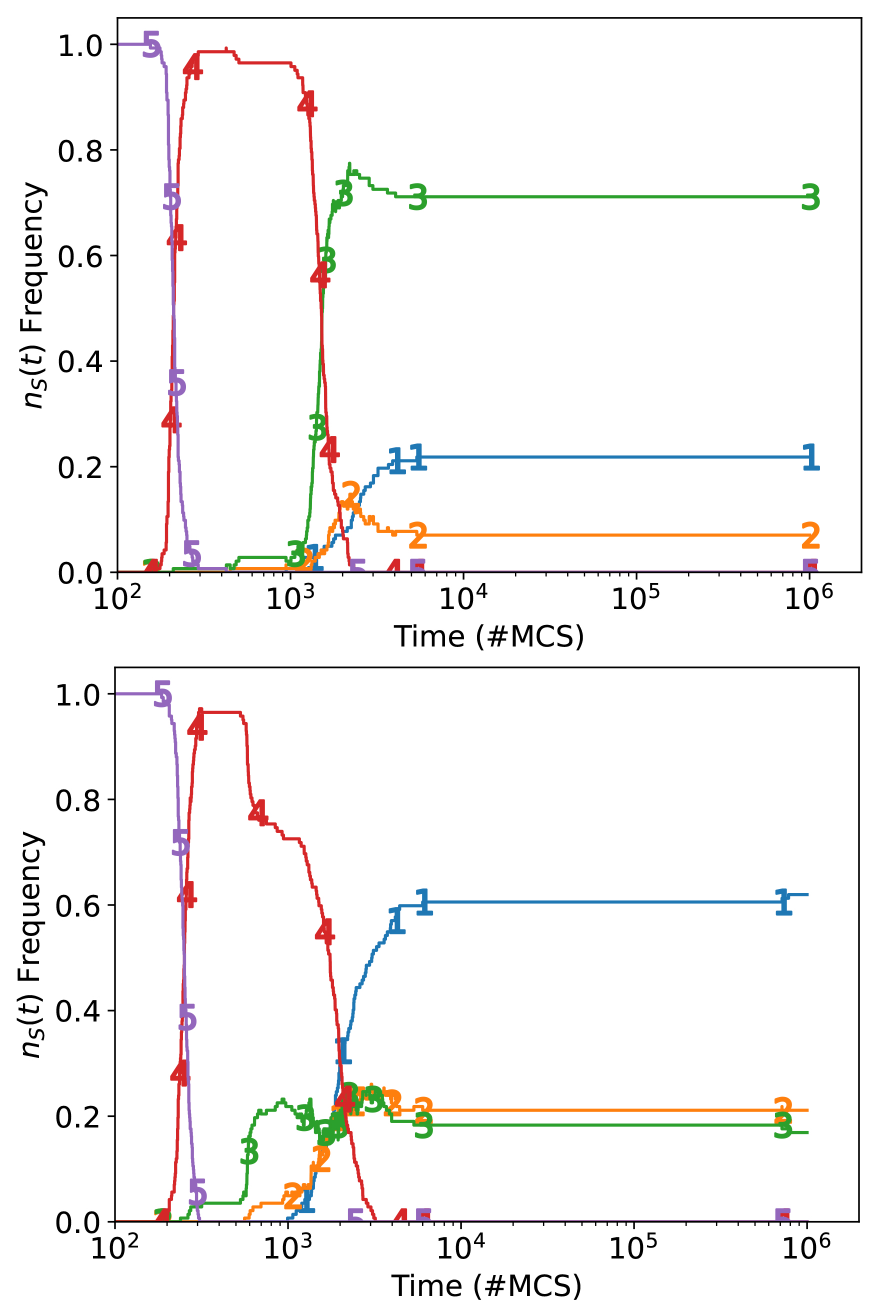}
\caption{
Plots of $FvsT$
for {\sc Rpsls} experiments with four ablations ($N_a$$ = 4$): $L$$ =
200$; $\mu$ $=\sigma$$ = 1.0$; 
$M$$=2.51$$\times$$10^{-5}$ (upper),
$M$$=3.98$$\times$$10^{-5}$ (lower).
Format as for  Figure~\ref{fig:Z2a_NST_M1e-07}.\label{fig:OES_NS05_NA4_L200_G}}
\end{figure}

\begin{figure}
\centering \includegraphics[width=0.7\textwidth] {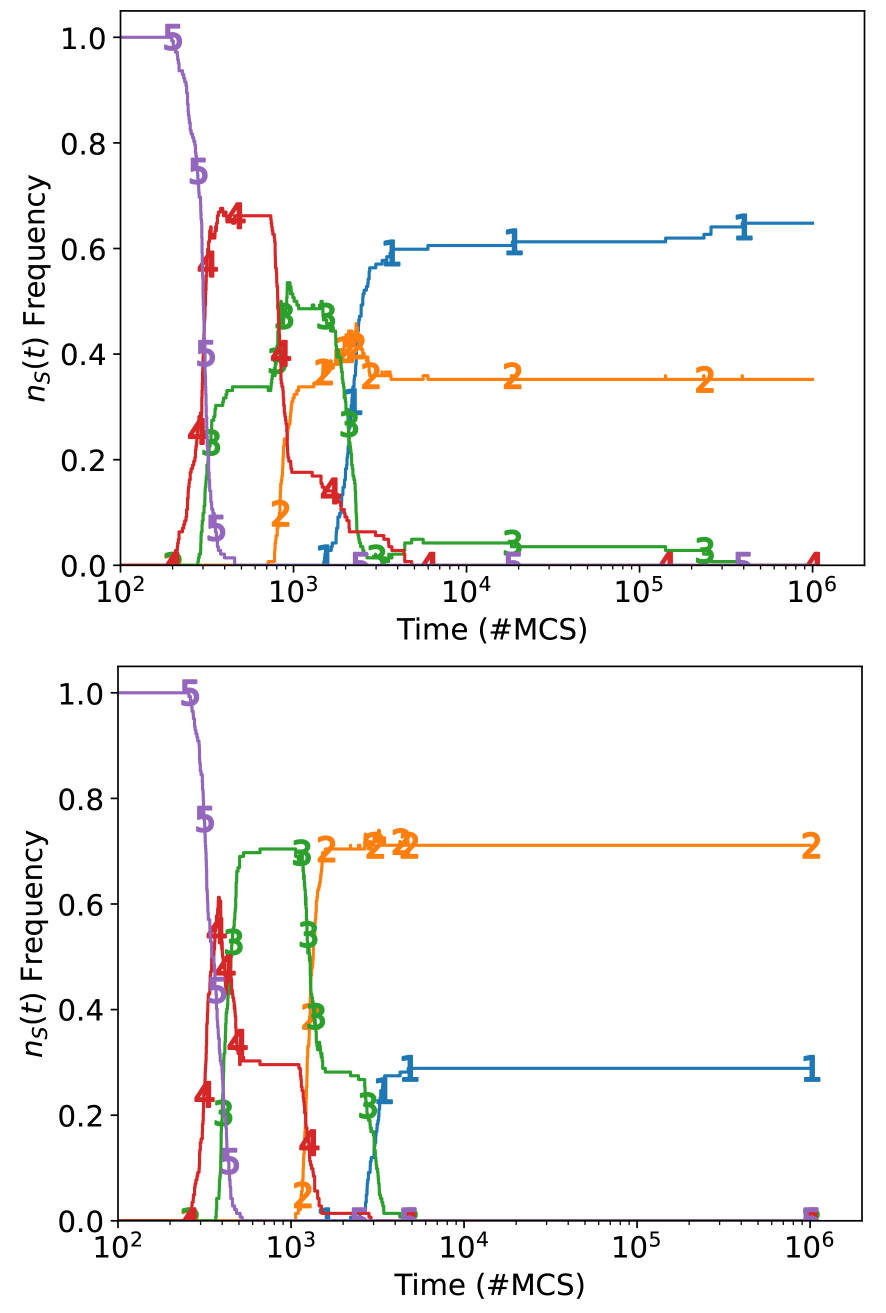}
\caption{
Plots of $FvsT$
for {\sc Rpsls} experiments with four ablations ($N_a$~$=~4$): $L$$~=~200$; $\mu$~$=~\sigma$=$1.0$; 
$M$~$=~6.31$$\times$$10^{-5}$ (upper),
$M$~$=~10^{-4}$ (lower).
Format as for  Figure~\ref{fig:Z2a_NST_M1e-07}.\label{fig:OES_NS05_NA4_L200_H}}
\end{figure}

\begin{figure}
\centering \includegraphics[width=0.7\textwidth] {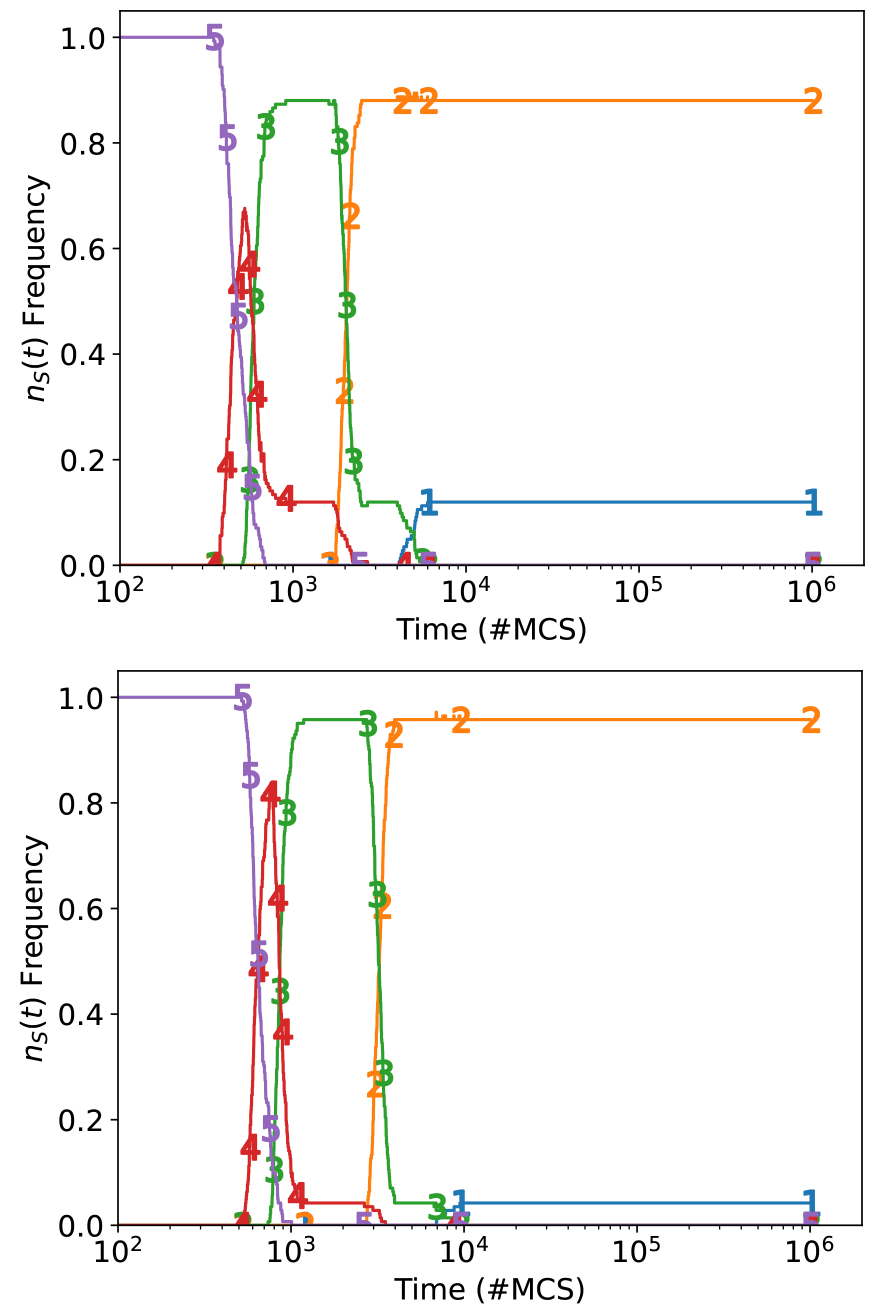}
\caption{
Plots of $FvsT$
for {\sc Rpsls} experiments with four ablations ($N_a$$=4$): $L$$=200$; $\mu$$=\sigma$$=1.0$; 
$M$$=1.58$$\times$$10^{-4}$ (upper),
$M$$=2.51$$\times$$10^{-4}$ (lower).
Format as for  Figure~\ref{fig:Z2a_NST_M1e-07}.\label{fig:OES_NS05_NA4_L200_I}}
\end{figure}

\begin{figure}
\centering \includegraphics[width=0.7\textwidth] {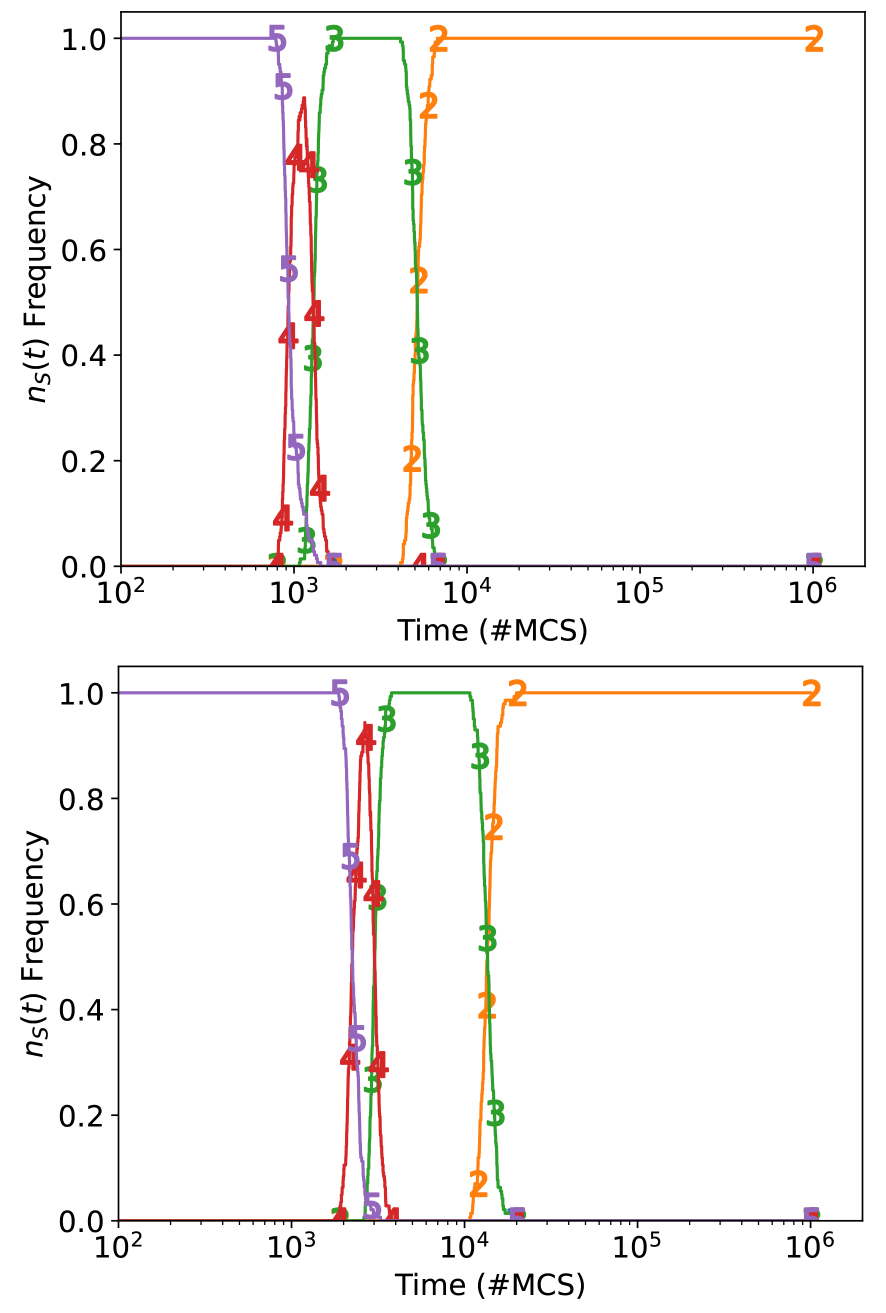}
\caption{
Plots of $FvsT$
for {\sc Rpsls} experiments with four ablations ($N_a$$=4$): $L$$=200$; $\mu$$=\sigma$$=1.0$; 
$M$$=3.98$$\times$$10^{-4}$ (upper),
$M$$=10^{-3}$ (lower).
Format as for  Figure~\ref{fig:Z2a_NST_M1e-07}.\label{fig:OES_NS05_NA4_L200_J}}
\end{figure}

Figures~\ref{fig:OES_NS05_NA4_L200_A} to~\ref{fig:OES_NS05_NA4_L200_J} show $FvsT$ plots of time-series of the evolution of the number of surviving species $n_s(t)$ over $10^6${\sc mcs} for the {\sc Rpsls} {\sc Escg} with four-ablation ($N_a
$$
=
$$
4$) 
dominance network Z4 (as illustrated in  Figure~\ref{fig:ZhongDomNets}),
$L
$$
=
$$
200$ and $\mu
$$
=
$$
\sigma
$$
=
$$
1.0$ for various values of $M
$$
\in
$$
[ 10^{-7}, 10^{-3}]$, with 150 {\sc iid} simulations executed for each value of $M$ sampled.

\clearpage
\newpage

\bibliographystyle{elsarticle-num} 

\bibliography{../../dc_bibliography}

\begin{thebibliography}{10}
\expandafter\ifx\csname url\endcsname\relax
  \def\url#1{\texttt{#1}}\fi
\expandafter\ifx\csname urlprefix\endcsname\relax\def\urlprefix{URL }\fi
\expandafter\ifx\csname href\endcsname\relax
  \def\href#1#2{#2} \def\path#1{#1}\fi

\bibitem{zhong_etal_2022_ablatedRPSLS}
L.~Zhong, L.~Zhang, H.~Li, Q.~Dai, J.~Yang, Species coexistence in spatial
  cyclic game of five species, Chaos, Solitons, and Fractals 156~(111806)
  (2022).

\bibitem{laird_schamp_2015}
R.~Laird, B.~Schamp, Competitive intransitivity, population interaction
  structure, and strategy coexistence, Journal of Theoretical Biology 365
  (2015) 149--158.

\bibitem{zhang_bearup_guo_zhang_liao_2022}
Z.~Zhang, D.~Bearup, G.~Guo, H.~Zhang, J.~Liao, Competition modes determine
  ecosystem stability in rock–paper–scissors games, Physica A 602~(128176)
  (2022).

\bibitem{cliff_2024_noops}
D.~Cliff, {Never Mind The No-Ops: Faster and Less Volatile Simulation Modelling
  of Co-Evolutionary Species Interactions via Spatial Cyclic Games}, in:
  {Proceedings of the 36th European Modelling and Simulation Symposium
  (EMSS2024)}, 2024.
\newblock \href {https://doi.org/DOI:
  https://doi.org/10.46354/i3m.2024.emss.011} {\path{doi:DOI:
  https://doi.org/10.46354/i3m.2024.emss.011}}.

\bibitem{reichenbach_mobilia_frey_2007_nature}
T.~Reichenbach, M.~Mobilia, E.~Frey, {Mobility promotes and jeopardizes
  biodiversity in rock-paper-scissors games}, Nature 448~(06095) (2007).

\bibitem{reichenbach_mobilia_frey_2007_physrevlet}
T.~Reichenbach, M.~Mobilia, E.~Frey, {Noise and Correlations in a Spatial
  Population Model with Cyclic Competition}, Physical Review Letters
  99~(238105) (2007).

\bibitem{reichenbach_mobilia_frey_2008_jtb}
T.~Reichenbach, M.~Mobilia, E.~Frey, {Self-Organization of Mobile Populations
  in Cyclic Competition}, Journal of Theoretical Biology 254~(2008) (2008)
  363--383.

\bibitem{may_leonard_1975}
R.~May, W.~Leonard, Nonlinear aspects of competition between species, {SIAM
  Journal of Applied Mathematics} 29 (1975) 243--253.

\bibitem{wolfram_2002_book}
S.~Wolfram, {A New Kind of Science}, Wolfram Media, 2002.

\bibitem{laird_schamp_2006_RPSLS}
R.~Laird, B.~Schamp, Competitive intransitivity promotes species coexistence,
  American Naturalist 168 (2006) 182--193.

\bibitem{laird_schamp_2008_RPSLS}
R.~Laird, B.~Schamp, Does local competition increase the coexistence of species
  in intransitive networks, Ecology 89 (2008) 237--247.

\bibitem{laird_schamp_2009_coexistence_RPSLS}
R.~Laird, B.~Schamp, Species coexistence, intransitivity, and topological
  variation in competitive tournaments, J.\ Theoretical Biology 256 (2009)
  90--95.

\bibitem{avelino_bazeia_losano_menezes_oliveira_2022}
P.~Avelino, D.~Bastia, L.~Logan, J.~Menezes, B.~{de Oliveria}, Junctions and
  spiral patterns in generalized rock-paper-scissors models, Physical Review E
  86~(036112) (2012).

\bibitem{nagatani_ichinose_tainaka_2018_RPS}
T.~Nagatani, G.~Ichinose, K.~Tainaka, Metapopulation model for
  rock–paper–scissors game: Mutation affects paradoxical impacts, Journal
  of Theoretical Biology 450~(22--29) (2018).

\bibitem{kabir_tanimoto_2021_RPS}
K.~Kabir, J.~Tanimoto, The role of pairwise nonlinear evolutionary dynamics in
  the rock–paper–scissors game with noise, Applied Mathematics and
  Computation 394~(125767) (2021).

\bibitem{mohd_park_2021_RPS}
M.~Mood, J.~Park, The interplay of rock-paper-scissors competition and
  environments mediates species coexistence and intriguing dynamics, Chaos,
  Solitons and Fractals 153~(111579) (2021).

\bibitem{bazeia_bongestab_deoliveira_2022_RPS}
D.~Bazeia, M.~Bongestab, B.~{de Oliveira}, Influence of the neighborhood on
  cyclic models of biodiversity, Physica A 587~(126547) (2022).

\bibitem{park_2021_RPS}
J.~Park, Evolutionary dynamics in the rock-paper-scissors system by changing
  community paradigm with population flow, Chaos, Solitons and Fractals
  142~(110424) (2021).

\bibitem{menezes_batista_rangel_2022_RPS}
J.~Menezes, S.~Batista, E.~Rangel, Spatial organisation plasticity reduces
  disease infection risk in rock–paper–scissors models, Biosystems
  221~(104777) (2022).

\bibitem{menezes_rangel_moura_2022_RPS}
J.~Menezes, E.~Rangel, B.~Moura, Aggregation as an antipredator strategy in the
  rock-paper-scissors model, Ecological Informatics 69~(101606) (2022).

\bibitem{zhang_bearup_guo_zhang_liao_2022_RPS}
Z.~Zhang, D.~Bearup, G.~Guo, H.~Zhang, J.~Liao, Competition modes determine
  ecosystem stability in rock–paper–scissors games, Physica A 607~(128176)
  (2022).

\bibitem{menezes_barbalho_2023_RPS}
J.~Menezes, S.~Rodrigues, S.~Batista, Mobility unevenness in
  rock–paper–scissors models, Ecological Complexity 52~(101028) (2023).

\bibitem{park_jang_2023_RPS}
J.~Park, B.~Jang, Role of adaptive intraspecific competition on collective
  behavior in the rock–paper–scissors game, Chaos, Solitons and Fractals
  171~(113448) (2023).

\bibitem{kubyana_landi_hui_2024_RPS}
M.~Kubyana, P.~Landi, C.~Hui, Adaptive rock-paper-scissors game enhances
  eco-evolutionary performance at cost of dynamic stability, Applied
  Mathematics and Computation 468~(128535) (2024).

\bibitem{kass_bryla_1998}
S.~Kass, K.~Bryla, {Rock Paper Scissors Spock Lizard},
  \url{samkass.com/theories/RPSSL.html}, accessed 2024-05-12 (1998).

\bibitem{avelino_deoliveria_trintin_2022_RPS_bigN}
P.~Avelino, B.~{de Oliveria}, R.~Trintin, Parity effects in rock-paper-scissors
  type models with a number of species ns $\leq$ 12, Chaos, Solitons and
  Fractals 155~(111738) (2022).

\bibitem{park_jang_2019}
J.~Park, B.~Jang, Robust coexistence with alternative competition strategy in
  the spatial cyclic game of five species, Chaos 29~(051105) (2019).

\bibitem{nowak_may_1992}
M.~Nowak, R.~May, Evolutionary games and spatial chaos, Nature 359 (1992)
  826--–829.

\bibitem{huberman_glance_1993}
B.~Huberman, N.~Glance, Evolutionary games and computer simulations,
  Proceedings of the National Academy of Sciences 90 (1993) 7716--7718.

\bibitem{axelrod_1984}
R.~Axelrod, The Evolution of Cooperation, Basic Books, 1984.

\bibitem{axelrod_1997}
R.~Axelrod, Complexity of Cooperation: Agent-Based Models of Competition and
  Collaboration, Princeton University Press, 1997.

\bibitem{cliff_2024_RES_ablatedRPSLS}
D.~Cliff, {The Devil Lies in The Detail for Species Coexistence Stability in
  Ablated and Unablated Five-Species Evolutionary Spatial Cyclic Games},
  Manuscript submitted for peer-review (2024).

\bibitem{cliff_2024_circulants}
D.~Cliff, {Tournament versus Circulant: On Simulating 7-Species Evolutionary
  Spatial Cyclic Games with Ablated Predator-Prey Networks as Models of
  Biodiversity}, in: {Proceedings of the 36th European Modelling and Simulation
  Symposium (EMSS2024)}, 2024.
\newblock \href {https://doi.org/https://doi.org/10.46354/i3m.2024.emss.028}
  {\path{doi:https://doi.org/10.46354/i3m.2024.emss.028}}.

\bibitem{bloom_cliff_2024}
J.~Bloom, D.~Cliff, {Exploring Biodiversity through Evolutionary Spatial Cyclic
  Games: Tournament versus Non-Tournament Circulant Dominance Networks, With
  and Without Ablations, for Five Species and More}, Manuscript in preparation
  (2024).

\bibitem{dong_li_yang_2010}
L.-R. Dong, Y.-M. Li, G.-C. Yang, Coexistence and extinction pattern of
  asymmetric cyclic game species in a square lattice, Communications in
  Theoretical Physics (Bejing, China) 53 (2010) 1201--1204.

\bibitem{maxwell_lau_howard_2015}
S.~Maxwell, M.~Lau, G.~Howard, {Is Psychology Suffering from a Replication
  Crisis? {What} Does `Failure to Replicate' Really Mean?}, American
  Psychologist 70~(6) (2015) 487--498.

\bibitem{milkowski_hensel_hohol_2018}
M.~Milkowski, W.~Hensel, M.~Hohol, {Replicability or reproducibility? {O}n the
  replication crisis in computational neuroscience and sharing only relevant
  detail}, Journal of Computational Neuroscience 45 (2018) 163--–172.

\bibitem{moody_keister_ramos_2022}
J.~Moody, L.~Keister, M.~Ramos, {Reproducibility in the Social Sciences},
  Annual Review of Sociology 48 (2022) 65--85.

\bibitem{nosek_etal_2022}
B.~Nosek, T.~Hardwicke, H.~Moshontz, A.~Allard, K.~Corker, A.~Dreber,
  F.~Fidler, J.~Hilgard, M.~{Kilne Struhl}, M.~Nuijten, J.~Roher, F.~Romero,
  A.~Scheel, L.~Scherer, F.~Schonbrodt, S.~Vazire, {Replicability, Robustness,
  and Reproducibility in Psychological Science}, Annual Review of Psychology 73
  (2022) 719--748.

\bibitem{jensen_kelly_pedersen_2023}
T.~Jensen, B.~Kelly, L.~Pedersen, {Is There a Replication Crisis in Finance?},
  The Journal of Finance 78~(5) (2023) 2465--2518.

\bibitem{smith_2016_justso}
R.~J. Smith, Explanations for adaptations, just-so stories, and limitations on
  evidence in evolutionary biology, Evolutionary Anthropology 25~(6) (2016)
  276--287.

\end{thebibliography}

\end{document}